\documentclass[aip,jcp,twocolumn,superscriptaddress,longbibliography,10pt]{revtex4-1}

\usepackage{graphicx}
\graphicspath{ {./Figures/} }

\newcommand{\matX}[4]{\ensuremath{\left(\begin{array}{c}#1\\#3\end{array}\begin{array}{c}#2\\#4\end{array}\right)}}

\usepackage{amsmath}
\usepackage{upgreek}
\usepackage{textcomp}

\usepackage{afterpage}

\usepackage{tikz}
\newcommand{\tikzline}[2][red,solid]{\tikz[baseline=-0.5ex]{\draw[#1,line width=#2](0,0) -- (5mm,0);}}%
\newcommand{\tikzsquare}[2][red,fill=red]{\tikz[baseline=-0.2ex]\draw[#1] (0,0) rectangle ++(2*#2,2*#2) ;}%

\usepackage{xcolor}
\definecolor{Light-gray}{gray}{0.95}
\definecolor{Dark-gray}{gray}{0.90}
\definecolor{color1}{rgb}{0.9763, 0.9831, 0.0538}
\definecolor{color2}{rgb}{0.82654285715, 0.732024999975, 0.34635}
\definecolor{color3}{rgb}{0.19898095235000002, 0.72135, 0.63084761905}
\definecolor{color4}{rgb}{0.078996428525, 0.5160142857000001, 0.83297857145}
\definecolor{color5}{rgb}{0.2081, 0.1663, 0.5292}

\newcommand{\bvec}[1]{{\mathbf{\string#1} }}

\begin{document}

\title{\large Particle-resolved topological defects of smectic colloidal liquid crystals in extreme confinement}

\affiliation{Institut f\"ur Theoretische Physik II: Weiche Materie, Heinrich-Heine-Universit\"at D\"usseldorf, D-40225 D\"usseldorf, Germany}
\affiliation{Department of Chemistry, Physical and Theoretical Chemistry Laboratory, University of Oxford, South Parks Road, Oxford OX1 3QZ, United Kingdom}
\affiliation{\mbox{School of Applied and Engineering Physics, Cornell University,~Ithaca,~NY~14853,~USA}}
\affiliation{These two authors contributed equally: Ren\'{e} Wittmann and Louis B.\ G.\ Cortes}
\affiliation{Correspondence and requests for materials should be addressed to R.W.\ (email: rene.wittmann@hhu.de), to H.L.\ (email: Hartmut.Loewen@uni-duesseldorf.de) or to D.G.A.L.A.\ (email: dirk.aarts@chem.ox.ac.uk)}

\author{Ren\'{e} Wittmann}
\affiliation{Institut f\"ur Theoretische Physik II: Weiche Materie, Heinrich-Heine-Universit\"at D\"usseldorf, D-40225 D\"usseldorf, Germany}
\affiliation{These two authors contributed equally: Ren\'{e} Wittmann and Louis B.\ G.\ Cortes}
\affiliation{Correspondence and requests for materials should be addressed to R.W.\ (email: rene.wittmann@hhu.de), to H.L.\ (email: Hartmut.Loewen@uni-duesseldorf.de) or to D.G.A.L.A.\ (email: dirk.aarts@chem.ox.ac.uk)}

\author{Louis B.\ G.\ Cortes}
\affiliation{Department of Chemistry, Physical and Theoretical Chemistry Laboratory, University of Oxford, South Parks Road, Oxford OX1 3QZ, United Kingdom}
\affiliation{\mbox{School of Applied and Engineering Physics, Cornell University,~Ithaca,~NY~14853,~USA}}
\affiliation{These two authors contributed equally: Ren\'{e} Wittmann and Louis B.\ G.\ Cortes}

\author{Hartmut L\"owen}
\affiliation{Institut f\"ur Theoretische Physik II: Weiche Materie, Heinrich-Heine-Universit\"at D\"usseldorf, D-40225 D\"usseldorf, Germany}
\affiliation{Correspondence and requests for materials should be addressed to R.W.\ (email: rene.wittmann@hhu.de), to H.L.\ (email: Hartmut.Loewen@uni-duesseldorf.de) or to D.G.A.L.A.\ (email: dirk.aarts@chem.ox.ac.uk)}

\author{Dirk G.\ A.\ L.\ Aarts}
\affiliation{Department of Chemistry, Physical and Theoretical Chemistry Laboratory, University of Oxford, South Parks Road, Oxford OX1 3QZ, United Kingdom}
\affiliation{Correspondence and requests for materials should be addressed to R.W.\ (email: rene.wittmann@hhu.de), to H.L.\ (email: Hartmut.Loewen@uni-duesseldorf.de) or to D.G.A.L.A.\ (email: dirk.aarts@chem.ox.ac.uk)}


\begin{abstract}
 Confined samples of liquid crystals are characterized by a variety of topological defects and can be
exposed to external constraints such as extreme confinements with nontrivial topology.
 Here we explore the intrinsic structure of smectic colloidal layers dictated by the interplay between entropy and an imposed external topology. Considering an annular confinement as a basic example,
 a plethora of competing states is found with nontrivial defect structures ranging from laminar states to  multiple smectic domains and arrays
 of edge dislocations which we refer to as Shubnikov states in formal analogy to the characteristic of type-II superconductors. 
Our particle-resolved results, gained by a combination of real-space microscopy of thermal colloidal rods and fundamental-measure-based density functional theory
of hard anisotropic bodies, agree on a quantitative level.
\end{abstract}

\maketitle


\section{Introduction}

Liquid crystals consist of particles that possess both translational and orientational degrees of freedom
and exhibit a wealth of mesophases with partial orientational or positional order
such as nematic, smectic and columnar states \cite{gennes1995physics}. As such, these phases are highly susceptible to
 external topological and geometrical influences\cite{LeferinkopReinink2013}.
This opens a fascinating new research realm on the internal structural response to such externally imposed constraints
with various highly relevant applications 
 in technology and material science \cite{Lagerwall2012ANE,katoREV2018}.
While these perspectives have been extensively exploited for spatially homogeneous mesophases, such as nematics,
there is much yet undisclosed potential stemming from the complex interplay between external constraints and internal order
emerging in more complex mesophases, such as the layered smectic phase.

One of the main research goals in liquid crystals is focused on topological defects.
These not only represent fingerprints of singularities and discontinuities in the ordering
but also naturally link topology to condensed matter physics.
The general importance of defects of liquid crystals is further fueled by 
the possibility to directly visualize the inherent orientational frustration on the macroscopic scale 
through the schlieren texture between two crossed polarizers \cite{gennes1995physics}.
Different orientational defect structures have therefore been explored a lot in the homogeneous nematic phase 
\cite{kamienREVnem,delasheras2009,delasheras2014,lewis2014colloidal,garlea2016finite,shin2008,lopez2011,li2016,fialho2017,yao2018,walters2019_machine,yaochen2020,kil2020,ravnik2020,urbanskiREV2017} 
with recent digressions to active systems \cite{baer2020REVSPR,keber2014,activenem2}.
Due to their simultaneous orientational and positional ordering,
defects in the smectic phase naturally exhibit an even higher degree of complexity.
 The main emphasis has been put hitherto on the positional layering \cite{poenaru1981,mosna2012,xing2012,machon2019,repula2018} or  
 orientational textures \cite{liang2011,lopez2012,urbanskiREV2017} alone, as well as, on coarse-grained calculations \cite{gennes1995physics,de1972analogy,xing2012,lopez2012,pevnyi2014,urbanskiREV2017,tang2020} 
and computer simulation of particle models \cite{garlea2014square,geigenfeind2015confinement,schlotthauer2017,allahyarov2017}.

Here we approach the defect structure of smectic liquid crystals from the
most fundamental particle-resolved scale and quantify both their positional and orientational disorder simultaneously in theory and experimnent.
In doing so we study two-dimensional smectics composed of lyotropic colloidal rods whose size enables a direct observation \cite{kuijk2011,kuijk2012phase,cortes2016colloidal}, while they have the advantage over granulates \cite{galanisNEMATICcircularGRANULATE2010,grannulus} 
that they are fully thermally equilibrated. The colloidal samples are exposed to extreme confinements possessing an annular shape and dimensions of a few particle lengths.
This combination of curved geometry and nontrivial topology is triggering certain characteristic defect patterns.
In our study we uniquely combine real-space microscopy of colloidal samples with modern first-principles density functional theory (DFT) \cite{Evans1979}
based on geometric fundamental measures \cite{Rosenfeld1989} which provides a full microscopic description of inhomogeneous and orientationally disordered smectics \cite{wittmann2014fundamental,wittmann2017phase}.
A plethora of different states with characteristic defect topologies is observed
in perfect agreement between theory and experiment up to the microscopic nuances in the defect shape and wall alignment.

\begin{figure*}[t]
\includegraphics[scale=0.918750645]{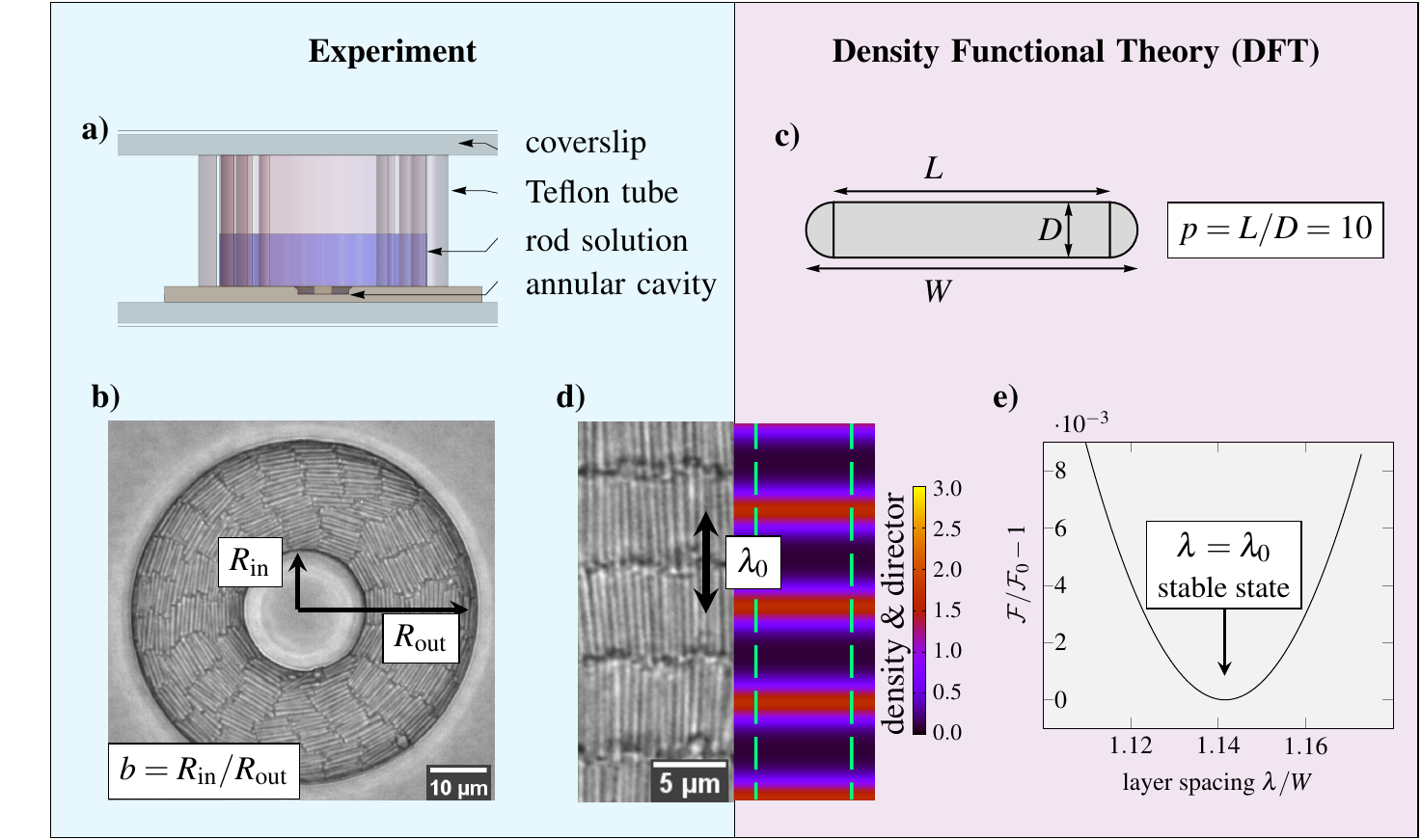} 
\caption{ 
 \textbf{Overview of experimental and theoretical methods.}
        \textbf{a)} Schematic illustration of the experimental cell. 
        \textbf{b)} Particle-resolved bright-field microscopy snapshot imaged in direct vicinity of the cavity bottom wall. The annular geometry is
        determined by the outer radius $R_\text{out}$ and the inclusion size
        ratio $b=R_\text{in}/R_\text{out}$ with the inner radius $R_\text{in}$.
        \textbf{c)} A discorectangle of rectangular
        length $L$, circular diameter $D$, total length $W=L+D$ and area $a=LD+D^2\pi/4$ used as the
        theoretical model particle. 
        \textbf{d)} Smectic bulk phase. Left: experimental
        snapshot showing individual particles. Right: theoretical density profile $\rho(\bvec{r},\phi)$ represented by a heat map of the orientationally
        averaged center-of-mass density $\bar{\rho}=\frac{a}{2\pi}\int_0^{2\pi}\!\mathrm{d}\phi\,\rho(\bvec{r},\phi)$
        and green arrows indicating the average local director orientation. 
         The large black arrow marks the spacing $\lambda_0$ between two layers.
        \textbf{e)} Bulk smectic free energy $\mathcal{F}$ as a function of the layer spacing $\lambda$, determined by density functional theory (DFT). The minimum at $\mathcal{F}=\mathcal{F}_0$
        corresponds to the optimal bulk layer spacing $\lambda_0$ in equilibrium. 
}\label{fig_setup}
\end{figure*}

Our study explores the intriguing competition between the internal liquid crystal properties 
and the extrinsic topological and geometrical constraints. In annular confinement this gives rise to
  three essential types of smectic defect configurations.
  Each of these defects is characteristic for a unique state with a discrete rotational symmetry in the orientation field,  
  which we refer to as follows.
In the {laminar states}, the smectic layers in a large defect-free domain (bounded by two parallel disclination lines)
resemble the flow lines around a circular obstacle (inclusion).
The {domain states} are governed by individual smectic domains, separated by radially oriented disclination lines, in three sectors of the annulus.
Finally, there are the {Shubnikov states}, named \cite{de1972analogy} in formal analogy between the typical arrays of edge dislocations and the flux quantization in type-II superconductors.
In addition, we observe peculiar symmetry-breaking {composite states} which
unite different types of defects in a single structure. 
 A locally adaptable layer spacing is found here to play the key role regarding the stability and distribution of defect strucutres in extreme confinement.

\section{Results}

 \subparagraph{Overview.}

Our complementary experimental and theoretical strategies (see appendices~\ref{SN0} and~\ref{SN1} for more details) to study smectic liquid crystals on the particle scale are illustrated in Fig.~\ref{fig_setup}.
 Experimentally, we directly observe fully equilibrated silica rods at the bottom of customized
confinement chambers (Fig.~\ref{fig_setup}a) through particle-resolved bright-field microscopy (Fig.~\ref{fig_setup}b).
 On the theoretical side, we analyze the microscopic density profiles, obtained from a free minimization of our geometrical DFT for two-dimensional hard discorectangles (Fig.~\ref{fig_setup}c).
We strive a direct comparison of experimental snapshots and theoretical density profiles, as illustrated for the bulk case in Fig.~\ref{fig_setup}d,
 where the DFT is minimized by the optimal layer spacing $\lambda_0$ (Fig.~\ref{fig_setup}e).
 To create the required overlapping parameter space, we employ both precise lithography to create robust confinement chambers with dimensions of only a few rodlengths 
and an efficient hard-rod density functional to tackle these system sizes. 
 The theoretical aspect ratio $p=10$ is chosen to closely match the effective value $p_\text{eff}=10.6$ in the experiment and
the density of rods is chosen to be slightly above the bulk nematic--smectic transition in each case.

\begin{figure*}[!!!!!!!!!!!!t]
		\includegraphics[scale=0.8325]{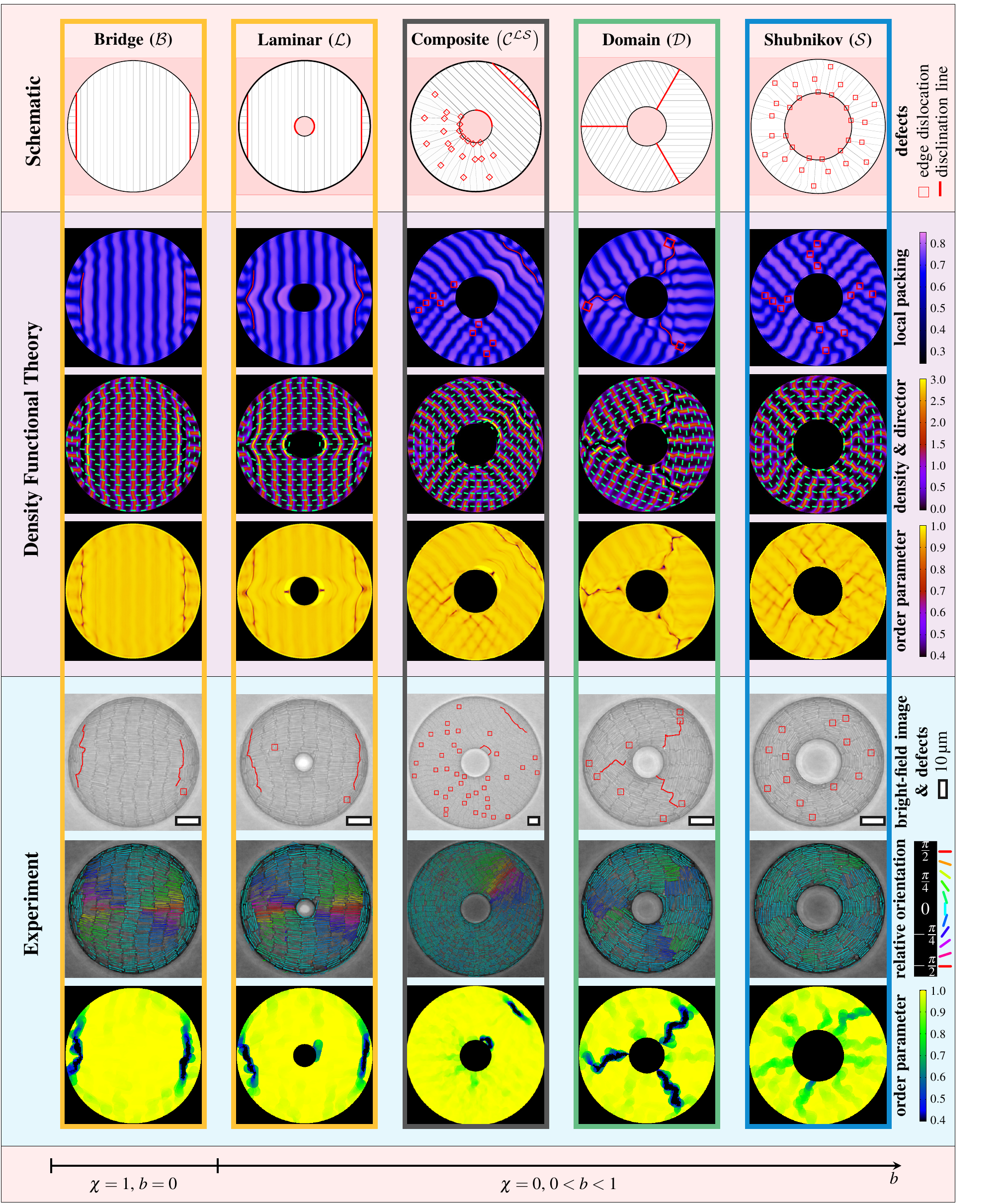}
		\caption{\label{fig_phases} 
 \textbf{Defects structures in smectic colloidal liquid crystals.} 
The columns represent the different states (as labeled), arranged from left to right by their occurrence 
in circular (Euler characteristic $\chi=1$) and annular ($\chi=0$) confinement
with increasing inclusion size ratio $b=R_\text{in}/R_\text{out}$. 
\textbf{First row:} idealized schematic representation of the mesoscopic arrangement of smectic layers (solid gray lines \tikzline[gray]{1})
 and defects (red squares \tikzsquare[thick,red]{2.75pt} and lines \tikzline[red]{1.5pt} as labeled).
\textbf{Rows 2-4:} local packing fraction with marked defects, density profiles with orientational director field (as in Fig.~\ref{fig_setup}d) and local order parameter from theory. 
\textbf{Rows 5-7:} typical particle-resolved snapshots with defects or color denoting the orientation relative to the nearest wall and local order parameter from experiment. 
   }
\end{figure*}

\begin{figure}
		\includegraphics[scale=0.955]{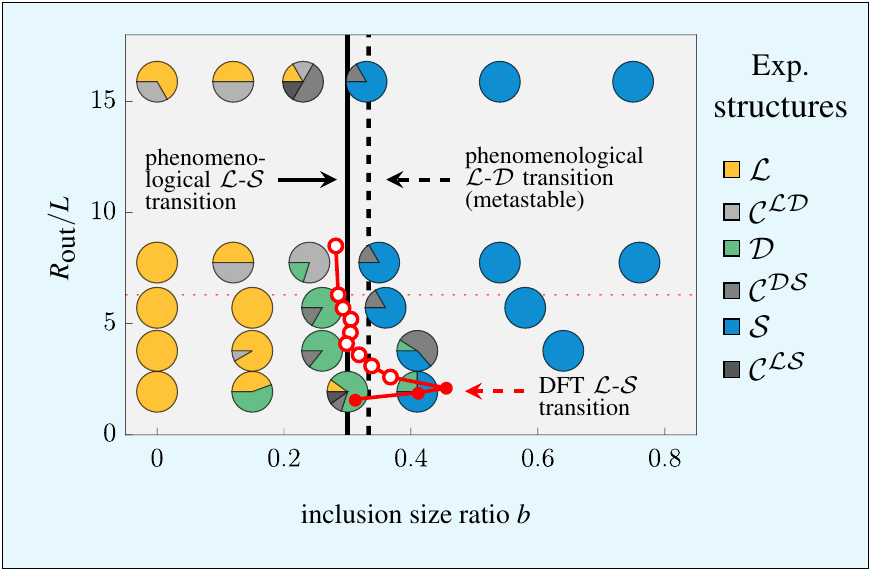}
\caption{  \textbf{Topological state diagram.} Shown are the stable states for different outer radii $R_\text{out}$ and inclusion size ratios $b=R_\text{in}/R_\text{out}$.
The large pie charts indicate the percentage at which each state occurs in the experiment according to the legend.
The small circles denote the theoretical laminar--Shubnikov ($\mathcal{LS}$) transition (the numerical error is of the order of the symbol size; filled circles indicate that no laminar state can exist for larger inclusions).
The vertical lines represent a possible scenario (see labels) predicted by a phenomenological model based on defect energies. 
\label{fig_statediagram}}
\end{figure}

\begin{figure}
		\includegraphics[scale=1.05]{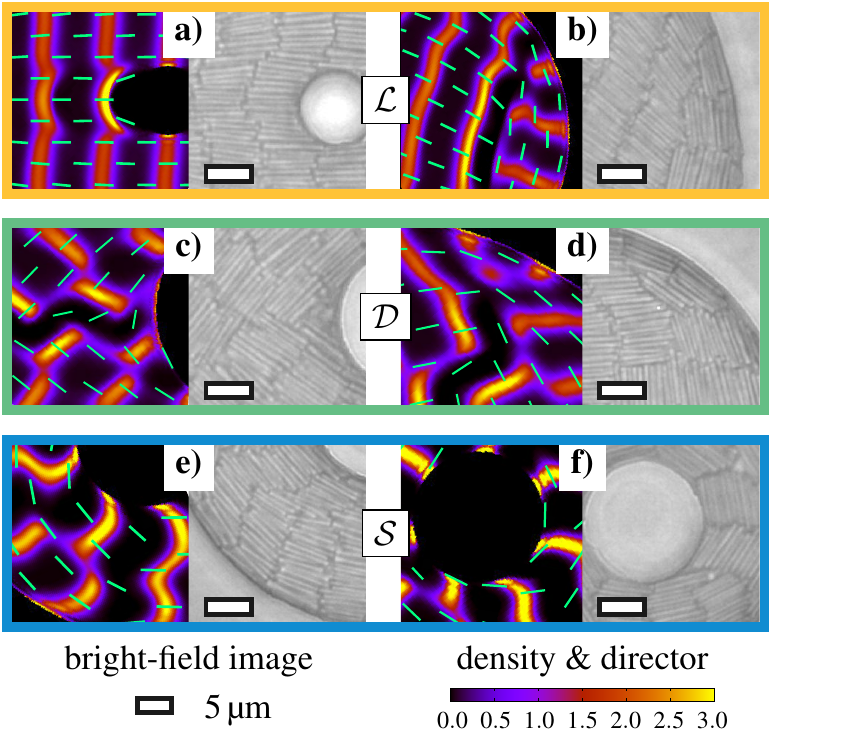}
		\caption{\label{fig_compare} \textbf{Structural details on the particle scale.} According observations from theory (left, density plots as in Fig.~\ref{fig_setup}d) and experiment (right): 
		\textbf{a)} bent layers adjacent to the inclusion and \textbf{b)} deformed layer near curved outer wall in the laminar state ($\mathcal{L}$),
		and
		\textbf{c)} gap between two domains 
	close to the inclusion in the domain state ($\mathcal{D}$) and
	\textbf{d)} planar alignment of rods at the outer wall 
		\textbf{e)} deformed layers adjacent to an edge dislocation and
		\textbf{f)} tilted alignment of some layers at the inclusion in the Shubnikov state ($\mathcal{S}$).
		}
\end{figure}

\begin{figure}
		\includegraphics[scale=0.55]{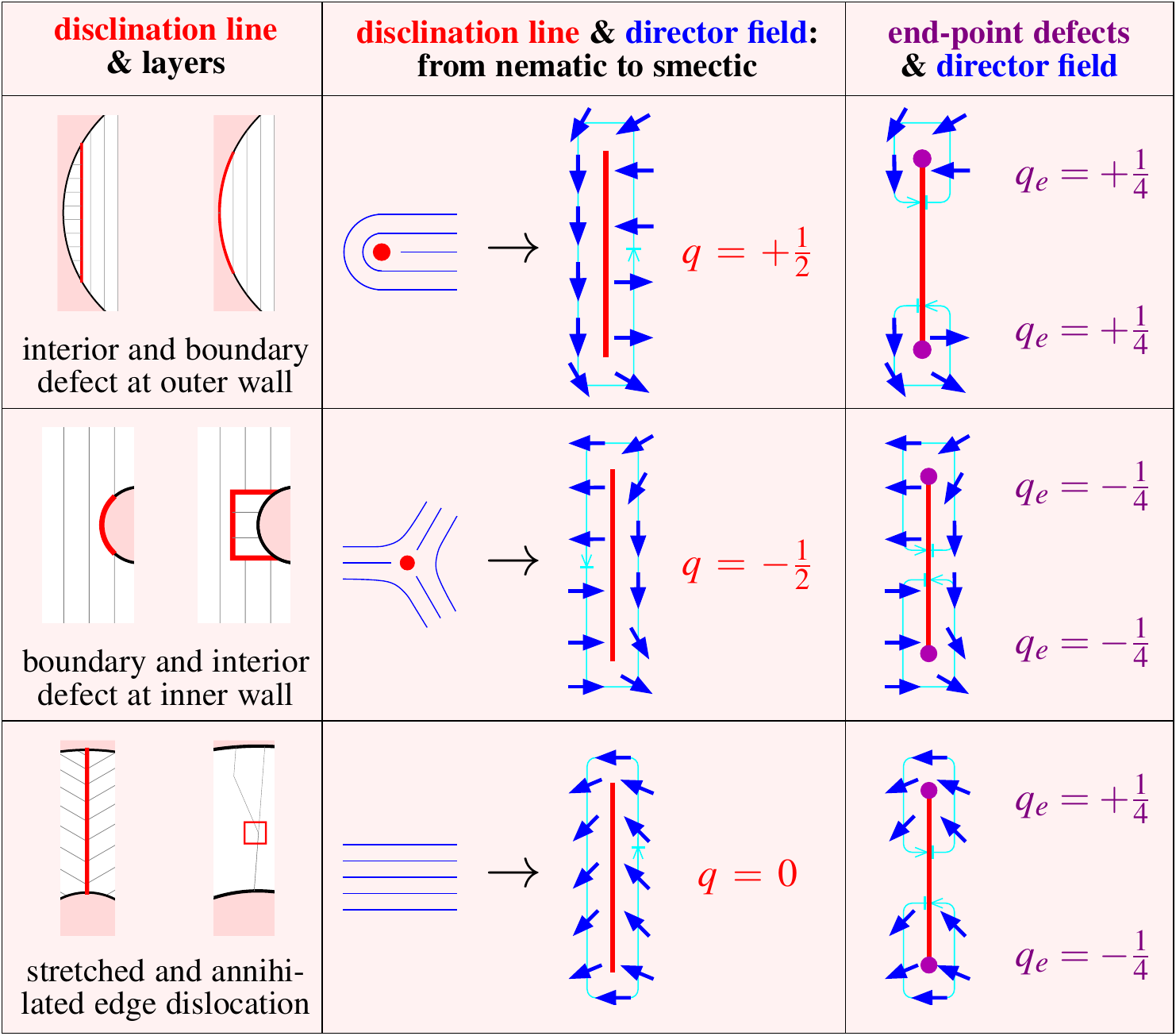}
		\caption{\label{fig_topo} 
		 \textbf{Topological defects in confined smectics.} 
				The anti-radial line disclinations at the outer (top row) and inner (middle row) wall are located in the interior or attached to the boundary.
		The radial line disclinations are topologically equivalent to edge dislocations (bottom row) and can be interpreted as a stretched or annihilated defect, respectively.
		\textbf{Left column:} schematic illustration as in Fig.~\ref{fig_phases}.
		\textbf{Middle column:} orientational director field (blue lines/arrows) around line defects (red) with topological charge $q$ according to the drawn closed integration path (cyan circular arrow). 
		The disclination lines can be interpreted as stretched nematic point defects, shown for comparison.
		\textbf{Right column:} director field at the end-point defects (violet) of disclination lines with charges ${q_\text{e}}$. Here the integration path is not closed but rather begins and ends on two sides of the disclination.}
\end{figure}

\begin{figure*}
		\includegraphics[scale=1.095]{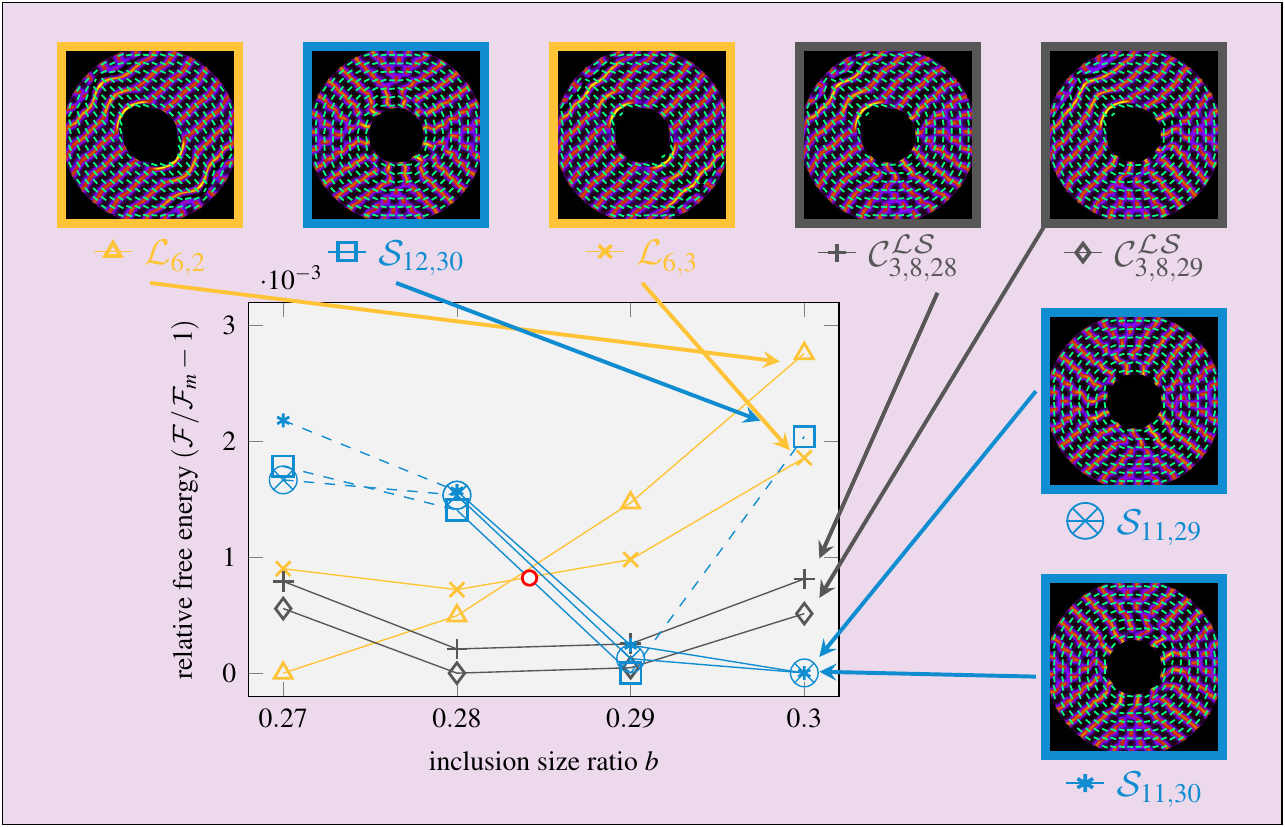}
\caption{  \textbf{Stable and metastable structures for different inclusion sizes.} Shown is the free energy $\mathcal{F}$ relative to that $\mathcal{F}_\text{m}$ of the global minimum for $R_\text{out}=6.3L$ and different inclusion size ratios $b$ close to the laminar--Shubnikov transition (red circle). 
	The legend depicts the theoretical density profiles (as in Fig.~\ref{fig_setup}d) of different states (color) with different microscopic structure (symbols) for $b=0.3$.
	The structures with alike symbols are created by subsequently equilibrating the density for slightly smaller inclusions, where the solid lines serve as a guide to the eye and 
	the dashed lines indicate a structural change regarding the number of layers in contact with the inclusion. The numerical error is of the order of the symbol size. 
\label{fig_statediagramX}}
\end{figure*}

\begin{figure*}
		\includegraphics[scale=1.05]{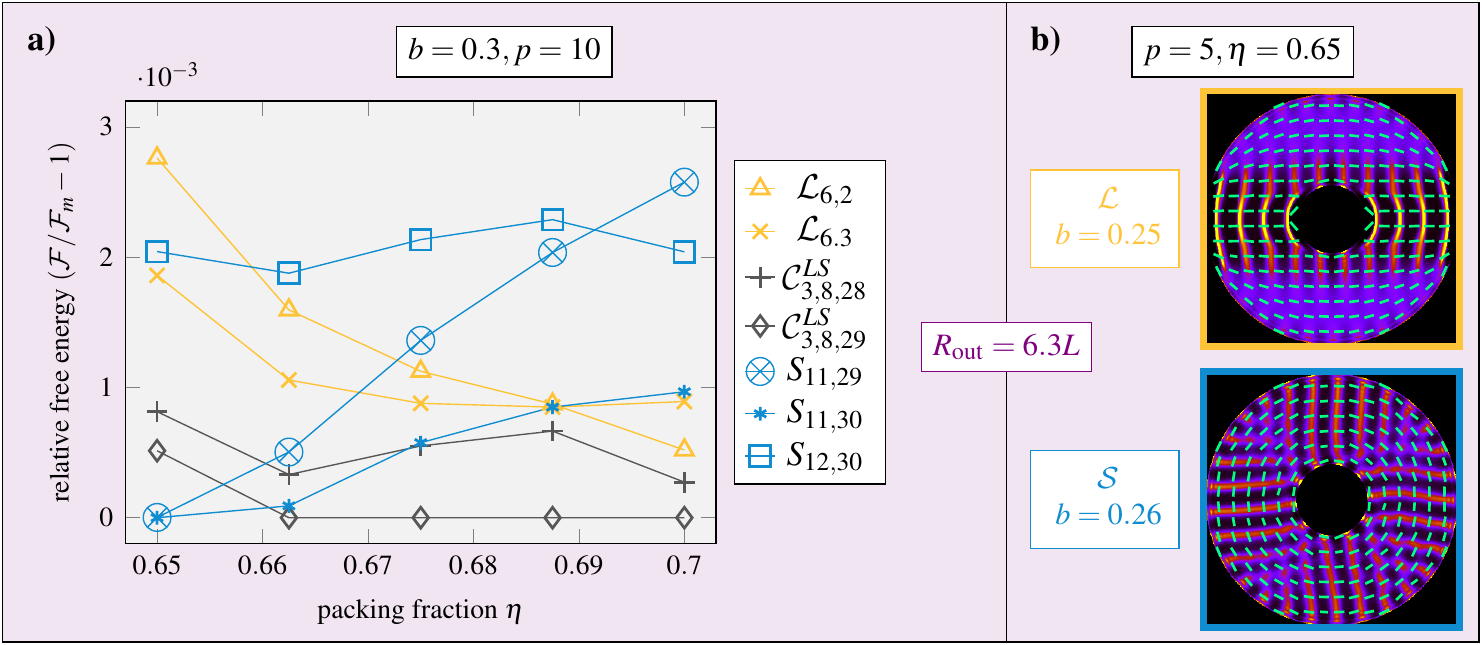}
\caption{ \textbf{Stable and metastable structures for different intrinsic parameters.} The geometry is fixed by $R_\text{out}=6.3L$ and $b=0.3$. 
\textbf{a)}~Dependence of the relative free energy on the density. 
The data for area fraction $\eta=0.65$ (at $b=0.3$) and the estimated numerical error are the same as in Fig.~\ref{fig_statediagramX}.
Here, alike symbols denote the structures obtained by gradually increasing the area fraction in steps of $0.0125$.
\textbf{b)} Stable structures (as in Fig.~\ref{fig_setup}d) for shorter rods with $p=5$ with inclusion size ratios $b=0.25$ and $b=0.26$. The laminar--Shubnikov transition is thus located in between. 
\label{fig_parametersM}}
\end{figure*}

The main results of our joint experimental and theoretical study of colloidal liquid crystals in annular confinement (see Fig.~\ref{fig_setup}b for the relevant geometrical parameters)  
are summarized as follows.
First, we identify a plethora of distinct smectic states (laminar, domain, Shubnikov and composite), shown in Fig.~\ref{fig_phases}, 
which possess a unique defect structure, layer arrangement and director topology.
Second, we predict in Fig.~\ref{fig_statediagram} a transition from the laminar state for small inclusion sizes to the Shubnikov state for large inclusion sizes,
 which nonmonotonically depends on the total confinement size.
 Third, analyzing the characteristic microscopic (Fig.~\ref{fig_compare}) and topological (Fig.~\ref{fig_topo}) details of each state
explains the stability of the observed structures.
 Finally, we illustrate in Fig.~\ref{fig_statediagramX} the full microscopic variety of different structures, including a stable composite state.
 We further argue with the aid of Fig.~\ref{fig_parametersM} how the state diagram changes with varying density and rod length.

\newpage\mbox{}\newpage\mbox{}\newpage

\subparagraph{Smectic states.}
 Figure~\ref{fig_phases} illustrates our central observation of different competing smectic states, each coping with the externally imposed constraints in a distinct way.
For each experimental structure in a given geometry, we find a perfectly matching theoretical density profile.
This depicted structural variety 
 results from a confinement with curved walls and a nontrivial topology,
 represented here by an annular cavity (Fig.~\ref{fig_setup}b) with Euler characteristic $\chi=0$.
The typical structure of these states is determined by the arrangement of the smectic layers (see microscopic details below) and the shape of topological defects with total charge of $Q=\chi=0$ (see topological details below).
 A detailed classification of the observed smectic states is given in appendix~\ref{SN2}.

To demonstrate the plain behavior of smectics in simply-connected domains,
we first remove the inclusion and consider circular confinement. 
In this reference case,
the only structure observed is the {{bridge state}} ($\mathcal{B}$).
It is characterized by a large domain of parallel layers spanning the system 
and frustration of the orientational order at the domain boundary. 
The latter is either due to the formation of two anti-radial disclination lines \cite{geigenfeind2015confinement,cortes2016colloidal} or, 
if the layers are directly adjacent to the outer wall, due to homeotropic (perpendicular to the wall) alignment~\cite{garlea2014square}, contrasting the preferred planar alignment at hard walls. 

 When adding a small inclusion, the layer arrangement resembles a laminar flow field around an obstacle,
which we refer to as the {{laminar state}} ($\mathcal{L}$).
The structure associated with the large bridging domain is identical to that of the bridge state,
 but the internal boundary may additionally disconnect some of the central layers and induce orientational frustration in the two tangential layers.
 The bridge state can thus be considered as a special undeformed case of a laminar state in the limit $b\rightarrow0$.
 
 At intermediate inclusion sizes, we observe a smectic {{domain state}} ($\mathcal{D}$) with three radially oriented disclination lines,  exhibiting a characteristic zig-zag pattern on the particle scale.

 Following de Gennes \cite{de1972analogy}, we refer to
the smectic structure at large inclusions as the {{Shubnikov state}} ($\mathcal{S}$).
It is characterized by layers spanning between the two disconnected system boundaries and
 an array of edge dislocations, which stabilizes the uniform orientational bend deformation 
 imposed here by the confining geometry.
This structural response is mathematically analogous to the magnetic vortices emerging in superconductors of type II subject to an external magnetic field~\cite{gennes1995physics,de1972analogy}.

 All smectic states introduced so far possess a discrete rotational symmetry.
In addition, {{composite states}} ($\mathcal{C^{LD}}$, $\mathcal{C^{DS}}$ or $\mathcal{C^{LS}}$) with two distinct regions, 
displaying characteristic order phenomena of either laminar, domain or Shubnikov states emerge at inermediate inclusion sizes. 
 A key paradigmatic example shown in Fig.~\ref{fig_phases} is the laminar--Shubnikov composite state $\mathcal{C^{LS}}$.

\subparagraph{State diagram.}
To answer the question about the stability of each state,
we illustrate in Fig.~\ref{fig_statediagram} the probability of its occurrence in our experiments.
For all state points considered, laminar states 
and Shubnikov states 
clearly dominate for $b\lesssim0.25$ and $b\gtrsim0.35$, respectively.
This provides compelling experimental evidence for the existence of a topological laminar--Shubnikov ($\mathcal{LS}$) transition around $b\approx 0.3$.
The state diagram is complemented by the intermediate domain 
state and several composite states.
  We observe that
 upon shrinking the system for a fixed intermediate inclusion size ratio $b\approx0.3$,
both the laminar state and the Shubnikov state become less stable, while 
the probability to find the domain state 
drops for larger systems.

 \begin{figure}
		\includegraphics[scale=1.475]{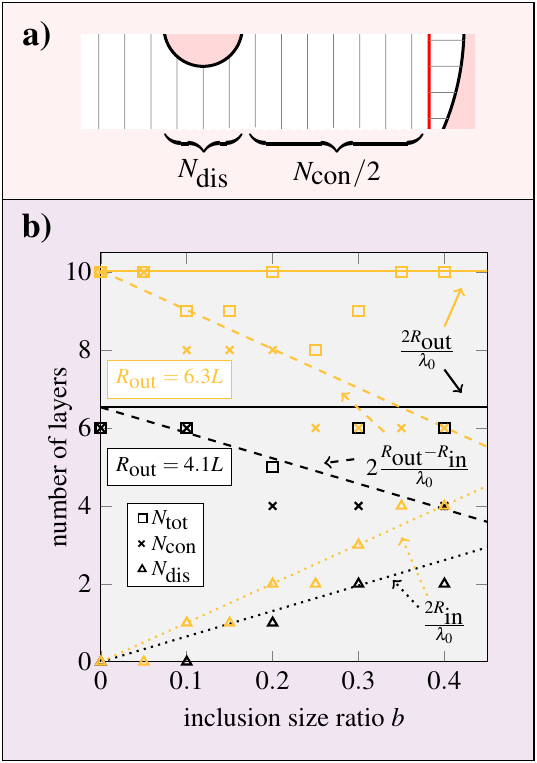}
		\caption{\label{fig_comparel} 
\textbf{Geometry dependence of the layers in the laminar state.} We show
		\textbf{a)} a schematic definition of connected/disconnected layer numbers $N_\text{con/dis}$, in total
		$N_\text{tot}=N_\text{dis}+N_\text{con}$, and
		\textbf{b)} the nonmonotonic dependence of layer numbers (symbols at integer values) 
		on the inclusion size ratio $b$ for two outer radii $R_\text{out}$ (colors as labeled). 
		For comparison, the lines indicate 
		the characteristic geometrical dimensions of the annulus divided by the bulk layer spacing $\lambda_0$ (as labeled).  
		}
\end{figure}

 \begin{figure}
		\includegraphics[scale=1.475]{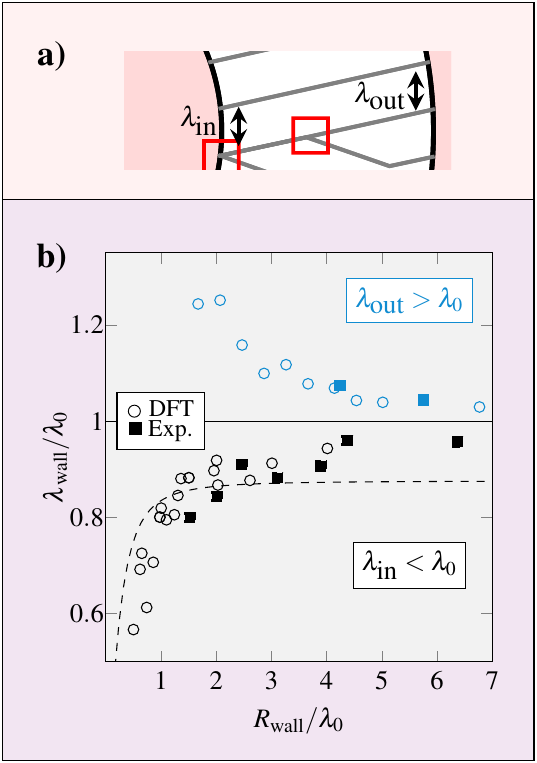}
		\caption{\label{fig_compares} 
\textbf{Geometry dependence of the layers in the Shubnikov state.} We show
		\textbf{a)} a schematic definition of the local layer spacing $\lambda_\text{in}$ and $\lambda_\text{out}$ at the inner and outer wall and
		\textbf{b)} the local layer spacing $\lambda_\text{wall}$ 
		as a function of the respective radius $R_\text{wall}$ of the inner ('wall' $=$ 'in') or outer ('wall' $=$ 'out') wall, normalized by $\lambda_0$ to values smaller or larger than one, respectively,
in different theoretical (circles) and experimental (squares) geometries.
		The dashed line represents a lower bound for rods packed at the inner wall with a perfectly planar alignment.
		}
\end{figure}

The described state diagram can be qualitatively understood in terms of a minimalistic phenomenological model, see appendix~\ref{SN3},
accounting solely for the length of the disclination lines ($\mathcal{L}$ and $\mathcal{D}$) or the number of edge dislocations ($\mathcal{S}$), cf.\ Fig.~\ref{fig_phases}. 
Only in the domain state the length of the disclination lines depends on the inclusion size ratio $b$, such that 
the laminar--domain transition can be located at a fixed $b=1/3$, independently of the unknown defect energy $\delta$ per unit length.
The energy of the Shubnikov state also depends on the inclusion size ratio $b$,
which is required to estimate for the total number of edge dislocations of energy $u_\text{ed}$.
The model thus predicts an alternative laminar-Shubnikov transition, depending on the ratio of $u_\text{ed}$ and $\delta$.
This fit parameter can be estimated by localizing the transition at the observed $b=0.3$, which gives rise to
 the scenario depicted in Fig.~\ref{fig_statediagram}a, where the domain state is only metastable.

To corroborate our experimental findings in full depth, we compute the free energies corresponding to the microscopic density profiles as a direct measure for their stability. 
Focusing on the precise localization of the $\mathcal{LS}$ transition,
 we observe in Fig.~\ref{fig_statediagram} a clear trend that the transition line shifts to larger inclusions for smaller $R_\text{out}$
in the experimentally accessible range of this parameter.
This agrees well with the distribution of the observed structures in our experiments.
For even more extreme confinements with $R_\text{out}\leq2.1L$ we locate the $\mathcal{LS}$ transition
close to the maximal inclusion size where a laminar state can form. 
Hence, the inclusion size ratio $b$ of the transition becomes smaller upon further decreasing $R_\text{out}$ below that threshold.
 For large confinements, the transition seems to approach the continuum limit with $b\approx0.27$.

 Further theoretical analysis shows that the composite state $\mathcal{C}^\mathcal{LS}$ is globally stable in a small but distinct region around the predicted $\mathcal{LS}$ transition,
 which we can understand on a microscopic level.

\begin{figure}
		\hspace*{0.6cm}\includegraphics[scale=1.265]{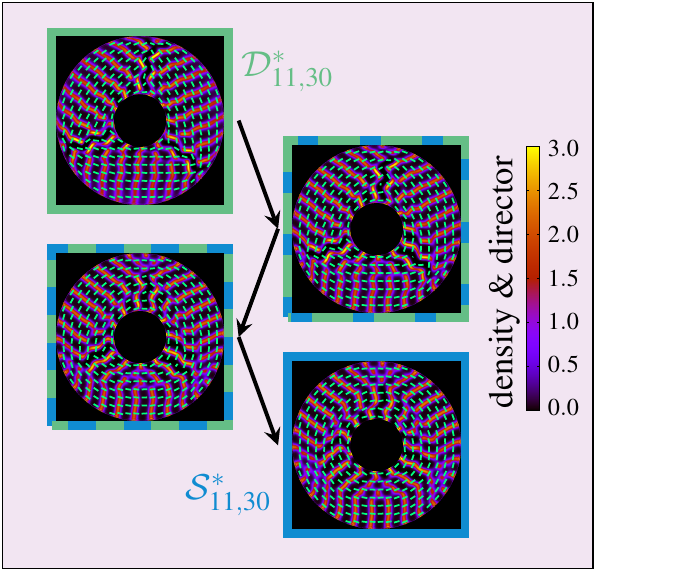}
\caption{ \textbf{Equilibration of a density profile initialized as a domain structure.}
The sequence of density profiles (as in Fig.~\ref{fig_setup}d) indicated by the arrows
 shows a continuous evolution from a domain state into a Shubnikov state with the same numbers of layers. The geometrical parameters are $R_\text{out}=6.3L$ and $b=0.3$.
\label{fig_statediagramDOM}}
\end{figure}

\subparagraph{Microscopic details.} 
The entropically optimal equilibrium structure of each state 
 emerges from a complex balance
 between several competing driving forces which aim to (i) remove all sorts of defects, (ii) achieve planar wall alignment, (iii) minimize the deformation energy 
and (iv) maintain the intrinsic layer spacing $\lambda_0$ in bulk. 
Our particle-resolved methods naturally provide an optimal account of points (i)-(iv) in the course of equilibration.
The quantitative agreement of the experimental and theoretical density profiles, both generated by the subtle interplay of these fundamental principles, 
allows us to unveil in Fig.~\ref{fig_compare} the characteristic microscopic structural details of each state.

From our microscopic insights, 
detailed below and further elaborated in appendix~\ref{SN4},
we draw the following conclusions regarding the state diagram in Fig.~\ref{fig_statediagram}.
Increasing the wall curvature (by decreasing $R_\text{out}$) increasingly distorts the layer spacing in the Shubnikov state, 
such that it becomes destabilized compared to the laminar state, for which the relative energy penalty arising from homeotropic wall alignment decreases.
 The nonmonotonic behavior of the resulting $\mathcal{LS}$ transition line is related to the varying compatibility of each state with the particular confining geometry.
 The indirect $\mathcal{LS}$ transition via an intermediate composite state $\mathcal{C}^\mathcal{LS}$
 can be explained by the increased number of possibilities, compared to the $\mathcal{L}$ and $\mathcal{S}$ states, to relax the geometrical constraints.

In detail, we observe that the layers and defect lines surrounding the inclusion in the observed {{laminar states}}
are typically deformed according to the shape of the inner wall, where the wall alignment is homeotropic, see Fig.~\ref{fig_compare}a.
 In contrast, due to the planar alignment at the outer wall, the disclination lines end on point defects,
 recognizable by the modulation of the adjacent layer, see Fig.~\ref{fig_compare}b.
  This has not been reported for straight walls \cite{geigenfeind2015confinement,cortes2016colloidal}.
  The optimal number of layers depends nonmonotonically on the geometric parameters, as shown in Fig.~\ref{fig_comparel}.

The particularly deformed microstructure of the {{domain states}}
is due to the competing angles $2\pi/3$ between the domain boundaries and $\pi/2$ between the intersecting layers.
The positional order is most frustrated in the vicinity to the inclusion, as reflected by the some dilute regions, shown in Fig.~\ref{fig_compare}c,
 in which the rods try to align with the wall.
Taking a closer look at the outer boundary, however,
we observe in Fig.~\ref{fig_compare}d some additional layers and edge dislocations between two adjacent domains, 
ensuring again an overall planar wall alignment.

The layers in the {{Shubnikov states}} are deformed in the vicinity of edge dislocations, see Fig.~\ref{fig_compare}e.
Although an overall planar wall alignment is generally possible, we frequently observe in small systems
that one or more layers are tilted with respect to the inner wall, as in Fig.~\ref{fig_compare}f.
 This reflects a strongly position-dependent layer spacing, as shown in Fig.~\ref{fig_compares}, which further distinguishes the Shubnikov from the laminar and domain states.
  In fact, the deviations in the local layer spacing reduce the number of point defects, such that
 we even observe some extreme structures without any defects for sufficiently small distances between the walls. 

\subparagraph{Topological details}

Having resolved the microscopic details of the topological defects, emerging due to the rigidity of the smectic layers, we are in a position to 
associate in Fig.~\ref{fig_topo} a topological charge $q$ with each occurring disclination line.
We further identify pairs of end-point defects to these lines, which formally carry peculiar quarter-integer charges ${q_\text{e}}$, such that $q=\sum{q_\text{e}}$.
Then one can easily verify from the sketches in Fig.~\ref{fig_phases} that, in each state, the total charge $Q=\sum q$ equals the Euler characteristic of the confining domain, as required by topology \cite{kamienREVnem,bowickREVTOP2009}.
 The topological protection due to charge conservation is discussed in appendix~\ref{SN5}.

The anti-radial disclination lines in the laminar (and $\mathcal{C^{LS}}$ composite) state,
can be understood as an expanded $q=+1/2$ point defect with two ${q_\text{e}}=+1/4$ charges at the end, see Fig.~\ref{fig_topo} (top).
For the laminar structures without any disclination lines, there exists an equivalent $q=+1/2$ defect attached to the boundary, recognizable by the misalignment at the wall.
Likewise, misalignment near the inclusion implies a negatively charged boundary defect with $q=-1/2$, see Fig.~\ref{fig_topo} (middle).
Some structures (compare, e.g., the experimental $\mathcal{C^{LS}}$ in Fig.~\ref{fig_phases}), 
display an explicit disclination line close to the inclusion ending on two ${q_\text{e}}=-1/4$ points.

The radial disclination lines are always in the interior of the system carrying the opposite end-point charges ${q_\text{e}}=-1/4$ and ${q_\text{e}}=+1/4$, close to the inner and outer end, respectively, see Fig.~\ref{fig_topo} (bottom).
Hence this type of line defect is nothing more than an expanded edge dislocation with topological charge $q=0$,
which unveils the true topological nature of the domain states.
They possess the same orientational topology 
 as the Shubnikov states into which they can evolve upon pair annihilation, compare Fig.~\ref{fig_statediagramDOM}. 
 The observation of domain states is thus owed to packing effects.

\subparagraph{Free energy landscape}

As apparent from the multitude of observed structures in some geometries, 
the system does not always equilibrate towards the global energy minimum.
It is thus important to understand the full free energy landscape generated by the described competing driving forces.
To this end, we additionally calculate the free energy associated with various theoretical density profiles, to directly assess their stability.
As a representative example, we choose $R_\text{out}=6.3L$ and
compare in Fig.~\ref{fig_statediagramX} seven sets of structures obtained by smoothly decreasing $b$ in the vicinity of the laminar--Shubnikov transition.
We draw four important conclusions.

First, we explicitly see that the laminar--Shubnikov transition is not sharp.
Instead, over a significant range $\Delta b\approx0.015$ of inclusion size ratios, a composite state of both structures is energetically favorable.
%
%
Second, the free-energy differences between two distinct structures are extremely small 
and the optimal microscopic structure changes multiple times upon small modifications of the confinement. 
These observations explain the large number of different structures observed in the experiment for $b\approx 0.3$. 
Third, it is important to identify the optimal microscopic structure of each state
to make a proper statement about possible topological transitions.
For example, only considering for $b=0.3$ the metastable Shubnikov structure with the largest energy 
would lead to the false conclusion that the laminar state (or a composite state) is more stable in this geometry.
Finally, a smooth variation of the inclusion size leaves laminar structures invariant (topologically protected), 
while the Shubnikov structures gradually adapt to the change in geometry, sometimes following a small hysteresis loop, as discussed in appendix~\ref{SN6}.

\subparagraph{Dependence on density and rod length}

Apart from the external topological and geometrical constraints,
the formation and stability of the reported states also depends on different intrinsic parameters, which is detailed further in appendix~\ref{SN7}. 
The effect of the preferred bulk layer spacing $\lambda_0$ is illustrated
in Fig.~\ref{fig_parametersM}. We see that increasing the density (Fig.~\ref{fig_parametersM}a), resulting in a smaller $\lambda_0$, stabilizes laminar structures compared to Shubnikov structures,
while decreasing the aspect ratio to $p=5$ (Fig.~\ref{fig_parametersM}b), resulting in a larger relative $\lambda_0/p$, has the opposite effect. 
Extending our state diagram towards shorter rods at a fixed density,
we further anticipate the emergence of stable tetratic structures~\cite{grannulus}, since smectic order is generally destabilized \cite{sitta2018}.

\begin{figure}[t]
\hspace*{-0.6cm}\includegraphics[width=1.15\linewidth]{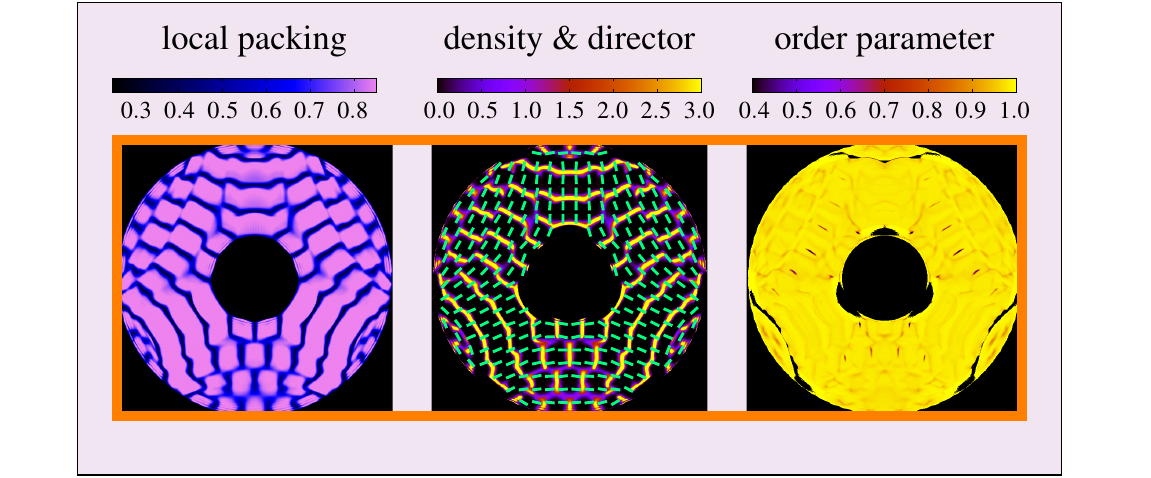}
\caption{ \textbf{Metastable smectic state with threefold symmetry.} Shown is the theoretical prediction of a structure analog to a $D_3$ nematic state for $R_\text{out}=6.3L$, $b=0.3$ and $\eta=0.75$.
We use the same representations as in Fig.~\ref{fig_phases} and depict the local packing fraction (mind the differences due to the higher overall density), 
the orientationally averaged density and director field and the local orientational order parameter.
Note that the free energy is much higher than for the other states reported in the main text using the same parameters, as there is a large number of deformations required to fit into the given geometry.
\label{fig_S3}}
\end{figure}

 For long rods at lower densities, different nematic states $D_n$ are found \cite{garlea2016finite}, classified by the number $n$ of $q=\pm1/2$ defect pairs.
  From a topological point of view, the laminar and Shubnikov states can be interpreted as the smectic analogy to $D_2$ and $D_\infty$, respectively,
 while possessing a distinct orientational director field (compare Fig.~\ref{fig_topo}), imprinted by the arrangement of smectic layers. 
  The smectic analogy to $D_3$ (three line disclination of charge $q=+1/2$ close to the outer wall) 
  is not stable here as the curved geometry requires too strong deformations, see Fig.~\ref{fig_S3},
while there is no direct nematic analogy to domain and composite states.
  In our experiment, we observe the formation of nematic states at the bottom of our chambers in course of the sedimentation process,
  which can be mimicked in DFT by subsequently increasing the density.
  From the latter approach, we predict in appendix~\ref{SN7} that the onset of smectic order depends on the underlying nematic state.
  In turn, the nematic order in sedimentation equilibrium is presumably dictated by the more rigid smectic structure observed at the bottom.

\section{Discussion}
We have performed a complementary particle-resolved experimental and theoretical study of hard colloidal rods in annular confinement.
Our observations emerge from the fundamental principle of globally maximizing the entropy 
subject to the constraints arising from the external influences of the confinement and the
 internal smectic layer structure, which depends on the particle shape and density.
All these competing driving forces are accounted for explicitly by our DFT data for the equilibrated structures.

In the future, it will be interesting to have a closer look at
the position dependence of particle diffusion between the layers \cite{chiappini2020}
or the formation dynamics of the different smectic structures, e.g., using dynamical DFT \cite{marconi1999,archer2004}.
Drawing phase-stacking diagrams \cite{delasheras2013stacking}
will provide vital information on how the coexistence of nematic and smectic structures affects their stability in the experiment.
 While some additional smecitc states could become stable in different geometries,
an even larger structural variety can be anticipated in more complex topologies, e.g., those with two holes.
 The next level of geometrical and topological complexity will be reached when 
immersing colloidal smectic liquid crystals in random porous media~\cite{sinha1998,iannacchione2004,araki2011,banerjee2018} and fractal confinement \cite{hashemi2017}.
 On the other hand, there is also a high intrinsic potential for finding novel structures when considering more exotic particle shapes \cite{avendano2016,sentker2018,schoenhoefer2018,chiappini2019,fernandez2019}.

In conclusion, we have shown that the topology and geometry of an externally imposed confinement
largely determine the preferred internal structure of a smectic liquid crystal.
Adjusting these screws allows to create a protocol for a guided self-assembly of a desired defect structure.
Owing to their robustness and large range of metastability, the described smectic states can then be smoothly transferred to any desired confining geometry and, if desired, solidified to unfold their potential for various microtechnological applications \cite{Lagerwall2012ANE,katoREV2018}.
 These possibly include  
 novel devices for information storage,
 templates for functional microstructured materials
 and channels for micro- or nanofluidics.
Regarding the recently flourishing research realm of living or self-motile particles,
a challenging extension of the present work could consist of
systematically studying the influence of activity on the predicted equilibrium state diagram \cite{bott2018}. 
  Finally, a fascinating connection with biology emerges from drawing the analogies 
 between colloidal liquid crystals and growing colonies of rod-shaped bacteria \cite{bacterialcolonies2018,bacterialcolonies2020a,bacterialcolonies2020b,bacterialcolonies2020c}.
  Our results thus lay the foundation for a deeper microscopic understanding of the structures emerging and persisting along the evolution dynamics
 when such living systems are subjected to extreme topological confinement.

\appendix

\section{\label{SN0} Summary of the methods}
\subparagraph{Sedimentation of silica rods \label{sec_exp}}

To experimentally create confined quasi-two-dimensional smectic structures,
we take advantage of the phase stacking of silica rods~\cite{kuijk2011,kuijk2012phase} in sedimentation equilibrium~\cite{cortes2016colloidal}.
 The bare dimensions of the rods are measured directly from scanning electron microscopy images. 
 The rods are dispersed into a $1\,\text{mM}$ NaCl water solution to ensure stability through double layer repulsion.
 Introducing effective dimensions (see appendix~\ref{SN1} for more details) to account for the Debye screening, our particles behave like hard rods of an effective aspect ratio $p_\text{eff}=10.6$.

 The confining cavities (see Fig.~\ref{fig_setup}a) in the shape of hollow cylinders are molded on the bottom coverslip using home made Polydimethylsiloxane (PDMS)  stamps and Norland Optical Adhesive~\cite{cortes2016colloidal}. 
In practice, several chambers are fitted in a single cell. 
After preparation, the rod solution is left in the tube to sediment for at least 12 hours. During sedimentation, the concentration of particles gradually increases along the direction of the gravity field
leading to successive isotropic, nematic and smectic order at the bottom. After a few hours, sedimentation diffusion equilibrium is reached and the three phases coexist in the cavities. 

The smectic structures are observed by means of bright-field microscopy in direct vicinity of the bottom wall. 
 We use a $1.42$ numerical aperture apochromat oil immersion objective mounted on an Olympus IX73 microscope and coupled to a Ximea CMOS xiQ camera,
 which allows an optical resolution comparable to the rod diameter.
Due to degenerate planar anchoring at the bottom wall, the system can be considered a quasi-two-dimensional fluid in annular confinement. 
 We choose the total amount of rods to obtain an effective volume fraction ${\phi_\text{eff}\approx45-50\%}$ close to the bottom.
 This ensures that there is no crystalline state and that the rods in direct contact with the bottom wall exhibit smectic order. 
To create some statistics, we repeat the measurements in a given geometry up to twelve times.

\subparagraph{Density functional theory (DFT) \label{sec_theo}}

We study by free minimization of a DFT \cite{Evans1979} in two dimensions hard discorectangles 
(see Fig.~\ref{fig_setup}c)
 with an aspect ratio $p=10$ that well reflects the experimental parameters. 
 The interaction between these particles are described by a
 free energy functional constructed as an extension of fundamental measure theory \cite{Rosenfeld1989,RothREVIEW2010,roth2012} to account for anisotropic particle shapes \cite{wittmann2017phase,wittmann2016fundamental}.
These geometrical functionals derived from first principles are exact in the low-density limit and have proven very reliable for highly packed systems.
 The annular confinement is included as an external hard-wall potential.

The key quantity in our theory is the one-body density profile $\rho(\bvec{r},\phi)$, providing the probability to find a particle with the center-of-mass position $\bvec{r}$ and its symmetry axis oriented along an angle $\phi$. 
Consider now a density functional $\Omega[\rho(\bvec{r},\phi)]=\mathcal{F}[\rho]+\!\int\!\mathrm{d}\bvec{r}\!\int_0^{2\pi}\!\frac{\mathrm{d}\phi}{2\pi}\,\rho(\bvec{r},\phi)(V_\text{ext}(\bvec{r},\phi)-\mu)$, 
where $\mathcal{F}[\rho]$ is the intrinsic Helmholtz free energy functional, $V_\text{ext}(\bvec{r},\phi)$ is the external potential and $\mu$ the chemical potential (see appendix~\ref{SN1} more details).
Then the density $\rho(\bvec{r},\phi)$ of a (meta-) stable state is found by iteratively solving the extremal condition $\delta\Omega[\rho]/\delta\rho=0$ 
starting from a particular initial guess for the density profile.
The average area fraction $\eta=0.65$ is kept fixed  throughout the iteration by adapting $\mu$ in each step.
 Then we compare the values of the free energy $\mathcal{F}[\rho]$ of the different structures to identify the global minimum 
 and quantify the likelihood to observe a metastable local minimum $\rho(\bvec{r},\phi)$ in a corresponding experiment.
Calculations are performed on a quadratic spatial grid with a high enough resolution $\Delta x=\Delta y=0.2$ and  
$N_\phi=96$ discrete orientational angles.
We iterate until the free energy differences between different structures can be sufficiently resolved.

\subparagraph{Overlapping parameter space}

Our experiment and theory are designed, such that they can both
tackle hard rods with a comparable anisotropy that is sufficiently high to ensure that the smectic phase is stable over a large range of densities \cite{bolhuis1997tracing,wittmann2017phase}.
Systems with variable inclusion size ratios $b$ and the radii $R_\text{out}$ of the circular outer wall ranging between $1.9L\leq R_\text{out}\leq5.7L$ are covered by both approaches.

\subparagraph{Data analysis and presentation}

 From the theoretical data we determine a local packing fraction by weighting the density with the local particle area to highlight the particle resolution within our data (second row of Fig.~\ref{fig_phases}).
As a standard representation of the full density field we use in the third row of Fig.~\ref{fig_phases} and for all other illustrations a plot of the
orientationally averaged density with the orientational director field, represented by green arrows of length given by the local order parameter.
Moreover, we also directly display the local order parameter field (fourth row of Fig.~\ref{fig_phases}).
 Note that the distorted appearance of the order parameter close to the inclusion in the laminar state reflects the very low but nonvanishing probability to find particles left and right of the symmetry axis that are perfectly aligned with the wall.
This underlines the similarity of interior and boundary defects illustrated in Fig.~\ref{fig_topo}.
We compare the free energies of the different structures in Figs.~\ref{fig_statediagram} and~\ref{fig_statediagramX}. Further details on numerical errors are given in appendix~\ref{SN1}.

The experimental snapshots (fifth row of Fig.~\ref{fig_phases}) are inspected visually and further processed using Wolfram Mathematica computing system.
This allows us to color each particle according to its orientation relative to the wall (sixth row of Fig.~\ref{fig_phases}).
From the measured center-of-mass positions and orientations we further extract a local field of the orientational order parameter (seventh row of Fig.~\ref{fig_phases}).
The different smectic states and their microscopic structures are identified according to the criteria described in appendix~\ref{SN2}.



\section{\label{SN1} Further details on the methods}

\subparagraph*{Sedimentation of silica rods.}

To experimentally create confined quasi-two-dimensional smectic structures,
we take advantage of the phase stacking of silica rods 
in sedimentation equilibrium~\cite{cortes2016colloidal,LCthesis}.
 The bare dimensions of the rods are measured directly from scanning electron microscopy images. The mean length is $W=5.3\,\mu\text{m}$ with a standard deviation $\sigma_W=0.5\,\mu\text{m}$. The rods are dispersed into a $1\,\text{mM}$ NaCl water solution to ensure stability through double layer repulsion,
  whose range is characterized by the Debye length $\kappa^{-1}=0.01 \,\mu\text{m}$.
 Moreover, the phase behavior of our charged rods can be mapped onto the phase behavior of hard rods \cite{onsager1949effects,grelet2014hard,stroobants1986effect,van2004liquid} by introducing the effective 
 rod length $W_\text{eff}=5.4\,\mu\text{m}$, diameter $D_\text{eff}=470\text{nm}$ and aspect ratio $p_\text{eff}=10.6$
 to account for the Debye screening.

 An illustration of the experimental cell can be found in Fig.~\ref{fig_setup}a. 
 The confining cavities in the shape of hollow cylinders are molded on the bottom coverslip using home made Polydimethylsiloxane (PDMS)  stamps and Norland Optical Adhesive~\cite{cortes2016colloidal}. 
In practice, several chambers are fitted in a single cell. 
After preparation, the rod solution is left in the tube to sediment for at least 12 hours. During sedimentation, the concentration of particles gradually increases along the direction of the gravity field
leading to the successive formation of isotropic, nematic and smectic phases. After a few hours, sedimentation diffusion equilibrium is reached and the three phases coexist in the cavities. 

The smectic structures are observed by means of bright-field microscopy in direct vicinity of the bottom wall. 
 We use a $1.42$ numerical aperture apochromat oil immersion objective mounted on an Olympus IX73 microscope and coupled to a Ximea CMOS xiQ camera,
 which allows an optical resolution comparable to the rod diameter.
Due to degenerate planar anchoring at the bottom wall, the system can be considered a quasi-two-dimensional fluid in annular confinement. 
 We choose the total amount of rods such that there is no crystalline state and the rods in direct contact with the bottom wall exhibit smectic order. 
The effective (three-dimensional) volume fraction ${\phi_\text{eff}\approx45-50\%}$ at the bottom of the cell is measured indirectly by comparing the height of the smectic region to the sedimentation equilibrium of our rod system in bulk \cite{LCthesis}
and assuming that the rods behave like hard spherocylinders \cite{bolhuis1997tracing}.
For this setup, we measure $\lambda_0\approx 1.3\,W_\text{eff}$ for the effective two-dimensional bulk layer spacing at the bottom wall.

\subparagraph*{Density functional theory (DFT).}

Classical DFT \cite{Evans1979} is a powerful and versatile tool to access the structure of inhomogeneous  
fluids on the particle scale.
Here, we study by free minimization of a DFT in two dimensions hard discorectangles with rectangular length $L$, total length $W=L+D$ and unit circular diameter $D$, see Fig.~\ref{fig_setup}c.
 If not denoted otherwise, we use $L=10D$, so that the aspect ratio $p=L/D=10$ well reflects the experimental parameters. 
 The energy unit is set by $\beta^{-1}:=k_\text{B}T$,
where $k_\text{B}$ and $T$ are Boltzmann's constant and the temperature, respectively.
 Note that the structural transition of perfectly hard particles are completely driven by entropy,
since $\beta$ only appears as a trivial scaling factor.
The key quantity in our theory is the one-body density profile $\rho(\bvec{r},\phi)$, providing the probability to find a particle with the center-of-mass position $\bvec{r}$ and its symmetry axis oriented along an angle $\phi$.

The general form of the density functional reads \cite{Evans1979}
\begin{align}
 \Omega[\rho]=\mathcal{F}[\rho]+\!\int\!\mathrm{d}\bvec{r}\!\int_0^{2\pi}\!\frac{\mathrm{d}\phi}{2\pi}\,\rho(\bvec{r},\phi)(V_\text{ext}(\bvec{r},\phi)-\mu)\,,
\end{align}
where the external potential $V_\text{ext}(\bvec{r},\phi)$ imposes the annular confinement through a hard-wall potential
and $\mu$ denotes the chemical potential.
The intrinsic Helmholtz free energy functional
\begin{align}\label{eq_freeenergy}
\beta\mathcal{F}[\rho]=\!\int\!\mathrm{d}\bvec{r}\!\int_0^{2\pi}\!\frac{\mathrm{d}\phi}{2\pi}\,\rho(\bvec{r},\phi) \left(\ln (\rho(\bvec{r},\phi)\Lambda^2)-1\right)+\beta\mathcal{F}_\text{ex}[\rho]
\end{align}
consists of an ideal-gas term ($\Lambda$ denotes the irrelevant thermal wave length)
and the excess free energy functional $\mathcal{F}_\text{ex}[\rho]$, which describes the interactions and correlations between the individual particles.
Here, the latter is constructed as an extension of fundamental measure theory 
to account for anisotropic particle shapes.
These geometrical functionals derived from first principles are exact in the low-density limit and have proven very reliable for highly packed systems.

The employed excess functional \cite{wittmann2017phase}
\begin{equation}
 \beta\mathcal{F}_\text{ex}[\rho]
 =\int\!\mathrm{d}\bvec{r}\left(-n_0\ln(1-n_2)+\frac{N}{2(1-n_2)}\right)\,,
 \label{eq_PhiRF}
\end{equation}
 of the two dimensional fundamental mixed measure theory is constructed as a function of weighted densities
\begin{equation}
 \label{eq_weighdenorient}
  n_{\nu}(\bvec{r}) = \!\int\!\mathrm{d}\bvec{r}_1\!\int_0^{2\pi}\!\frac{\mathrm{d}\phi}{2\pi}\,\rho(\bvec{r}_1,\phi)\, \omega^{(\nu)}(\bvec{r}-\bvec{r}_1,\phi) \,.
\end{equation}
 These are calculated by convolution of the density and the one-body geometrical measures $\omega^{(\nu)}$,
 representing a local area ($\nu=2$), circumference ($\nu=1$) and boundary curvature ($\nu=0$) of the particles.
The mixed weighted density $N(\bvec{r})$ generally depends on the geometry of two bodies.
To efficiently study long rods in large systems, we use the approximate representation 
\begin{equation}
 N\approx\frac{2+a}{6\pi}n_1n_1+\frac{a-4}{6\pi}n_{1,\alpha}n_{1,\alpha}
 +\frac{2-2a}{6\pi}n_{1,\alpha\beta}n_{1,\beta\alpha}
 \,,
 \label{eq_PhiRFapp}
\end{equation}
found by an expansion introducing vectorial $n_{1,\alpha}$ and tensor-valued weighted densities $n_{1,\alpha\beta}$ up to rank-two,
where we use the convention of summation over repeated indices $\alpha,\beta\in\{1,2\}$.
The correction parameter $a=4$ allows to qualitatively describe the smectic phase over the full range of aspect ratios \cite{wittmann2017phase,wittmann2014fundamental}.

Generally, the density $\rho(\bvec{r},\phi)$ of a (meta-) stable state is found by iteratively solving the extremal condition $\delta\Omega[\rho]/\delta\rho=0$.
{  In practice, an individual density profile $\rho_i(\bvec{r},\phi_i)$ is considered for each discrete orientation $\phi_i$.
After each iteration step $j$, the current densities $\rho^{(j)}_i$ are updated according to a Picard iteration scheme \cite{RothREVIEW2010,Evans1979}.
This involves a formal solution of 
\begin{equation}
 \frac{\delta\Omega[\rho^{(j)}_i]}{\delta\rho^{(j)}_i}=0
\end{equation}
 for a new $\tilde{\rho}_i^{(j)}$
and calculating $\rho_i^{(j+1)}=(1-\gamma)\rho_i^{(j)}+\gamma\tilde{\rho}_i^{(j)}$
with a dynamical mixing parameter $\gamma$, initially set to $\gamma=0.08$.
If the value of the local packing fraction $n_2(\bvec{r})$, see Eq.~\eqref{eq_weighdenorient}, exceeds one for any coordinate $\bvec{r}$, 
the step is rejected and $\gamma$ is decreased by a factor of $0.2$.
After each 200 steps, the program tries to gradually increase $\gamma$ by factors of $2$ to speed up the iteration.}
The average area fraction 
\begin{equation}
 \eta=N\,\frac{4LD+D^2\pi}{4\pi R^2_\text{out}(1-b^2)}
\end{equation}
 is kept fixed throughout the iteration by adapting $\mu$ in each step.
{ Hence, the stable state corresponds to a global minimum of the free energy $\mathcal{F}[\rho]$.}

{  The minimization is initialized by creating some appropriate random density profiles with different symmetries.
If the value of $n_2(\bvec{r})$ exceeds one for an initial guess, the density profiles are renormalized to obtain a valid profile. 
In the course of the minimization, the packing fraction $\eta$ is then gradually increased back to its input value.
We iterate until the free energy differences between different structures, which typically are of the order $10^{-5}$ in units of the thermal energy, can be sufficiently resolved.
This is usually the case when the free energy changes by less than $5\cdot10^{-7}$ in the last 1000 iteration steps.
We thus assume that the numerical error in the free energy is of the order $5\cdot10^{-6}$.
Since the true free energy is always smaller for each structure, most of this error occurs as a systematic shift, irrelevant when comparing relative free energies as we do here.
In some cases, as for $R_\text{out}=6.3L$ and $b=0.29$ or $b=0.3$ in Fig.~\ref{fig_statediagramX}, there are competing structures with free energy differences of $10^{-6}$ and smaller,
making it difficult to unambiguously determine the global minimum.
However, the systematic errors due to the approximations in the density functional can be larger.}

 After equilibration of multiple structures, we compare the values of the free energy $\mathcal{F}[\rho]$ to distinguish between local and global minima
 and quantify the likelihood to observe a particular state in a corresponding experiment.
Calculations are performed on a quadratic spatial grid with a high enough resolution $\Delta x=\Delta y=0.2$ and  
$N_\phi=96$ orientational angles (for $p=10$).
{ Some structures were created with a smaller resolution and then further equilibrated for the given parameters.}

The obtained density profiles $\rho(\bvec{r},\phi)$ are usually graphically represented by a color scheme on a range from zero to three 
denoting the dimensionless orientationally averaged density
\begin{equation}
 \bar{\rho}(\bvec{r}):=\left(LD+\frac{D^2\pi}{4}\right)\!\int_0^{2\pi}\!\frac{\mathrm{d}\phi}{2\pi}\,\rho(\bvec{r},\phi)\,.
 \label{eq_barrho}
\end{equation}
Here we also display lines of length given by the orientational order parameter and orientation indicating the local orientational director field.
{ These quantities can be extracted from the local order tensor 
\begin{equation}
 Q(\bvec{r},\phi)=\frac{2\,\rho(\bvec{r},\phi)}{\int_0^{2\pi}\mathrm{d}\phi\,\rho(\bvec{r},\phi)}\!\matX{\cos^2\phi-\frac{1}{2}}{\cos\phi\sin\phi}{\cos\phi\sin\phi}{\sin^2\phi-\frac{1}{2}}
 \label{eq_QQ}
\end{equation}
as its largest Eigenvalue and the corresponding Eigenvector, respectively.} 
An alternative graphical representation, used in Fig.~\ref{fig_phases}, depicts the local packing fraction $n_2(\bvec{r})$. 
As this quantity corresponds to the density weighted with the particle area, it
directly illustrates the locations of the particles, while ranging between zero and one. Since the behavior is fluid-like within smectic layers, the local packing fraction is generally smeared-out,
while the silhouettes of individual particles in the packed regions close to the walls are directly visualized.

Our benchmark calculation in bulk using the described DFT approximation locates the nematic--smectic transition around $\eta\approx0.62$.
{ As this two-dimensional area fraction is not directly comparable to the experimentally accesible volume fraction $\phi$, 
our theoretical calculations in the confined system are generally carried out for $\eta=0.65$.
With this choice of the smectic density, the relative deviation from the bulk nematic--smectic transition density in experiment and theory are similar.}
At this area fraction, the optimal bulk layer spacing is $\lambda_0=12.56D=1.142W$.

 To determine the inclusion size ratio $b_\text{t}$ at which a structural transition occurs in annular confinement,
we select two values $b_1$ and $b_2$, usually differing by $b_2-b_1=0.01$, 
and check whether the sign of the free-energy difference between two states of interest is different.
If so, we determine $b_1<b_\text{t}<b_2$ as the point where the linearly interpolated free energies are equal.

\section{\label{SN2} Classification and identification of the different smectic states}

In general, we use the following criteria to distinguish the different {{smectic states}} 
through occuring defects and the arrangement of smectic layers.
This is most easily done by observing the number and orientation of line disclinations and identifying connected layers, which span between two opposite sites of the outer boundary. 
Usually, there is a typical number of edge dislocations associated with Shubnikov structures, but a certain number of edge dislocations can also occur in other states.
Microscopic details are not relevant for this first classifications.

{{Laminar (or bridge) state $\mathcal{L}$ (or $\mathcal{B}$):}} there is at least one connected layer at each side of the inclusion. 
Usually, this is equivalent to observing two anti-radial disclination lines close to the outer boundary. 
{{Domain state $\mathcal{D}$:}} there are exactly three radially oriented disclination lines, separated by 120 degrees; the layers from different domains meet at an angle of 90 degrees.
{{Shubnikov state $\mathcal{S}$:}} there is neither a connected layer nor any disclination line; all layers span between the inner and outer boundary.
{{Laminar--domain composite state $\mathcal{C^{LD}}$:}} there are one or two radial disclination lines and at least one connected layer (or anti-radial disclination line) at only one side of the inclusion. 
{{Laminar--Shubnikov composite state $\mathcal{C^{LS}}$:}} there are no radial disclination lines and at least one connected layer (or anti-radial disclination line) at only one side of the inclusion.
{{Domain--Shubnikov composite state $\mathcal{C^{DS}}$:}} there are one or two radial disclination lines and no connected layer.

For each intact experimental chamber,
we identify the dominant state by visual inspection of our particle-resolved images according to the above criteria.
In the smallest experimental chambers considered it is not always possible to clearly associate an observed structure with one of these states.
Strongly deformed theoretical structures are not taken into further consideration, since it is clear from their large free energy that they are irrelevant in the search for the
most stable mesoscopic state.

\section{\label{SN3} Phenomenological model for defect energies \label{app_pheno}}

 The existence of the laminar--Shubnikov transition in Fig.~\ref{fig_statediagram} can be understood in the light of a phenomenological model.
 Making some minimal assumptions, we estimate the energy penalty resulting from the characteristic defects in each state. 
Let us first denote the energy of a disclination line with unit length by $\delta$ and of each edge discolation by $u_\text{ed}$.
  We then assume that the two anti-radial disclination lines in the laminar state have the length $R_\text{out}$ 
  and that each radial domain boundary is a straight line of length $R_\text{out}-R_\text{in}$. 
  The number 
  \begin{equation}
   N_\text{out}-N_\text{in}\approx\frac{2\pi R_\text{out}-2\pi R_\text{in}}{\lambda_0}
  \end{equation}  
  of edge dislocations in the Shubnikov state is assumed as the difference of the outer and inner wall perimeters, divided by the bulk layer spacing $\lambda_0$. 
  The resulting energies read
  \begin{align}
   U_\mathcal{L}&=2\,\delta\,R_\text{out}\,,\cr U_\mathcal{D}&=3\,\delta\,R_\text{out}\left(1-b\right)\,,\cr U_\mathcal{S}&=\frac{2\pi}{\lambda_0}R_\text{out}\left(1-b\right)u_\text{ed} 
   \label{eq_energies}
  \end{align}
for the laminar, domain and Shubnikov state, respectively.
  
  For all choices of the fit parameters $\delta$ and $u_\text{ed}$, the state diagram predicted by this simple model is independent of the size of the annulus, because all energies in Eq.~\eqref{eq_energies} scale with $R_\text{out}$.
  Moreover, there is exactly \emph{one} stable transition for increasing $b$, namely from a laminar state, at small inclusions, to either a domain or Shubnikov state at large inclusion.
 One of the two latter states is always metastable, since both $U_\mathcal{D}$ and $U_\mathcal{S}$ are proportional to $(1-b)$. These energies further decrease to zero for an infinitely thin annulus ($b\rightarrow1$),
while the energy $U_\mathcal{L}$ assumed for the laminar state remains constant when increasing $b$. Therefore, the laminar state always becomes metastable for large inclusion sizes. 
   A stable laminar--domain transition may occur at $b=\frac{1}{3}$ 
if $U_\mathcal{D}<U_\mathcal{S}$, which means that $3\delta\lambda<2\pi u_\text{ed}$. 
In the opposite case, we expect a stable laminar--Shubnikov transition at $b=1-\frac{\lambda\delta}{\pi u_\text{ed}}<\frac{1}{3}$, while the domain state is only metastable.

  Although this phenomenological model does not capture all theoretical and experimental observations, it nicely reflects the competition between domain- and Shubnikov states, which possess the same director topology at the outer wall,
  and rationalizes the laminar--Shubnikov transition, which is one of our main results. 
Using as an input to the presented model that this transition can be roughly observed at $b_0=0.3$, we can determine the ratio 
\begin{equation}
\frac{\lambda\delta}{u_\text{ed}}=\pi\left(1-b_0\right)\approx2.2\,
\end{equation}
of the two energy parameters entering in Eq.~\eqref{eq_energies}.
  The transition lines for these parameters are drawn into the state diagram in Fig.~\ref{fig_statediagram}.
 
This model calculation provides a rough interpretation of some experimental findings,
but has limited predictive power, e.g., it does not feature any dependence of the location of the transitions on the total confinement size.
What is neglected in this simple model are all types of elastic deformations, 
the interaction of line and point defects, 
preferences in the wall alignment, 
a locally adaptable smectic layer spacing 
and density differences within the smectic layers close to the walls or to defects. 
Altogether, these effects occur on the order of the correlation length, which is typically on the particle scale.
This underlines the importance of a fully microscopic theoretical treatment to comprehensively characterize the emerging structures in extreme confinement, which is illustrated in Fig.~\ref{fig_compare}.

\section{\label{SN4} Optimal microscopic arrangement of layers \label{sec_intrinsictopology}}

Associating an observed structure with a particular smectic state does not account for the full microscopic information.
In particular, different structures can mainly be distinguished by counting the explicit numbers of smectic layers.
The optimal microscopic structure which is most stable for a given state itself depends on the competition between different driving forces, which we detail below.

\subparagraph*{Laminar state $\mathcal{L}_{N_\text{con},N_\text{dis}}$.}

In a given laminar state $\mathcal{L}_{N_\text{con},N_\text{dis}}$,
there are precisely $N_\text{con}$ connected layers spanning throughout the confinement and $N_\text{dis}$ layers disconnected by the inclusion, summing up to a total number $N_\text{tot}=N_\text{con}+N_\text{dis}$ of layers (where $N_\text{tot}=N_\text{con}$ in the bridge state), see Fig.~\ref{fig_comparel}.
Depending on the geometry, there may further be two anti-radial disclination lines close to the boundary.
We consider the shape of the disclination line and the number of layers in the separated domains as structural details of next order,
since the theory automatically finds the optimal structure corresponding to $\mathcal{L}_{N_\text{con},N_\text{dis}}$ in a given confinement.

In general, the theory predicts only relatively small deviations 
of the laminar layer spacing from the bulk value $\lambda_0$.
 As a consequence, the total number $N_\text{tot}$ of smectic layers in the parallel domain,
 fluctuates as a function of the inclusion size ratio $b$, as shown in Fig.~\ref{fig_comparel},
 which also applies to the length and shape of the disclination lines. 
The {{optimal microscopic laminar structure}} is then determined by
 the integer multiples of $\lambda_0$ which are closest to the respective dimensions of the confinement.
For example, $N_\text{dis}$ is usually bounded by the diameter $2R_\text{in}/\lambda_0$ of the inclusion
to avoid additional defects or strong deformations near the wall.

\subparagraph*{Shubnikov state $\mathcal{S}_{N_\text{in},N_\text{out}}$.}

the most significant structural property of a given Shubnikov state $\mathcal{S}_{N_\text{in},N_\text{out}}$
is the number $N_\text{out}-N_\text{in}$ of point defects, which is
is determined by the numbers $N_\text{in}$ and $N_\text{out}$ of layers in direct contact with the inner and outer wall, respectively.
The exact radial (and relative) location of these defects are structural details of next order,
 which can be accounted for by creating and comparing a large number of different structures.
As indicated in Fig.~\ref{fig_compares}, we further consider a local layer spacing
\begin{align}\lambda_\text{in}=\frac{\pi (2R_\text{in}+D)}{N_\text{in}}\end{align}
 at the inner and
\begin{align}\lambda_\text{out}=\frac{\pi \sqrt{(2R_\text{out}-D)^2-L^2}}{N_\text{out}}\end{align}
at the outer wall, explicitly calculated here from the circumference of the line connecting the particle centers in direct vicinity of each wall. 
The effective layer spacing associated with the closest planar packing (assuming that all rods are aligned perfectly tangential to the wall) 
corresponds to the length of the arc through the particle center between the two radial lines tangential to the particle and reads
\begin{align}
 \lambda^*_\text{in}=(2R_\text{in}+D)&\left(\arctan\left(\frac{L}{2R_\text{in}+D}\right)\right.\cr&+\left.\arcsin\left(\frac{D}{\sqrt{(2R_\text{in}+D)^2+L^2}}\right)\right).\ \ \ \ \ \ 
\end{align}
This will serve as a reference value.

 To quantify the variation in the local layer spacing, we determine $\lambda_\text{wall}$ (the subscript 'wall' either stands for 'in' or 'out')
 for several theoretical and experimental structures.
According to Fig.~\ref{fig_compares}, we find that 
$\lambda_\text{in}$ is much smaller and $\lambda_\text{out}$ is much larger than the bulk value $\lambda_0$ in both theory and experiment.
This means that the
{{optimal microscopic Shubnikov structure}} results from the competition between optimizing the layer spacing and minimizing the number $N_\text{out}-N_\text{in}$ of point defects.
This deviation is most significant in extreme confinement.
For small inclusions, $N_\text{in}$ can even become larger than the maximal number of rods that can be packed around the wall with a planar orientation, i.e., $\lambda_\text{in}<\lambda^*_\text{in}$, 
 which results in the tilt shown in Fig.~\ref{fig_compare}f.
Moreover, there are extreme structures without any defects for sufficiently small distances between the walls,
 e.g., the theory predicts a stable $\mathcal{S}_{19,19}$ for $R_\text{out}=4.1L$ and $b=0.8$.

\subparagraph*{Laminar--Shubnikov composite state $\mathcal{C}^\mathcal{LS}_{N_\text{con},N_\text{in},N_\text{out}}$.}

In a given Laminar--Shubnikov composite state $\mathcal{C}^\mathcal{LS}_{N_\text{con},N_\text{in},N_\text{out}}$,
there is a precise number $N_\text{con}$ of connected layers, reflecting the characteristics of the laminar state.
The number of perpendicular layers separated by the adjacent disclination line (if present) constitutes here an important detail, 
since the location of the connected layers relative to the center of the annulus is not fixed by symmetry, in contrast to the laminar state.
Therefore, this composite state is characterized by two further numbers $N_\text{in}$ and $N_\text{out}$,
i.e., the total number of contacts with the two walls, reflecting the characteristics for the Shubnikov state.

The stability of such a laminar--Shubnikov composite state can be explained by a mutual relaxation of the external constraints, which we understand as follows.
The innermost connected layer in the laminar part can take the optimal position relative to the inclusion that minimizes the penalty arising from deformations and false wall alignment.
In contrast, the arrangement of the layers in an ordinary laminar state is dictated by its axial symmetry.
The same is true in view of the optimal layer spacing at the boundaries in the Shubnikov part of the structure, which is restricted to discrete values in the ordinary Shubnikov state.
Further note that the intersection of the two half-structures is smooth, i.e., it does not introduce additional domains or deformations.
\section{\label{SN5} Topological protection of mesoscopic defect structures \label{sec_topo}}

Here we make some more statements regarding the stability of the different states deduced from their topological details.
As elaborated in the main text, the total topological charge $Q$, which is the sum of the charge of all occurring defects, compare Fig.~\ref{fig_topo},
reflects the Euler characteristic $\chi$ of the bounding domain.
Hence, in annular confinement we have $Q=\chi=0$.
Due to the typically different types of defects and their spatial distribution (oppositely charged defects are separated by rigid layers),
the $\mathcal{L}$, $\mathcal{C^{LS}}$ and $\mathcal{S}$ states are topologically protected,
i.e., they cannot be transformed into one another upon a smooth variation of the particle distribution.

{ In contrast, we find that the theoretically created domain structures are not topologically protected.
 In fact, they are unstable with respect to topologically equivalent} Shubnikov structures, as depicted in Fig.~\ref{fig_statediagramDOM}.
In the course of the minimization of DFT, the 
{ end-point defects of the radial disclination lines gradually annihilate, while the layers at the boundary become larger, eventually resulting in a uniform}
 orientational order.
The chosen annular geometry thus prefers the Shubnikov state.
However, keeping in mind the experimental observations, we suspect that a domain state can become stable, e.g., 
by increasing the density or adding some polydispersity. 
In general, it should become more likely to observe a domain state in other confining geometries respecting the threefold symmetry.

 \begin{figure*}
\rotatebox{90}{\large$\ \ \ \ \ \ \ \mathcal{S}_{12,30}$}\hfill\includegraphics[width=0.17\linewidth]{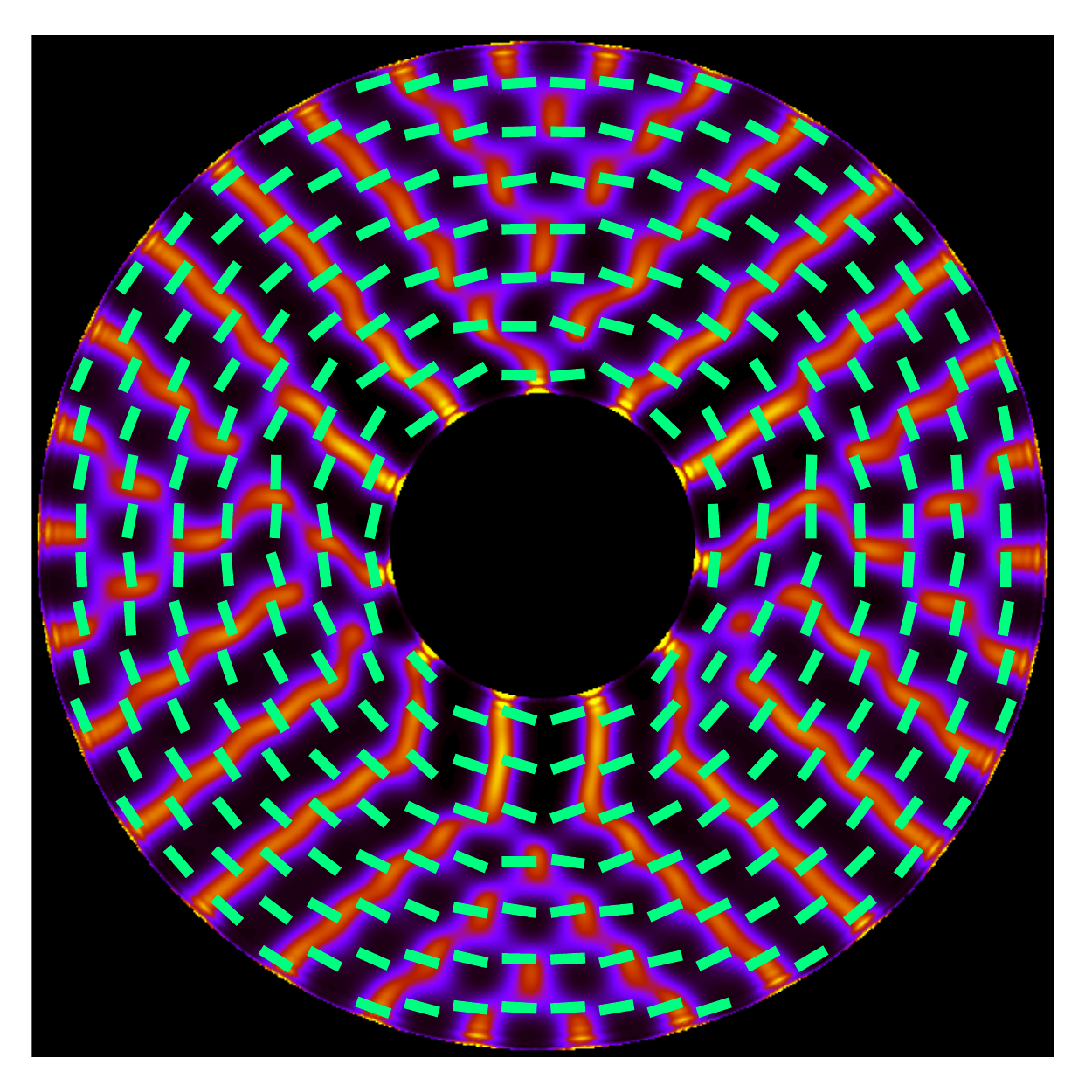}\hfill\parbox{0.015\linewidth}{\centering\vspace*{-2.85cm}$\rightarrow$}\hfill
\includegraphics[width=0.17\linewidth]{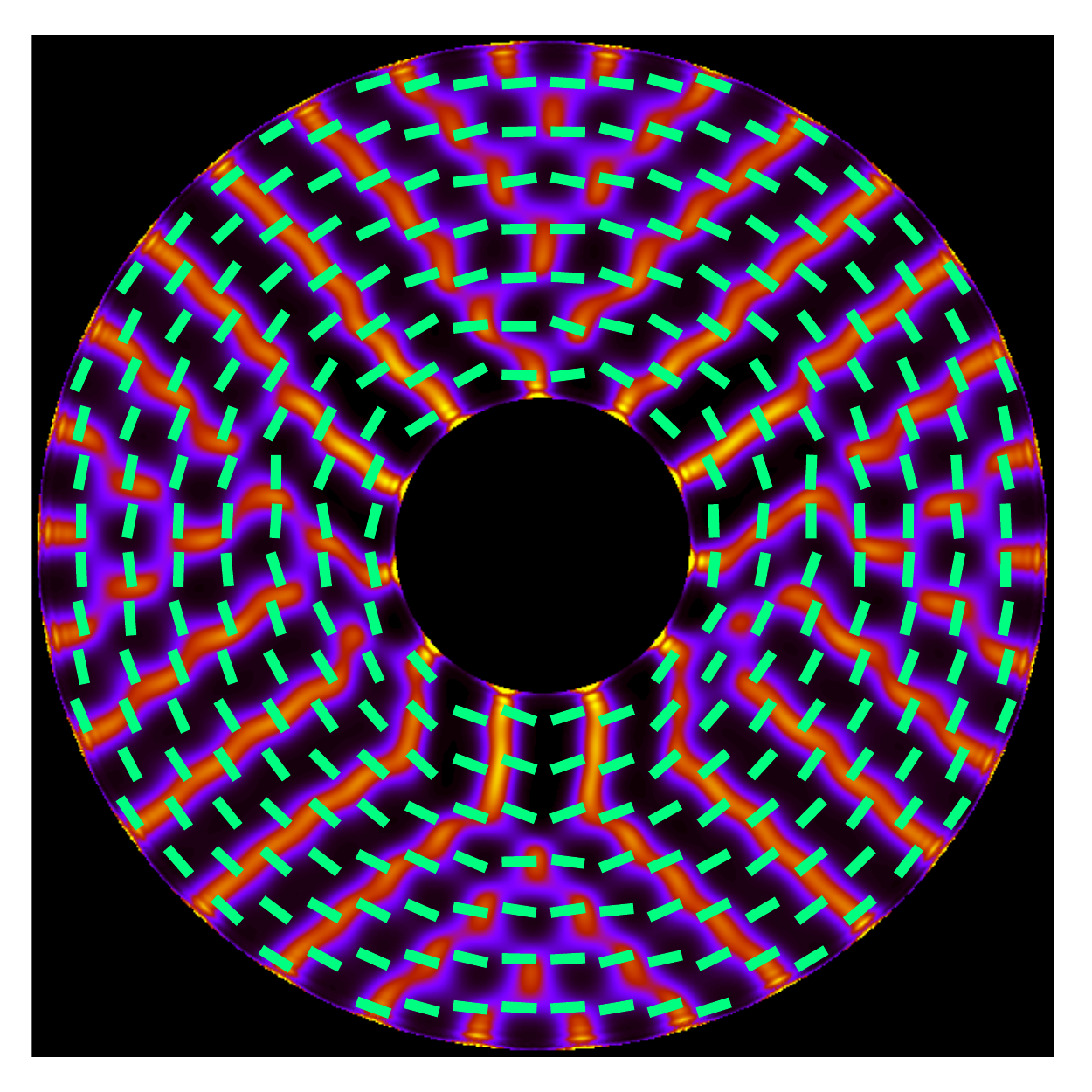}\hfill\parbox{0.015\linewidth}{\centering\vspace*{-2.85cm}$\rightarrow$}\hfill
\includegraphics[width=0.17\linewidth]{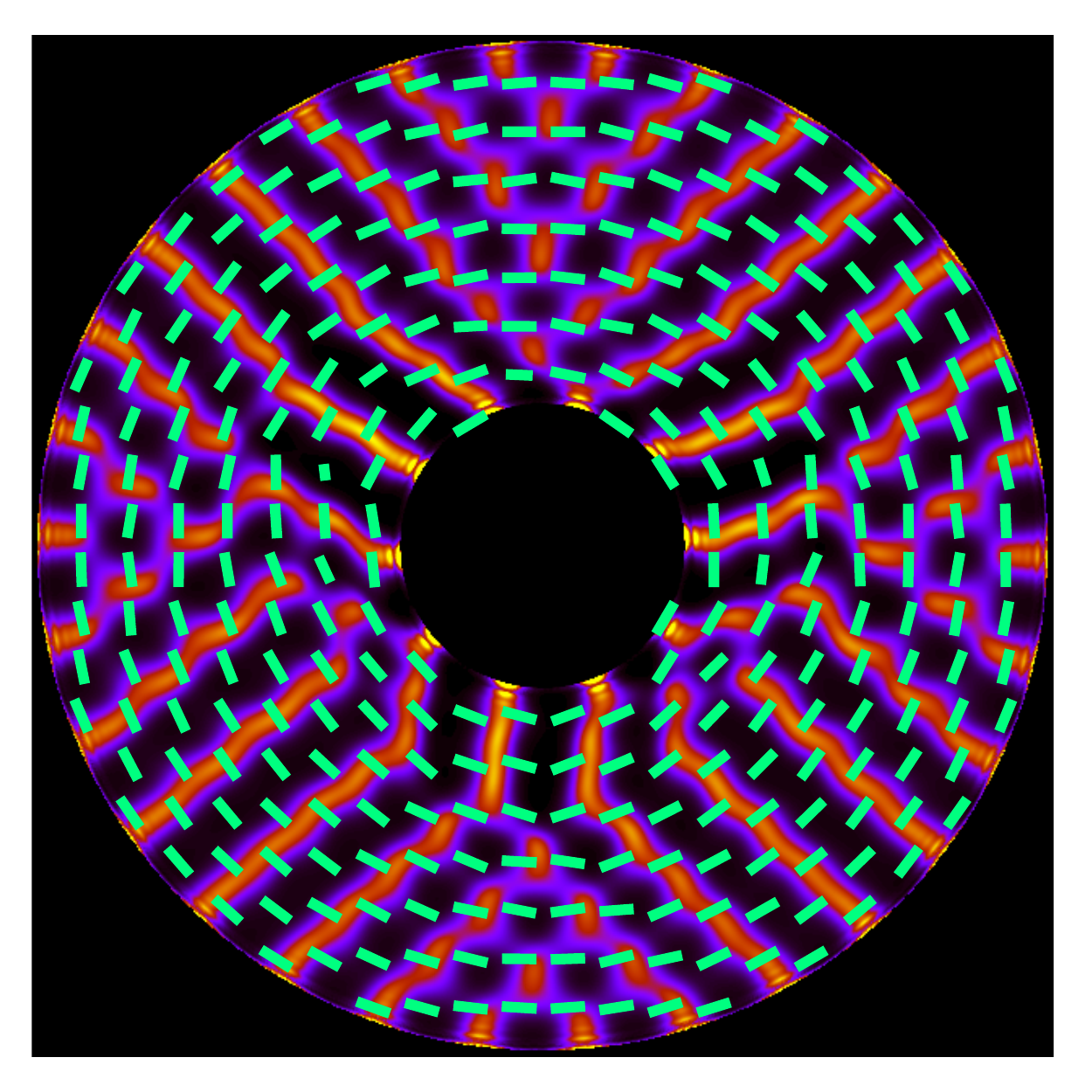}\hfill\parbox{0.015\linewidth}{\centering\vspace*{-2.85cm}$\rightarrow$}\hfill
\includegraphics[width=0.17\linewidth]{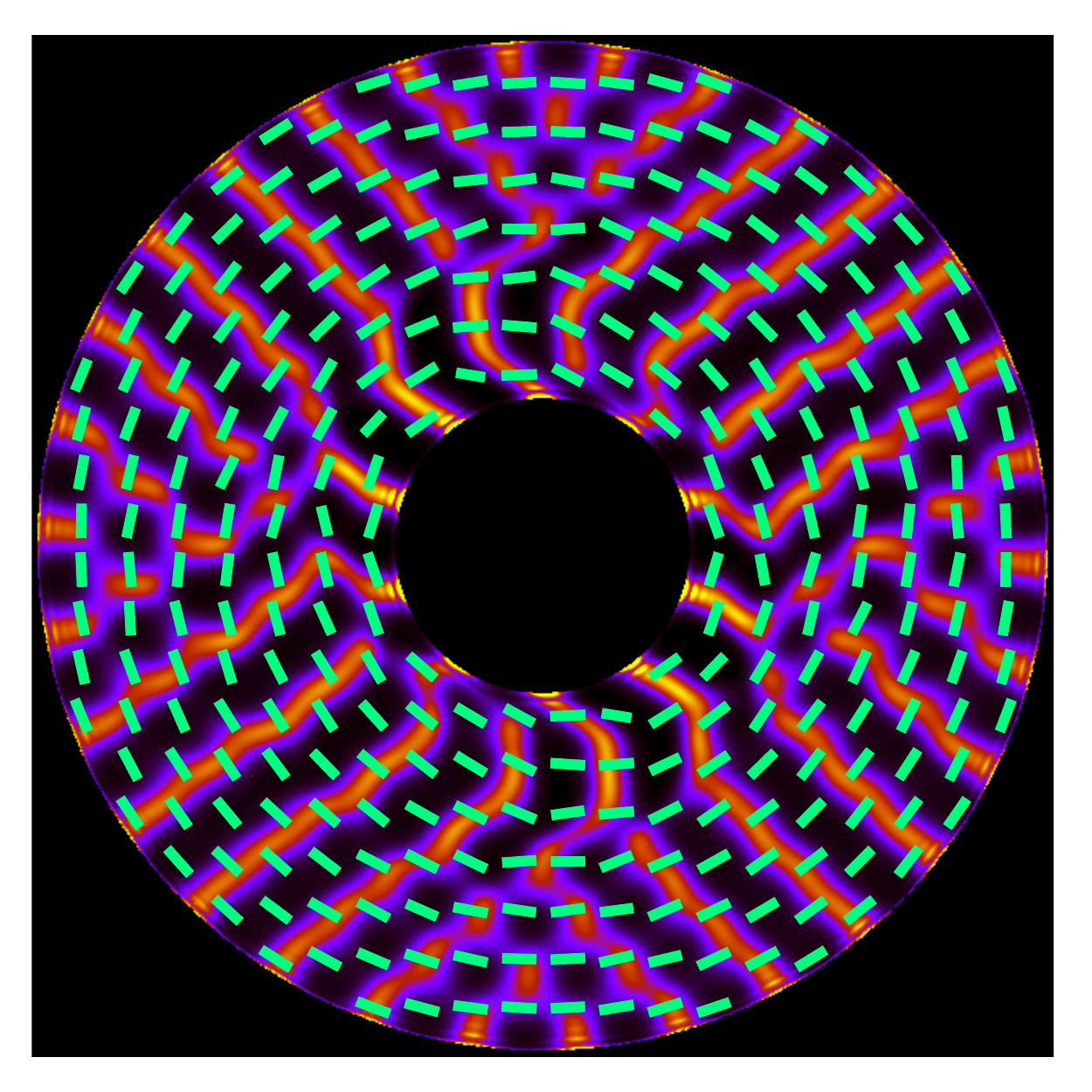}\hfill\parbox{0.015\linewidth}{\centering\vspace*{-2.85cm}$\rightarrow$}\hfill
\includegraphics[width=0.17\linewidth]{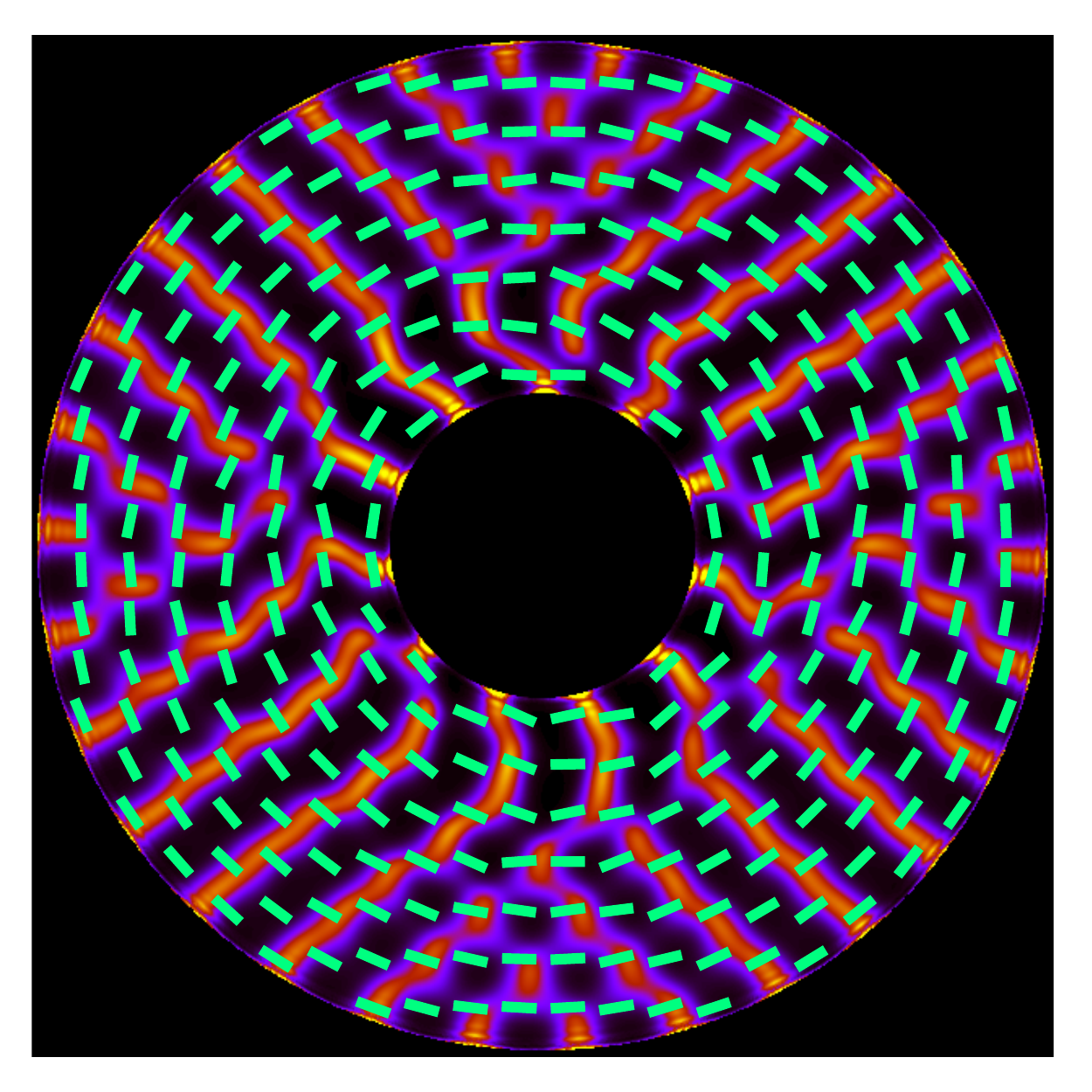}\hfill
\includegraphics[height=0.17\linewidth]{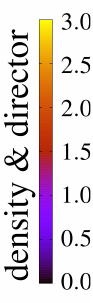}\\\vspace*{-0.3cm}
\flushleft\ \ \ \ \ \ \ \  $\mathcal{S}_{11,30}$, $b=0.29$ \ \ \ \  \ \ \ \ \ \ $\mathcal{S}_{11,30}$, $b=0.28$ \ \ \ \ \ \ \ \ \ \:\,$\mathcal{S}_{10,30}$, $b=0.27$ \ \ \ \ \ \ \ \ \ \ \;$\mathcal{S}_{10,30}$, $b=0.28$ \ \  \ \: \ \ \ \ \ \,$\mathcal{S}_{11,30}$, $b=0.29$\hspace*{0.2cm} \\\vspace*{0.2cm}
\rotatebox{90}{\large$\ \ \ \ \ \ \ \mathcal{S}_{11,29}$}\hfill\includegraphics[width=0.17\linewidth]{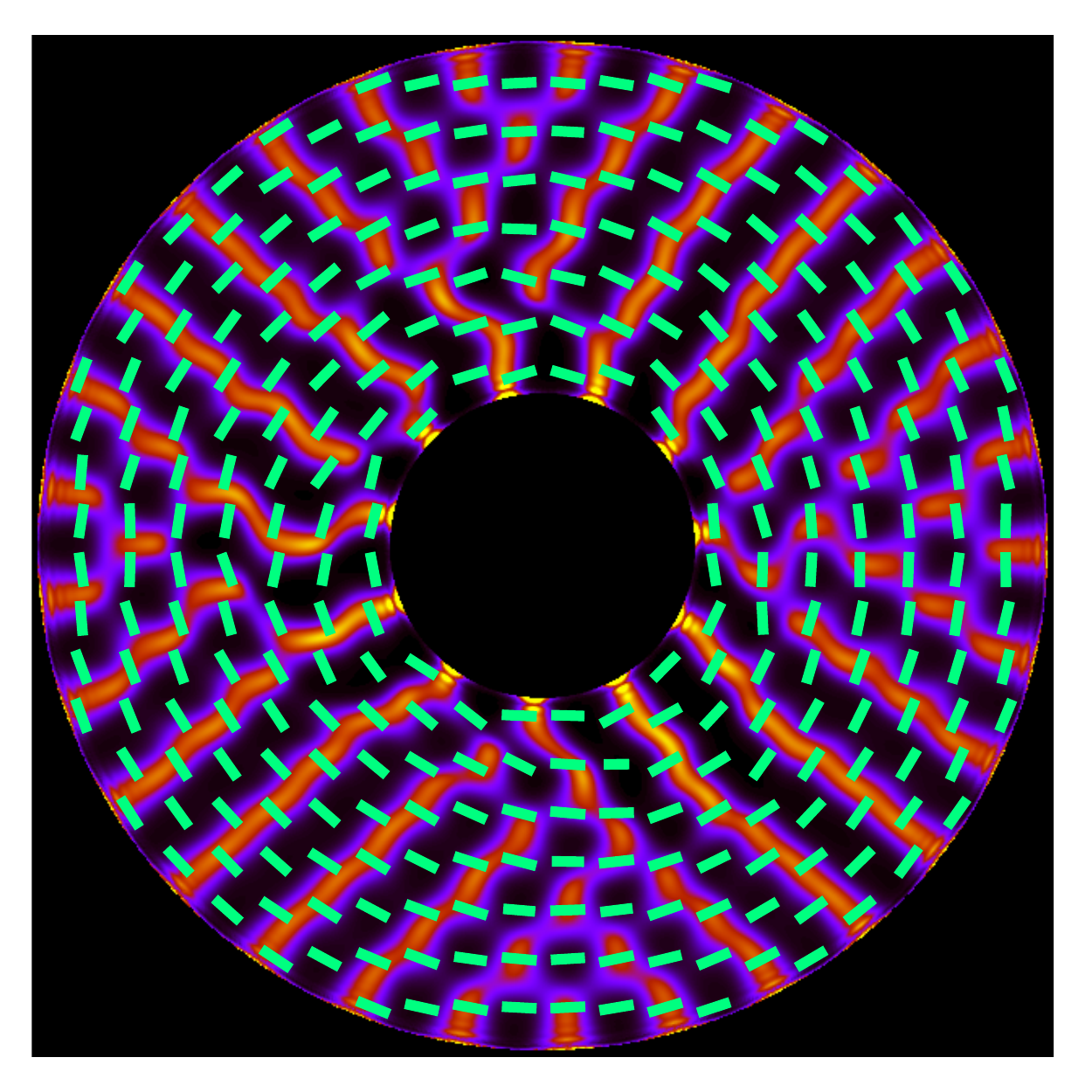}\hfill\parbox{0.015\linewidth}{\centering\vspace*{-2.85cm}$\rightarrow$}\hfill
\includegraphics[width=0.17\linewidth]{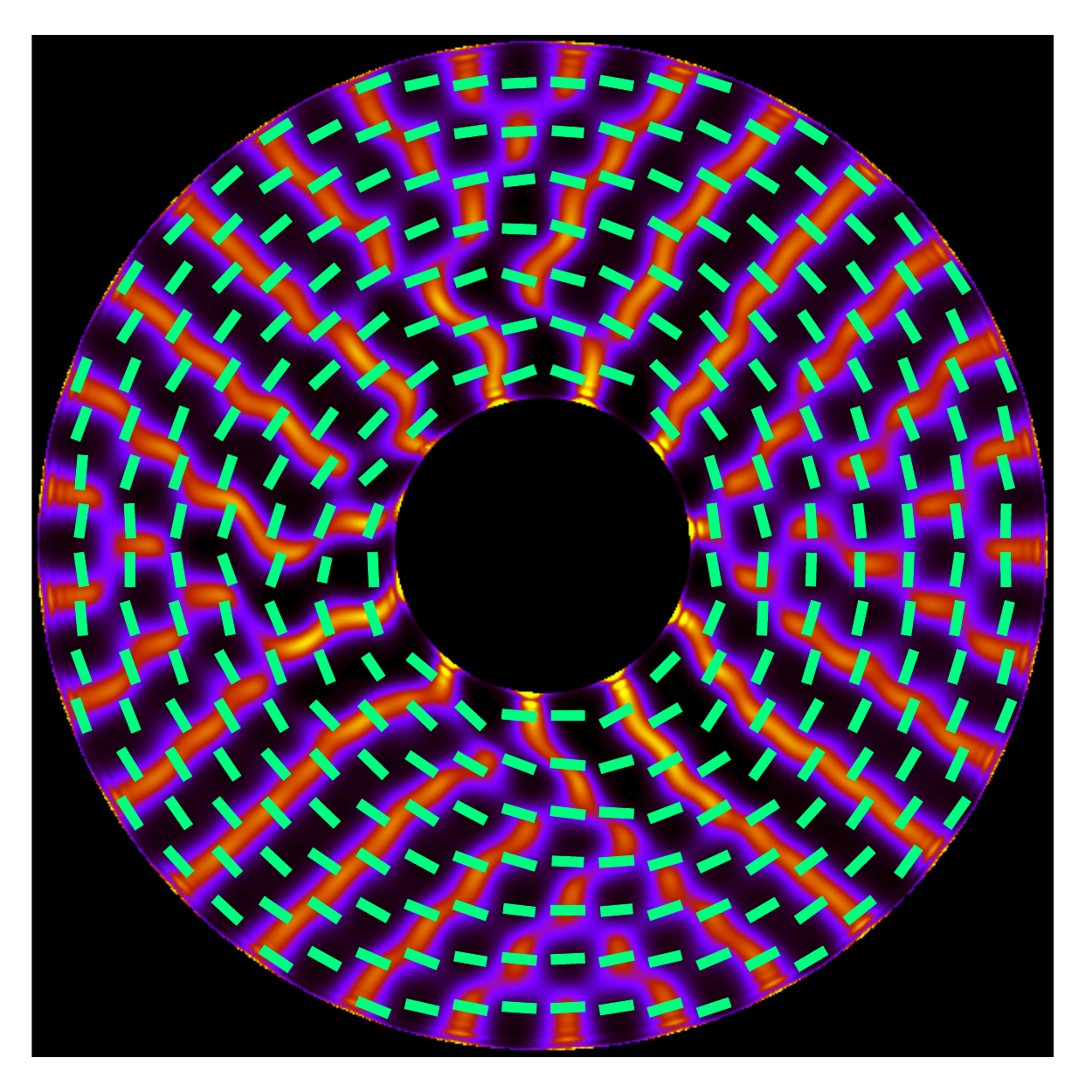}\hfill\parbox{0.015\linewidth}{\centering\vspace*{-2.85cm}$\rightarrow$}\hfill
\includegraphics[width=0.17\linewidth]{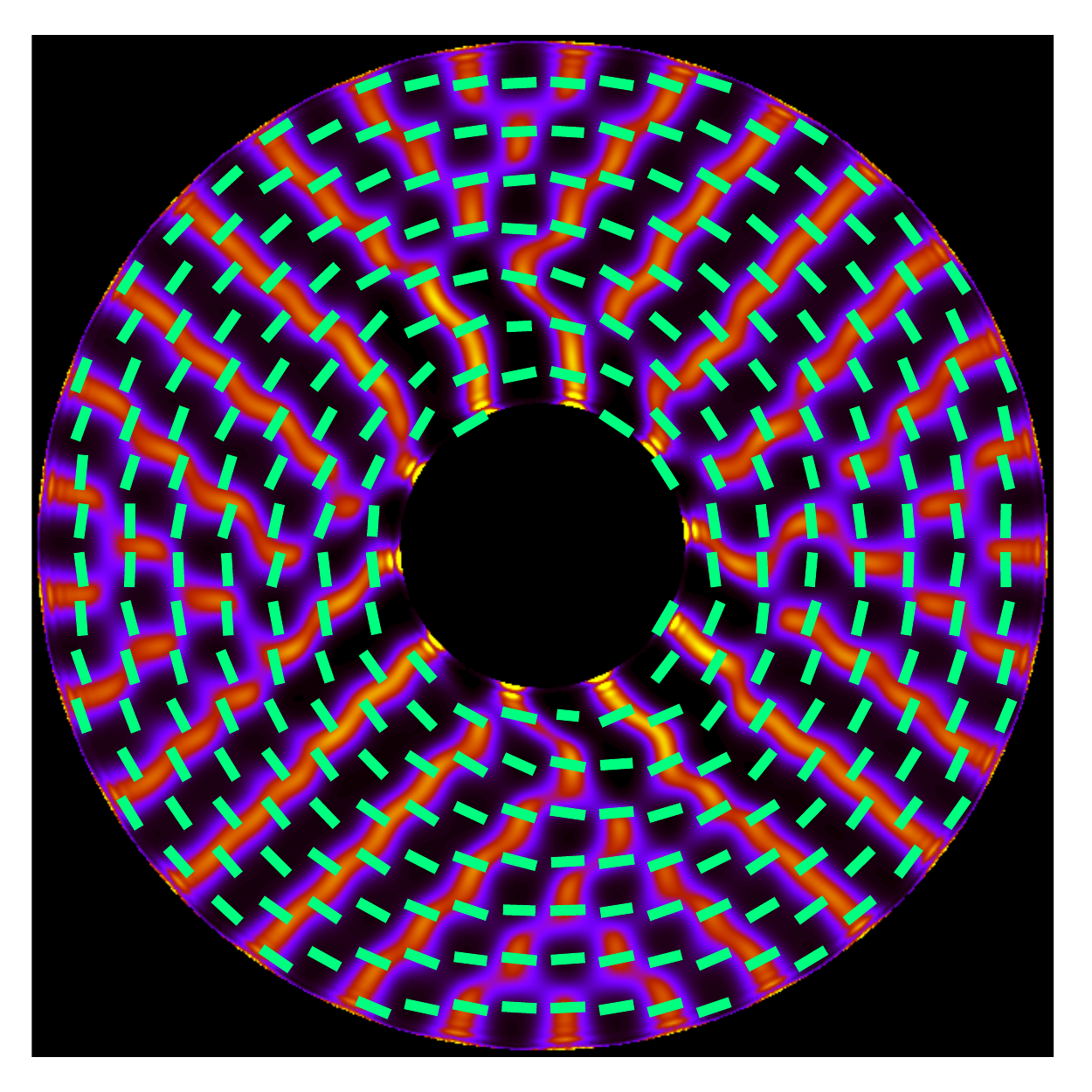}\hfill\parbox{0.015\linewidth}{\centering\vspace*{-2.85cm}$\rightarrow$}\hfill
\includegraphics[width=0.17\linewidth]{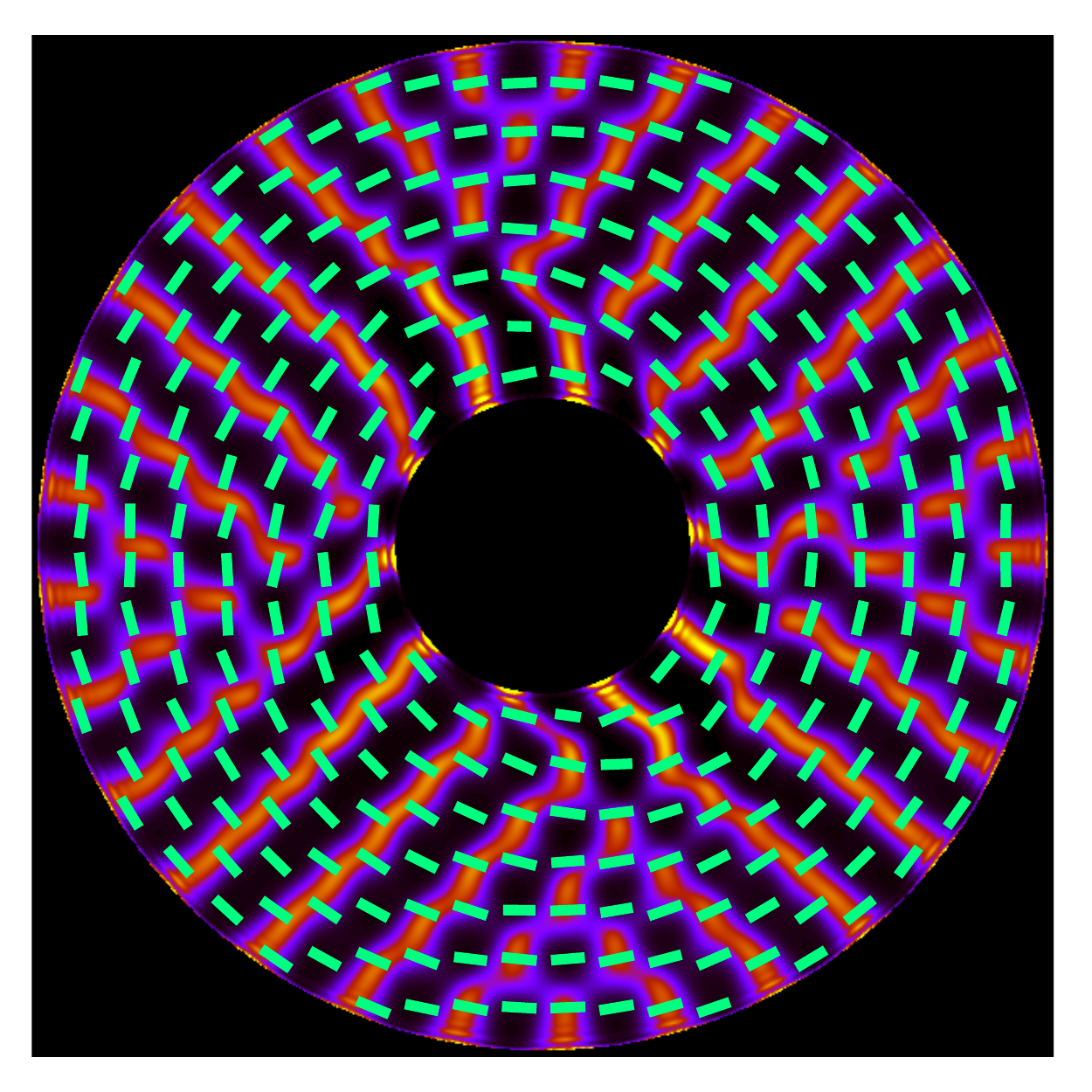}\hfill\parbox{0.015\linewidth}{\centering\vspace*{-2.85cm}$\rightarrow$}\hfill
\includegraphics[width=0.17\linewidth]{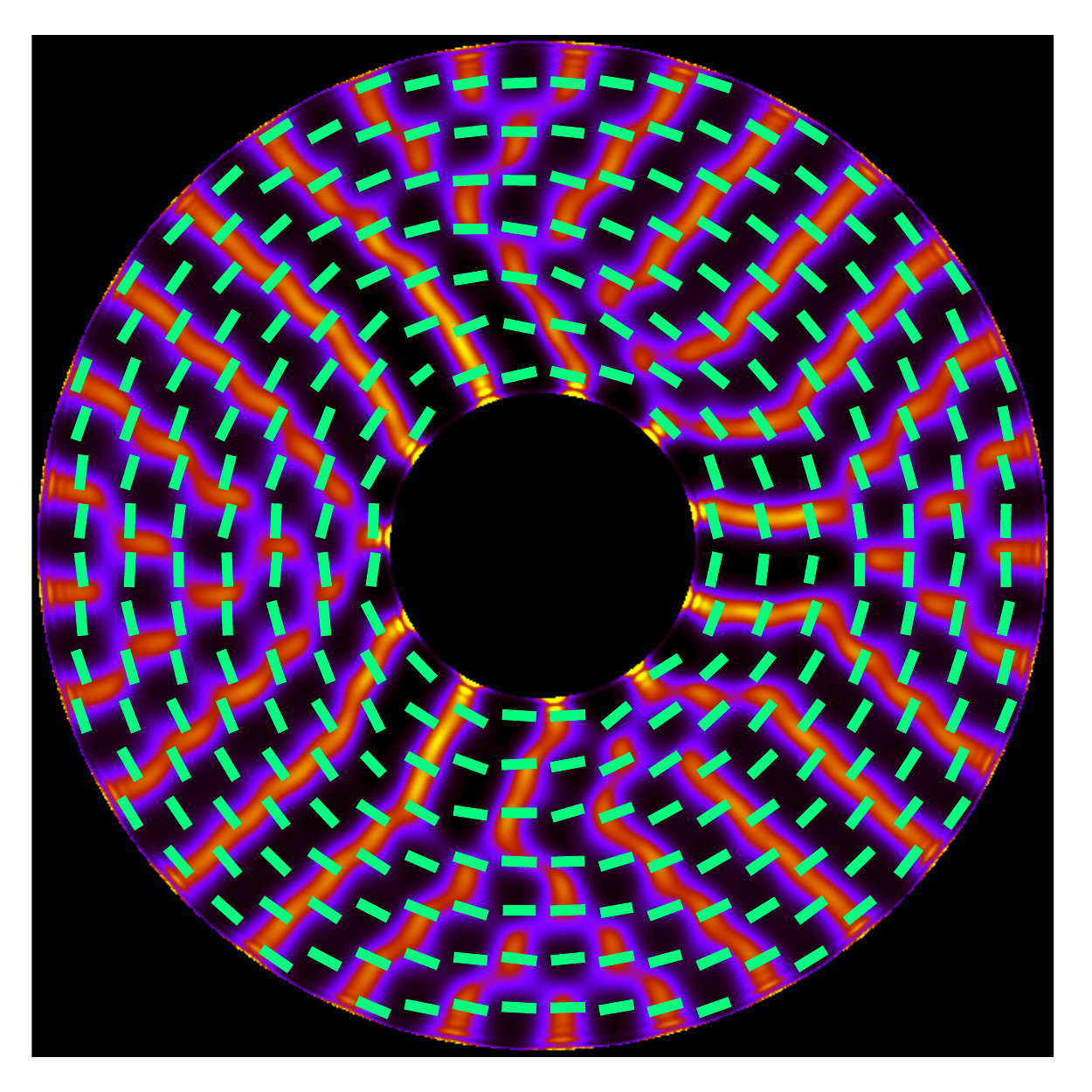}\hfill
\includegraphics[height=0.17\linewidth]{legend03newX.pdf}\\\vspace*{-0.3cm}
\flushleft\ \ \ \ \ \ \ \  $\mathcal{S}_{11,29}$, $b=0.29$ \ \ \ \  \ \ \ \ \ \ $\mathcal{S}_{11,29}$, $b=0.28$ \ \ \ \ \ \ \ \ \ \:\,$\mathcal{S}_{10,29}$, $b=0.27$ \ \ \ \ \ \ \ \ \ \ \;$\mathcal{S}_{10,29}$, $b=0.28$\ \ \: \ \: \ \ \ \ \ $\mathcal{S}_{11,29}$, $b=0.29$\hspace*{0.2cm} \\\vspace*{0.2cm} \rotatebox{90}{\large$\ \ \ \ \ \ \ \mathcal{S}_{11,30}$}\hfill\includegraphics[width=0.17\linewidth]{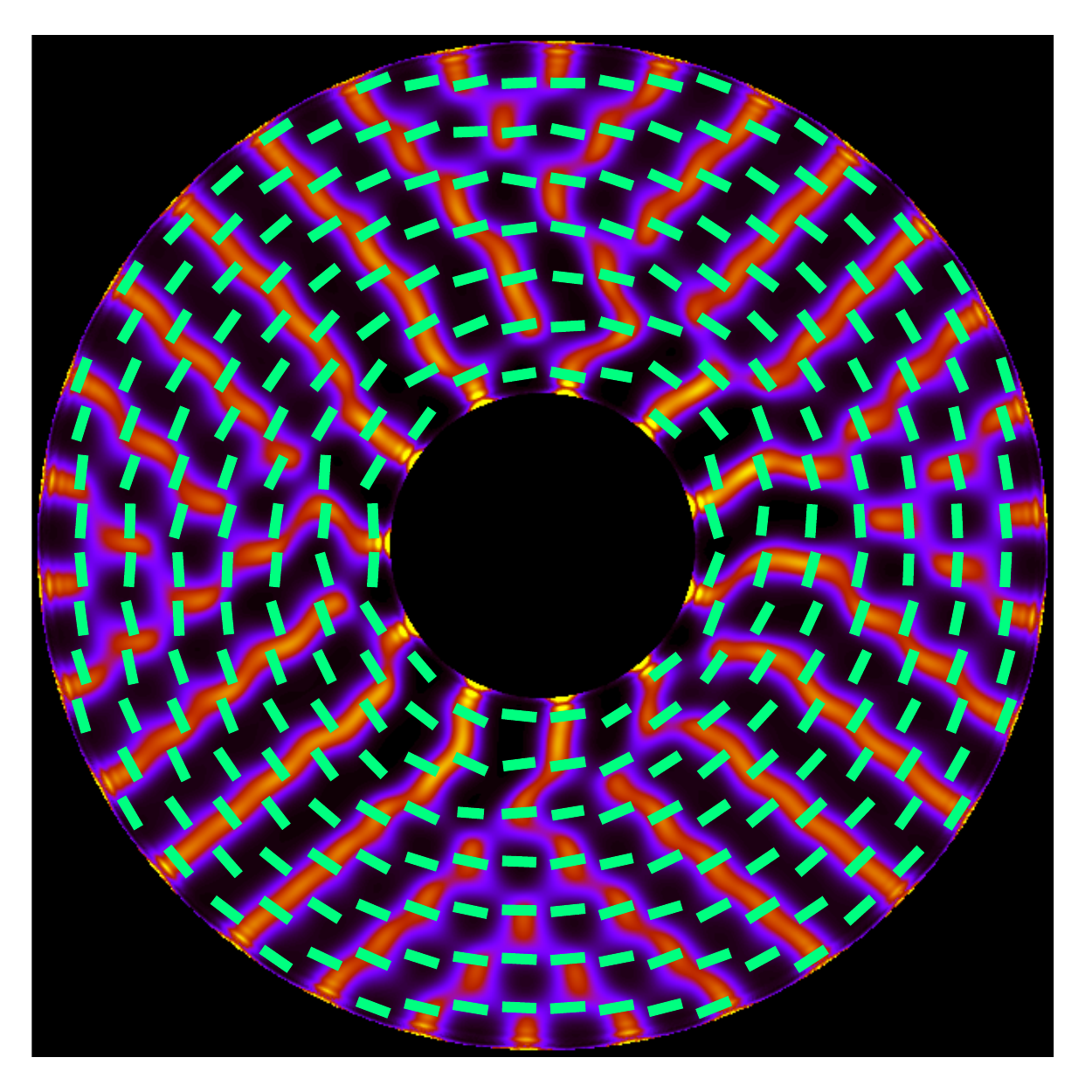}\hfill\parbox{0.015\linewidth}{\centering\vspace*{-2.85cm}$\rightarrow$}\hfill
\includegraphics[width=0.17\linewidth]{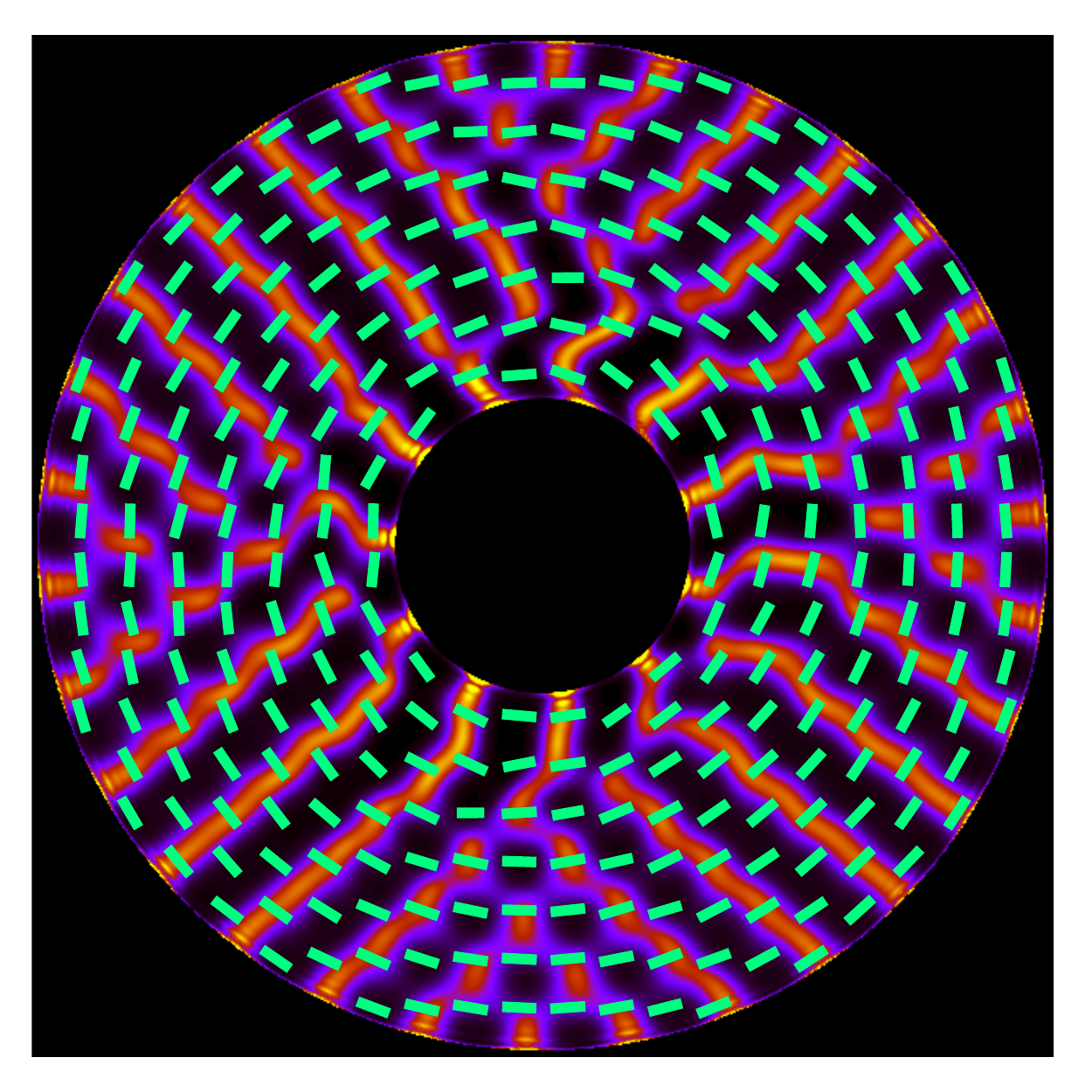}\hfill\parbox{0.015\linewidth}{\centering\vspace*{-2.85cm}$\rightarrow$}\hfill
\includegraphics[width=0.17\linewidth]{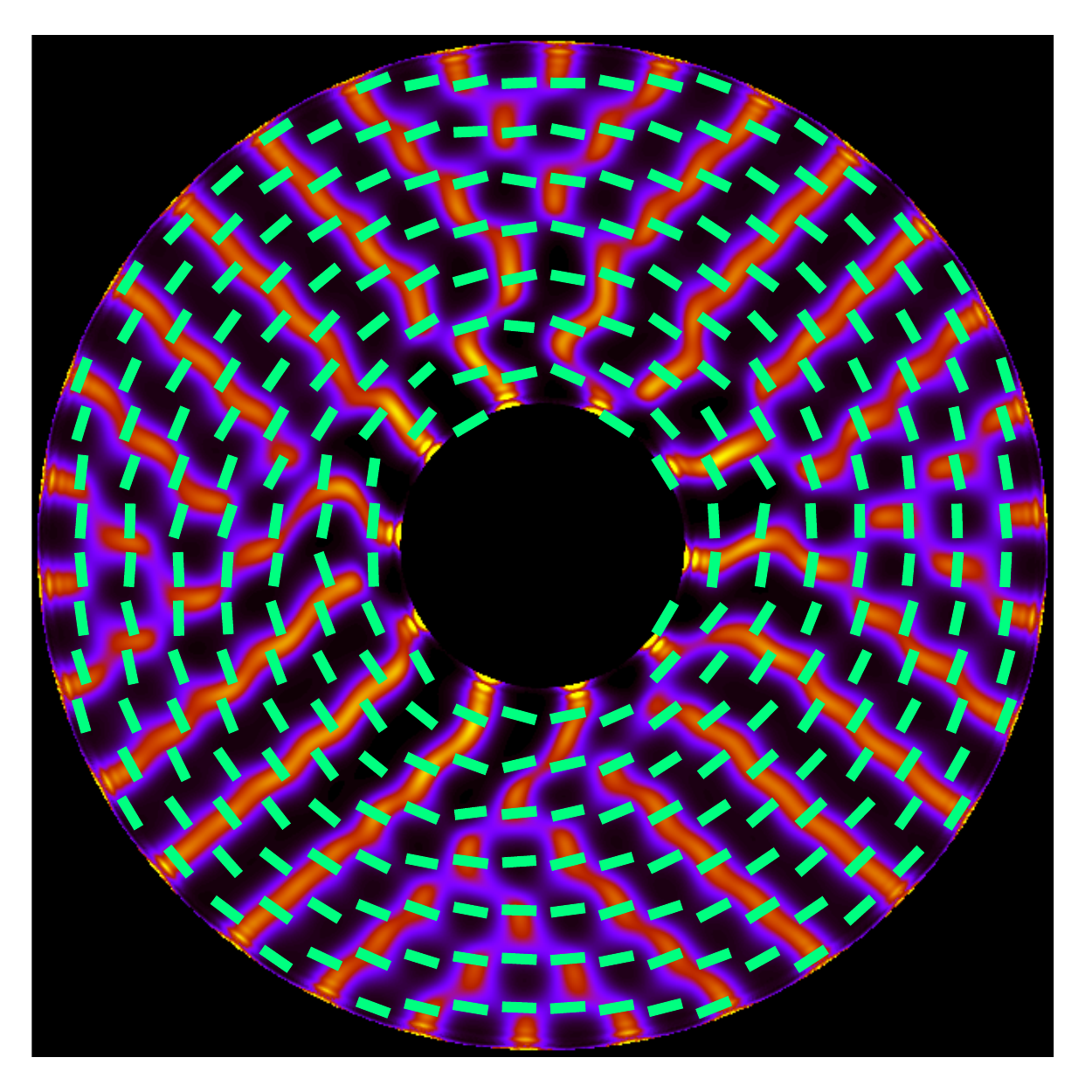}\hfill\parbox{0.015\linewidth}{\centering\vspace*{-2.85cm}$\rightarrow$}\hfill
\includegraphics[width=0.17\linewidth]{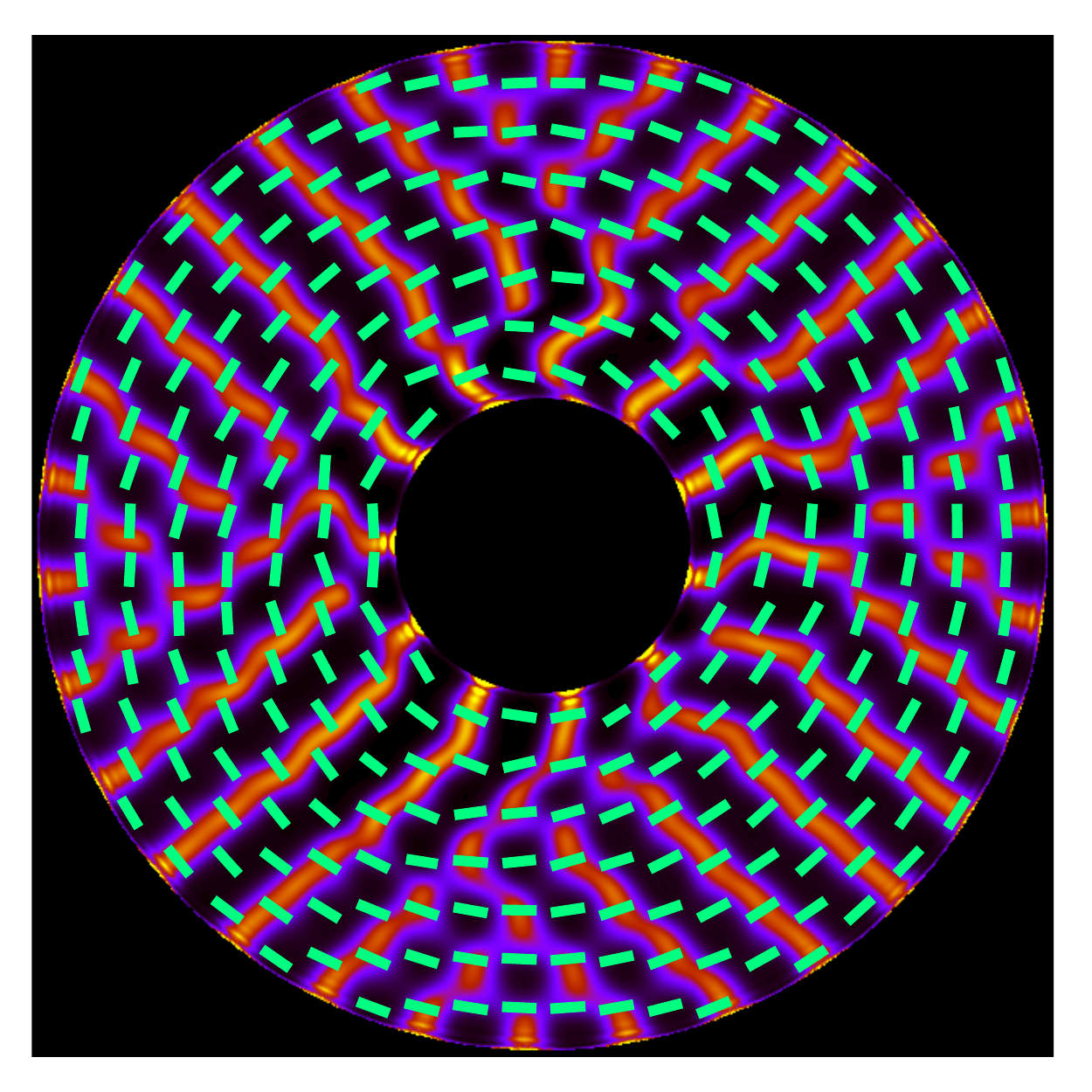}\hfill\parbox{0.015\linewidth}{\centering\vspace*{-2.85cm}$\rightarrow$}\hfill
\includegraphics[width=0.17\linewidth]{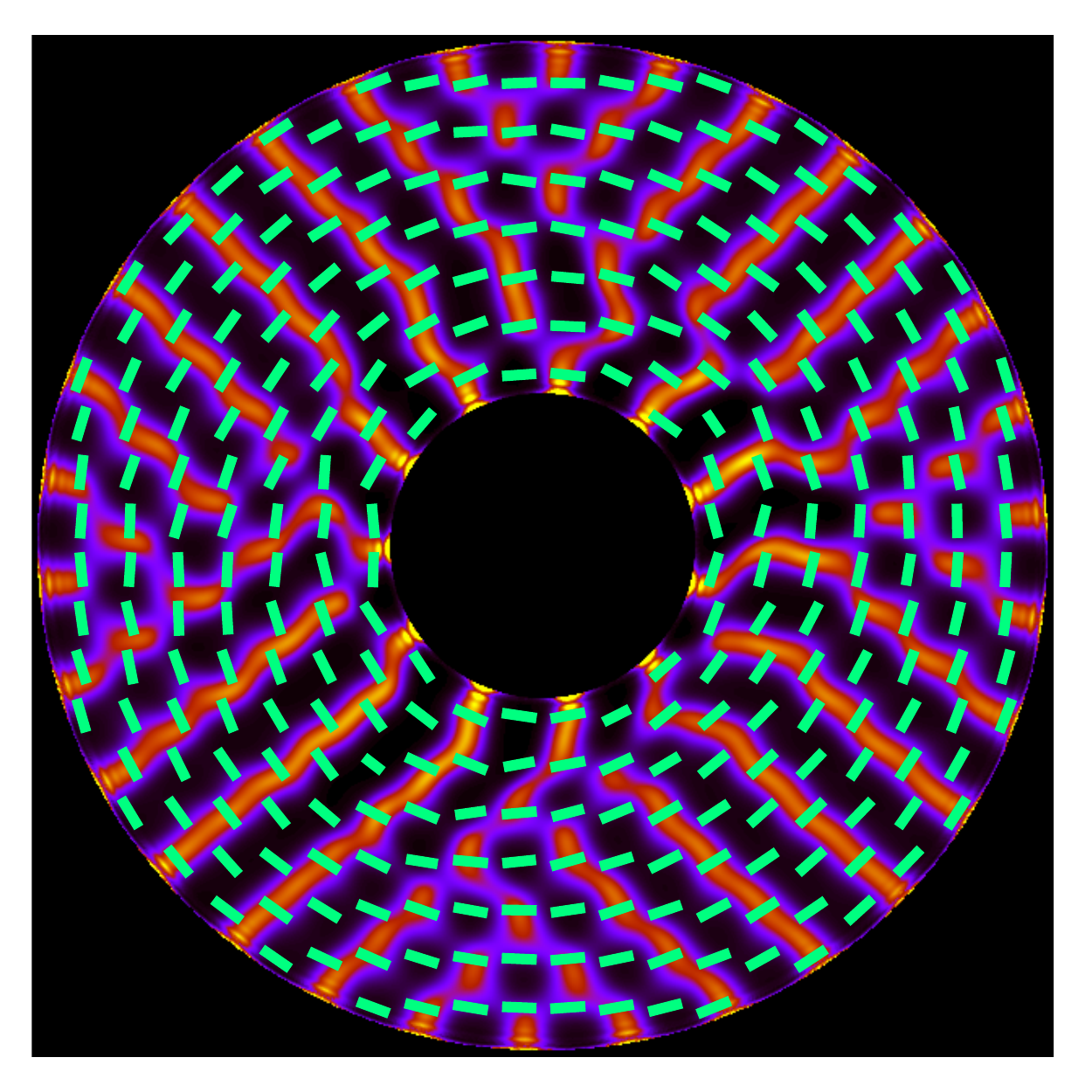}\hfill
\includegraphics[height=0.17\linewidth]{legend03newX.pdf}\\\vspace*{-0.3cm}
\flushleft\ \ \ \ \ \ \ \  $\mathcal{S}_{11,30}$, $b=0.29$ \ \ \ \  \ \ \ \ \ \ $\mathcal{S}_{11,30}$, $b=0.28$ \ \ \ \ \ \ \ \ \ \:\,$\mathcal{S}_{10,30}$, $b=0.27$ \ \ \ \ \ \ \ \ \ \ \;$\mathcal{S}_{11,30}$, $b=0.28$ \ \  \: \ \ \ \ \ \ \,$\mathcal{S}_{11,30}$, $b=0.29$ \hspace*{0.2cm}
\vspace*{0.025cm}\caption{ \textbf{Hysteresis in the Shubnikov state I.} Shown are the structures obtained by first decreasing the inclusion size ratio $b$ from $b=0.29$ down to $b=0.27$ and then increasing $b$ up to $b=0.29$, in steps of 0.01. 
The initial structures for $b=0.29$ are obtained from those for $b=0.3$, shown in Fig.~\ref{fig_statediagramX} and labeled on the left
(see Fig.~\ref{fig_hyster2} for the transition from $\mathcal{S}_{12,30}$ to $\mathcal{S}_{11,30}$ omitted in the first row). Color bar and arrows denote the orientationally averaged density and the director field, compare Eqs.~\eqref{eq_barrho} and~\eqref{eq_QQ}, respectively.
\label{fig_hyster}}
\end{figure*}

{ 
An alternative and somewhat more insightful interpretation of the topologically protected states
can be made by considering the inclusion as a topological defect itself, which possesses an own winding number $k$.
In the Shubnikov state $\mathcal{S}$, we have $k=1$ representing the uniformly bent director field parallel to the wall.
Whenever the rods are misaligned along the integration path around the inclusion
 $k$ is reduced by $-1/2$.
Thus we have $k=1/2$ for $\mathcal{C^{LS}}$ and $k=0$ for $\mathcal{L}$.
Thereby, the negative half-integer boundary charges described in the main text are absorbed into $k$ and do not contribute to the total charge $\tilde{Q}$, 
which is therefore different in these three states when choosing the current interpretation.
Here, the charge conservation resulting in topological protection becomes directly apparent.
However, the conservation law between $\tilde{Q}$ and the geometric quantity $\chi$
has to be generalized to $\chi=\tilde{Q}+k-h$,
where $h$ denotes the number of holes in the confining domain ($h=0$ in circular and $h=1$ in annular confinement) and thus still holds.
Further note that for the domain state, it is not possible to unambiguously define the winding number $k$ due to the presence of the radial line disclinations,
which underlines its topological instability.}

\section{\label{SN6} Response of microscopic layering to geometric changes \label{sec_hyster}}

 The identified topologically protected states ($\mathcal{L}$, $\mathcal{C^{LS}}$ and $\mathcal{S}$) even remain metastable over a large range of inclusion sizes as analyzed in Fig.~\ref{fig_statediagramX}.
However, we are further interested in the particular microscopic layer structure corresponding to a certain state,
whose stability is discussed below in more detail.

An initial intrinsic laminar structure $\mathcal{L}_{N_\text{con},N_\text{dis}}$ remains invariant upon smoothly changing $b$
until the layers do not fit the geometry any more, e.g., if $N_\text{con}\lambda_0> R_\text{out}-R_\text{in}$, which then results in an irreversible formation of defects.
Hence, there exist some (metastable) laminar--laminar transitions, as also observed in Fig.~\ref{fig_statediagramX}.

 \begin{figure*}
\rotatebox{90}{\large$\ \ \ \ \ \ \ \mathcal{S}_{12,30}$}\hfill\includegraphics[width=0.17\linewidth]{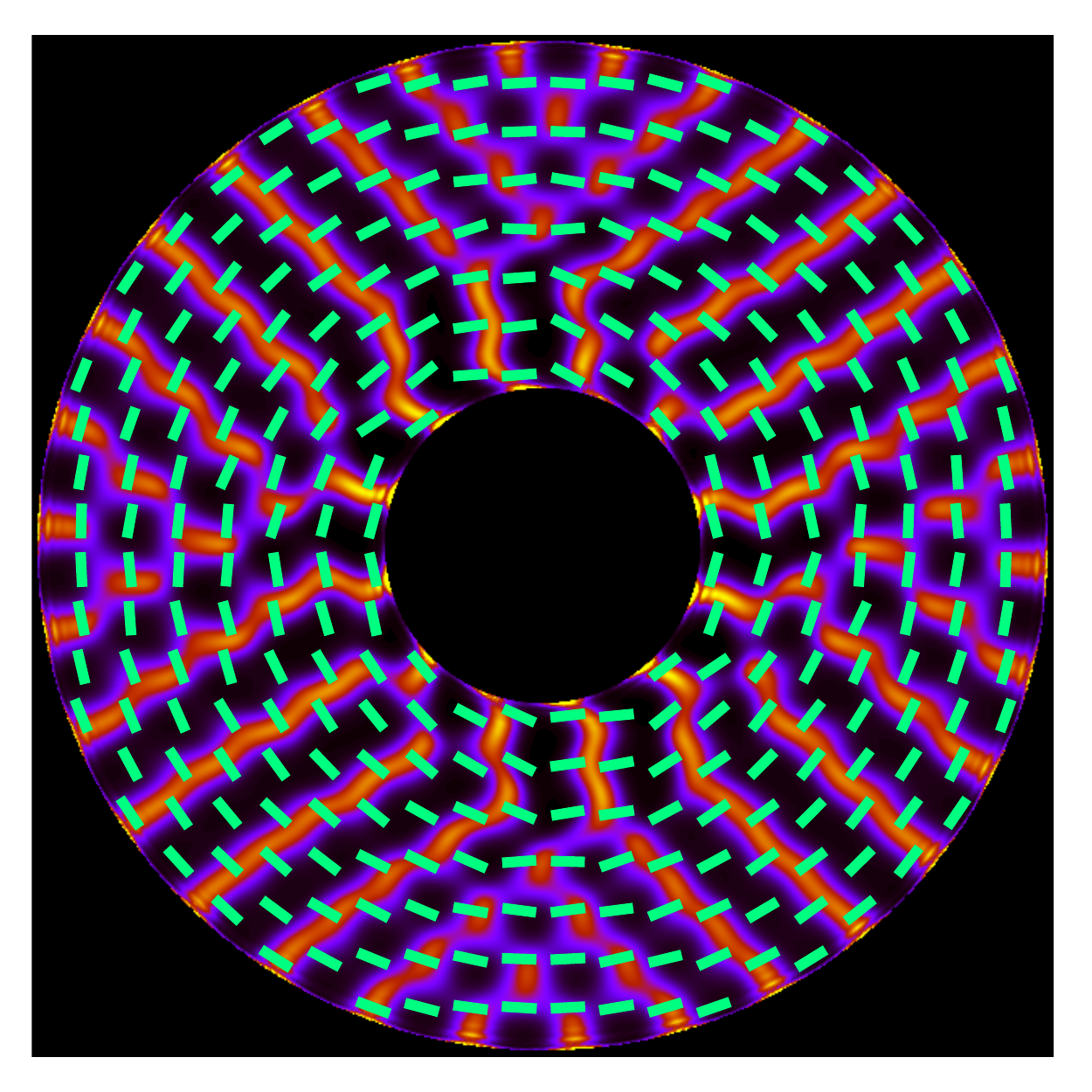}\hfill\parbox{0.015\linewidth}{\centering\vspace*{-2.85cm}$\rightarrow$}\hfill
\includegraphics[width=0.17\linewidth]{E7_M126p029S12_30.pdf}\hfill\parbox{0.015\linewidth}{\centering\vspace*{-2.85cm}$\rightarrow$}\hfill
\includegraphics[width=0.17\linewidth]{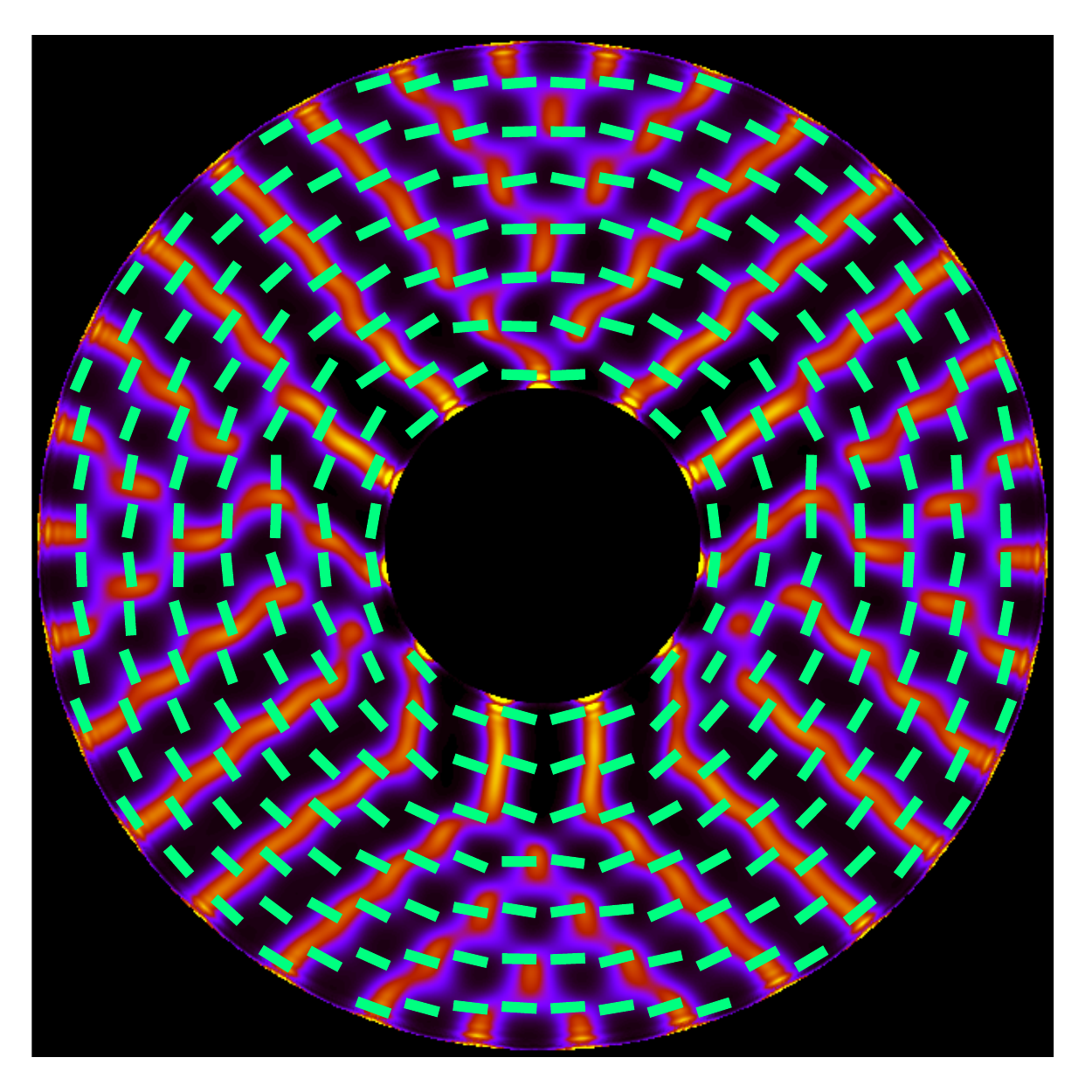}\hfill\parbox{0.015\linewidth}{\centering\vspace*{-2.85cm}}\hfill
\hspace*{-0.15cm}\includegraphics[height=0.17\linewidth]{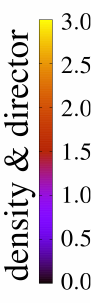}\hspace{0.15cm}\hfill
\includegraphics[width=0.17\linewidth]{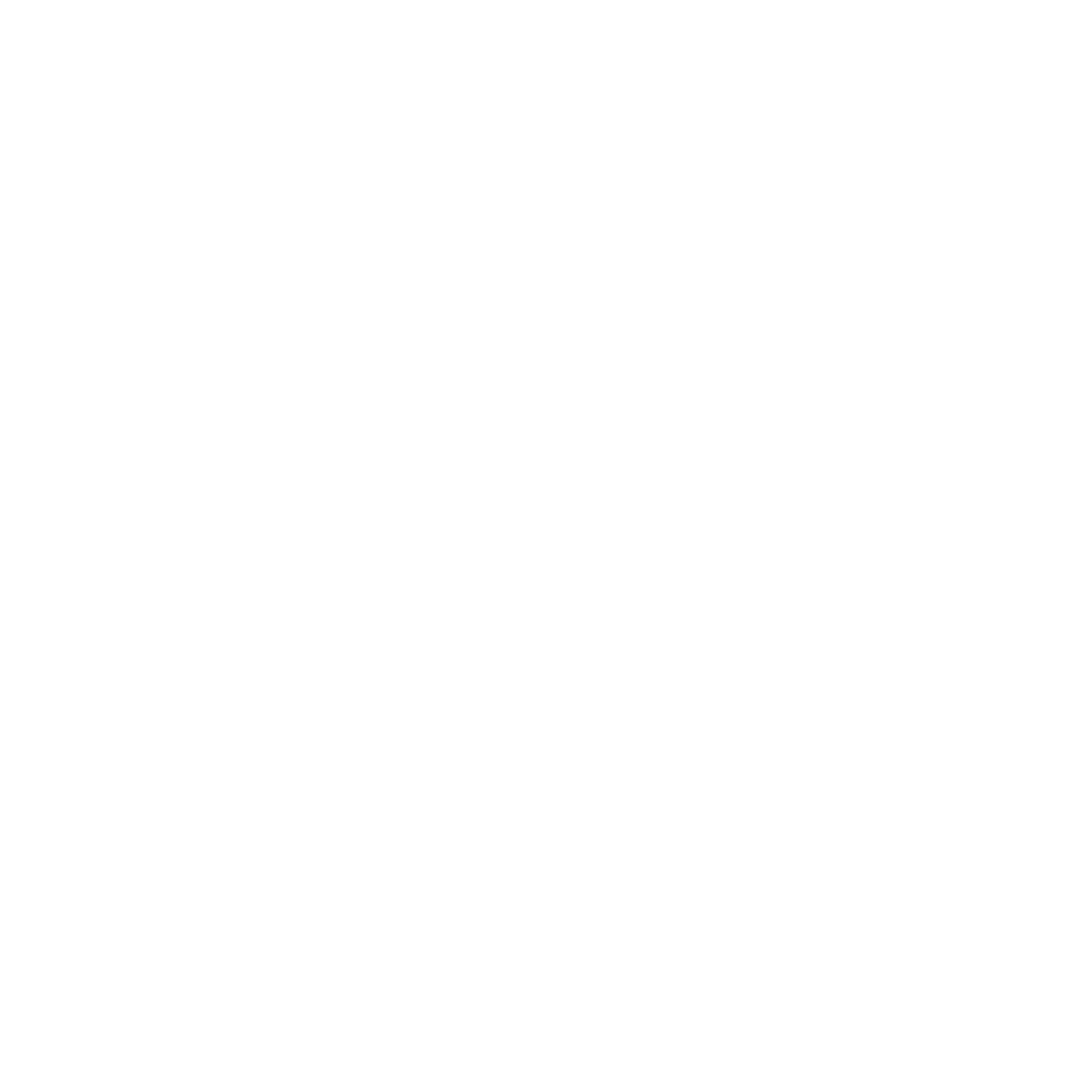}\hfill\parbox{0.015\linewidth}{\centering\vspace*{-2.85cm}}\hfill
\includegraphics[width=0.17\linewidth]{DUMMY.pdf}\hfill\hfill\hspace*{4.5cm}\\\vspace*{-0.3cm}
\flushleft\ \ \ \ \ \ \ \  $\mathcal{S}_{12,30}$, $b=0.30$ \ \ \ \ \ \ \ \ \ \ \ $\mathcal{S}_{11,30}$, $b=0.29$ \ \ \ \ \ \ \ \ \ \ \:\,$\mathcal{S}_{11,30}$, $b=0.30$ \hfill\hspace*{0.2cm}
\vspace*{0.025cm}\caption{ \textbf{Hysteresis in the Shubnikov state II.} Shown are the structures obtained from $\mathcal{S}_{12,30}$ by first decreasing the inclusion size ratio $b$ from $b=0.3$ down to $b=0.29$ and increasing $b$ back to $b=0.3$. Color bar and arrows denote the orientationally averaged density and the director field, compare Eqs.~\eqref{eq_barrho} and~\eqref{eq_QQ}, respectively.
\label{fig_hyster2}}
\end{figure*}

In contrast, the number of edge dislocations in a given Shubnikov structure $\mathcal{S}_{N_\text{in},N_\text{out}}$ is not topologically protected and thus
adapts to the geometrical changes 
 as soon as the local layer spacing near a boundary differs too much from the bulk value. .
 This comes along with some hysteresis effects upon first decreasing $b$ and then returning to its original value.
{  In particular, Fig.~\ref{fig_hyster} depicts that, for all structures considered in Fig.~\ref{fig_statediagramX},
 the number $N_\text{in}$ of layers in contact with the inclusion decreases from $N_\text{in}=11$ to $N_\text{in}=10$ upon decreasing the inclusion size ratio from $b=0.28$ to $b=0.27$, 
Upon a subsequent reversal from $b=0.27$ back to $b=0.28$ 
the contact number $N_\text{in}=10$ remains the same for two of the structures considered (shown in the two top rows), 
whose energy is then clearly larger than at the beginning ($b=0.28$).
The other structure (bottom row) gradually develops another contact and becomes similar to the original one with $N_\text{in}=11$.}
Further increasing the size of the inclusion to $b=0.29$, all structures have $N_\text{in}=11$ contacts, which closes all hysteresis loops.
{  In fact, their energy lies slightly below that of the original structures for $b=0.29$,
which indicates that a fluctuating confinement aids the equilibration.}
On the other hand, starting with the metastable structure $\mathcal{S}_{12,30}$ for $b=0.3$,
the hysteresis loop (shown in Fig.~\ref{fig_hyster2}) via $b=0.29$ 
 includes in a new structure $\mathcal{S}_{11,30}$ for $b=0.3$ which is much more stable than the original one.
This is not surprising, since the global minimum for $b=0.3$ also has $N_\text{in}=11$.

The composite states $\mathcal{C}^{\mathcal{L}\mathcal{S}}_{N_\text{con},N_\text{in},N_\text{out}}$ do not show any microscopic changes in the number of edge dislocations over the range of $b$ considered in Fig.~\ref{fig_statediagramX},
since the value of the local layer spacing $\lambda_\text{in}$ at the inclusion  in the Shubnikov part of the structure can be balanced by a deformation of the adjacent laminar layer.
If a hysteresis loop similar to the Shubnikov state exists here, it thus must be much larger.

\section{\label{SN7} Dependence on intrinsic parameters}

Apart from the external topological and geometrical constraints, 
the formation of smectic structures can also be controlled by indirectly tuning the
preferred layer spacing $\lambda_0$ in bulk.
This intrinsic structural property of the smectic liquid crystal
 depends on the size and aspect ratio of the particles, as well as, the total density.
In the following, we briefly explore this extended parameter space within DFT in more detail.

\subparagraph*{Changes in density and particle shape.}

Increasing the density, the bulk layer spacing decreases and packing effects become more important.
To explore the effect on the smectic states, we consider again the density profiles for $R_\text{out}=6.3L$ and $b=0.3$ shown in Fig.~\ref{fig_statediagramX}.
Increasing the density, the global minimum gradually shifts from a Shubnikov to a composite state, as shown in Fig.~\ref{fig_parametersM}a and then to a laminar state (not shown).
The optimal microscopic structure of the different states also changes for a higher density,
since, for example, a larger value of $N_\text{out}$ becomes more favorable.

The dependence of the state diagram on the absolute particle size can be extracted from the scaling used in Fig.~\ref{fig_statediagram}.
Decreasing now the aspect ratio of the rods to $p=5$ for a fixed density and relative system size $R_\text{out}=6.3L$, 
we find that the Shubnikov state is still stable for $b=0.26$, cf. Fig.~\ref{fig_parametersM}b,
although the absolute system size $R_\text{out}$ has accordingly decreased by a factor two.
The laminar--Shubnikov transition occurs thus at a smaller value $b\approx0.259$ than for more elongated rods.
Hence, we conclude that the Shubnikov state is most stable when considering systems of small and short rods at a low density within the smectic regime.

\begin{figure*}
\rotatebox{90}{\large$\ \ \ \ \ \ \ \ \ \: D_{0}$}\hfill
\includegraphics[width=0.17\linewidth]{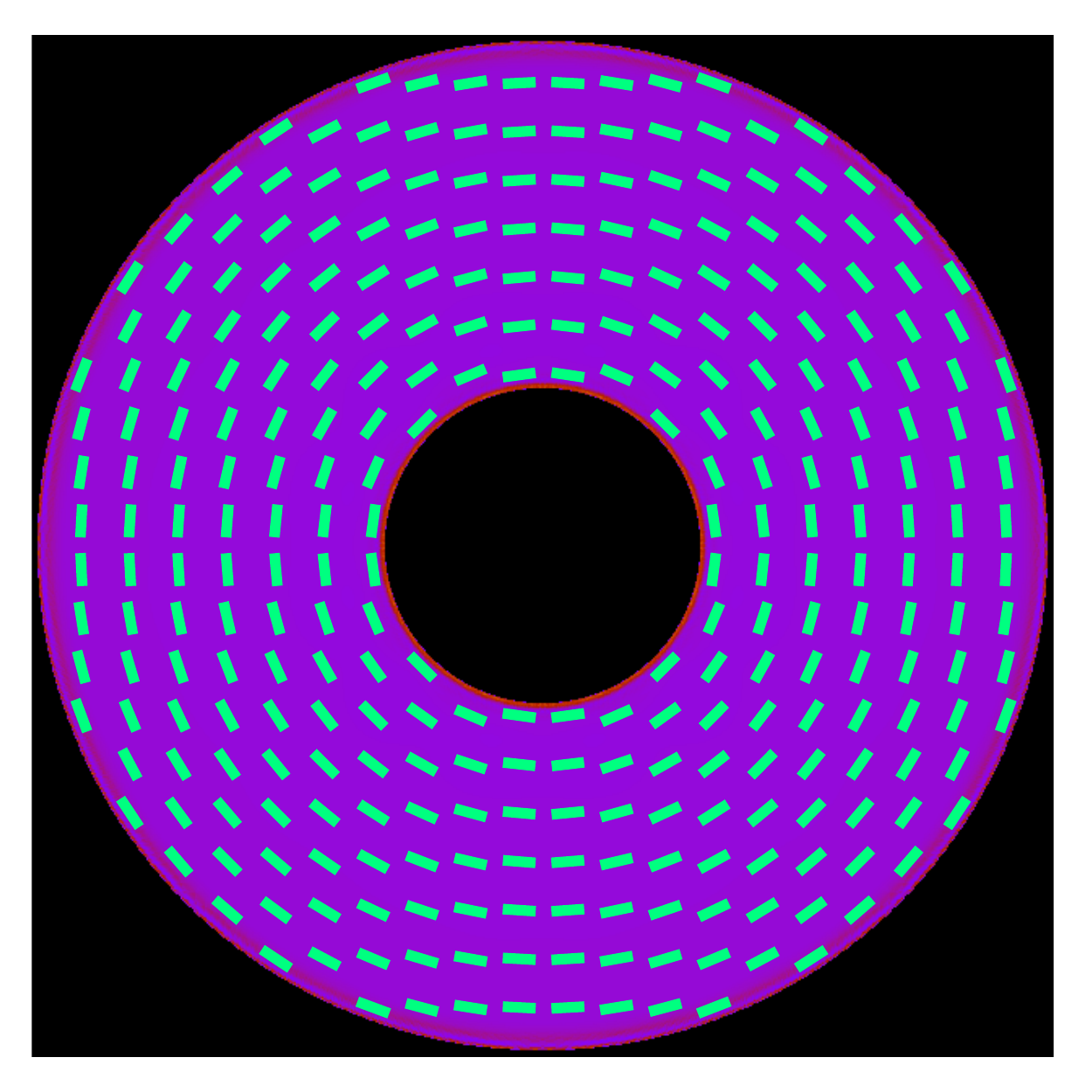}\hfill\parbox{0.015\linewidth}{\centering\vspace*{-2.85cm}$\rightarrow$}\hfill
\includegraphics[width=0.17\linewidth]{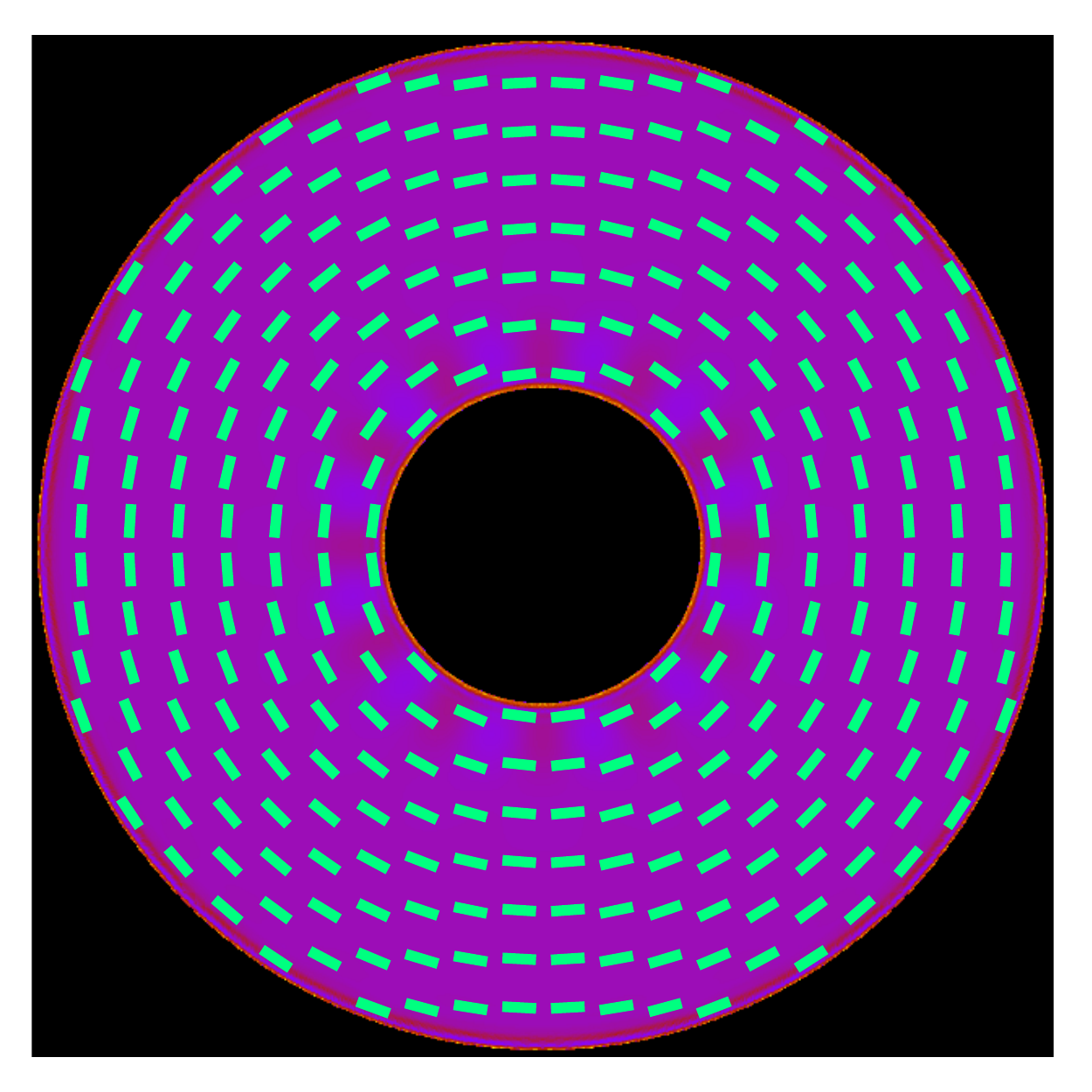}\hfill\parbox{0.015\linewidth}{\centering\vspace*{-2.85cm}$\rightarrow$}\hfill
\includegraphics[width=0.17\linewidth]{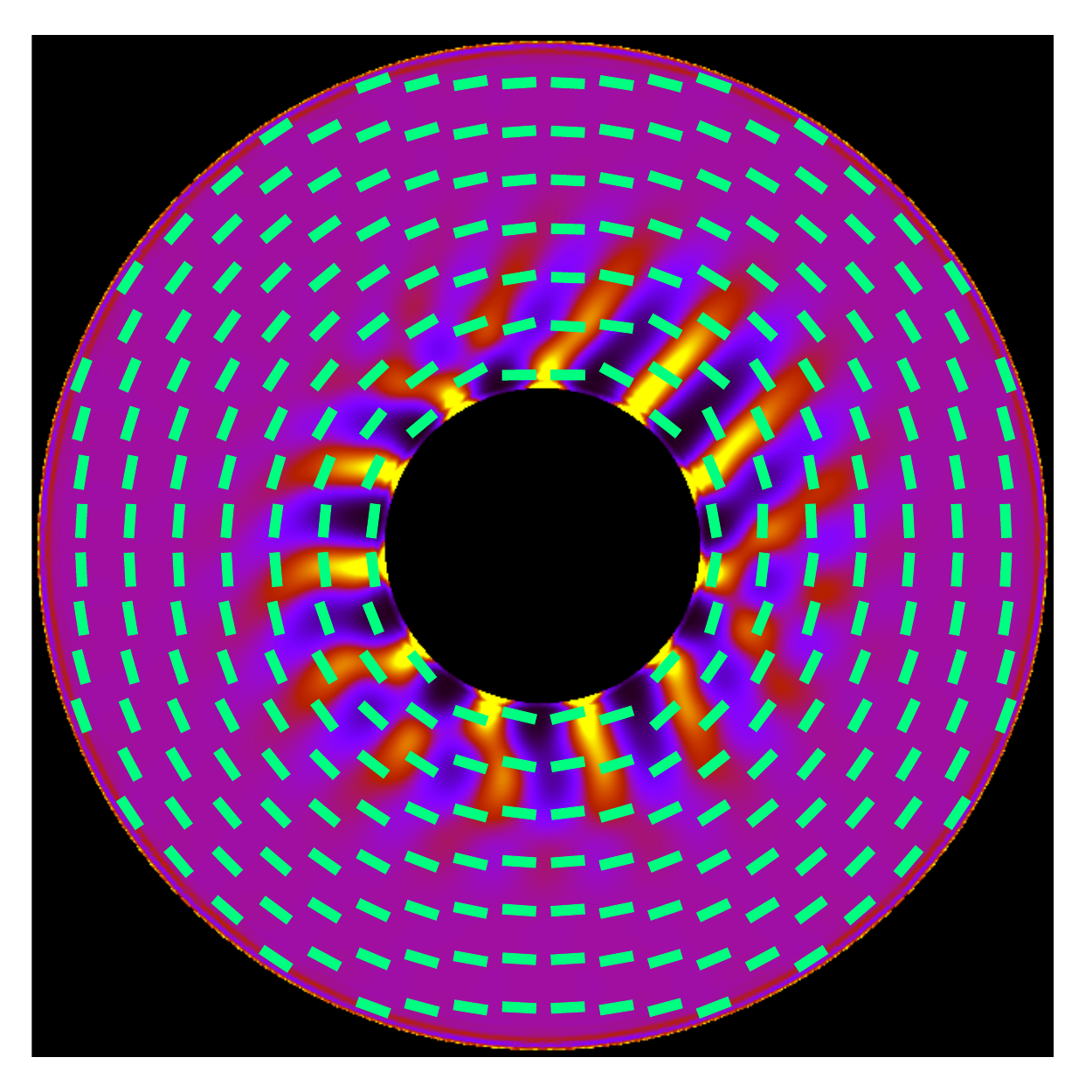}\hfill\parbox{0.015\linewidth}{\centering\vspace*{-2.85cm}$\rightarrow$}\hfill
\includegraphics[width=0.17\linewidth]{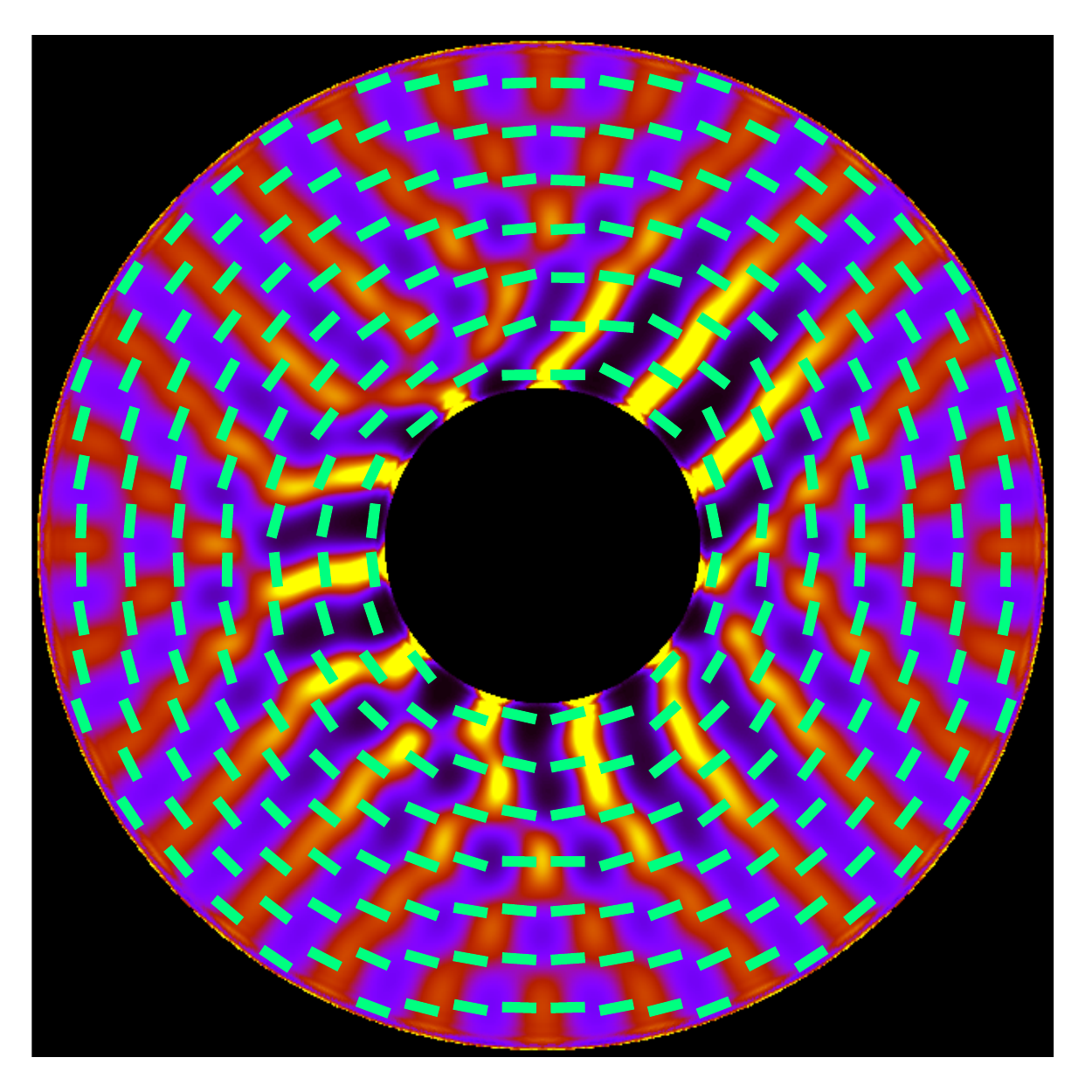}\hfill\parbox{0.015\linewidth}{\centering\vspace*{-2.85cm}$\color{white}\rightarrow$}\hfill
\hspace*{-0.3cm}\includegraphics[height=0.17\linewidth]{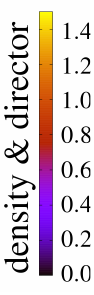}\hspace*{0.3cm}\hfill
\includegraphics[width=0.17\linewidth]{DUMMY.pdf}\\\rotatebox{90}{\large$\ \ \ \ \ \ \ \ \ \: D_{2}$}\hfill
\includegraphics[width=0.17\linewidth]{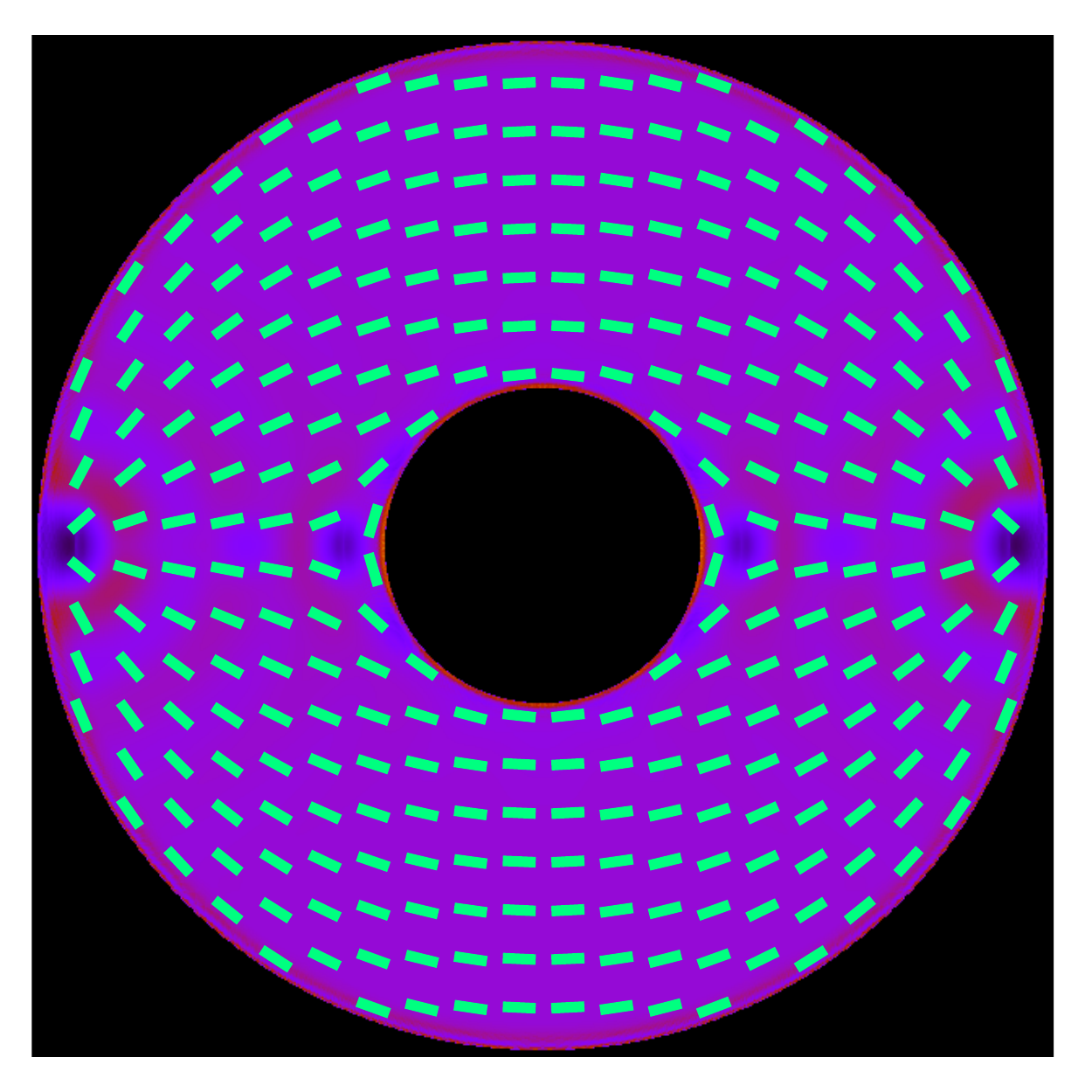}\hfill\parbox{0.015\linewidth}{\centering\vspace*{-2.85cm}$\rightarrow$}\hfill
\includegraphics[width=0.17\linewidth]{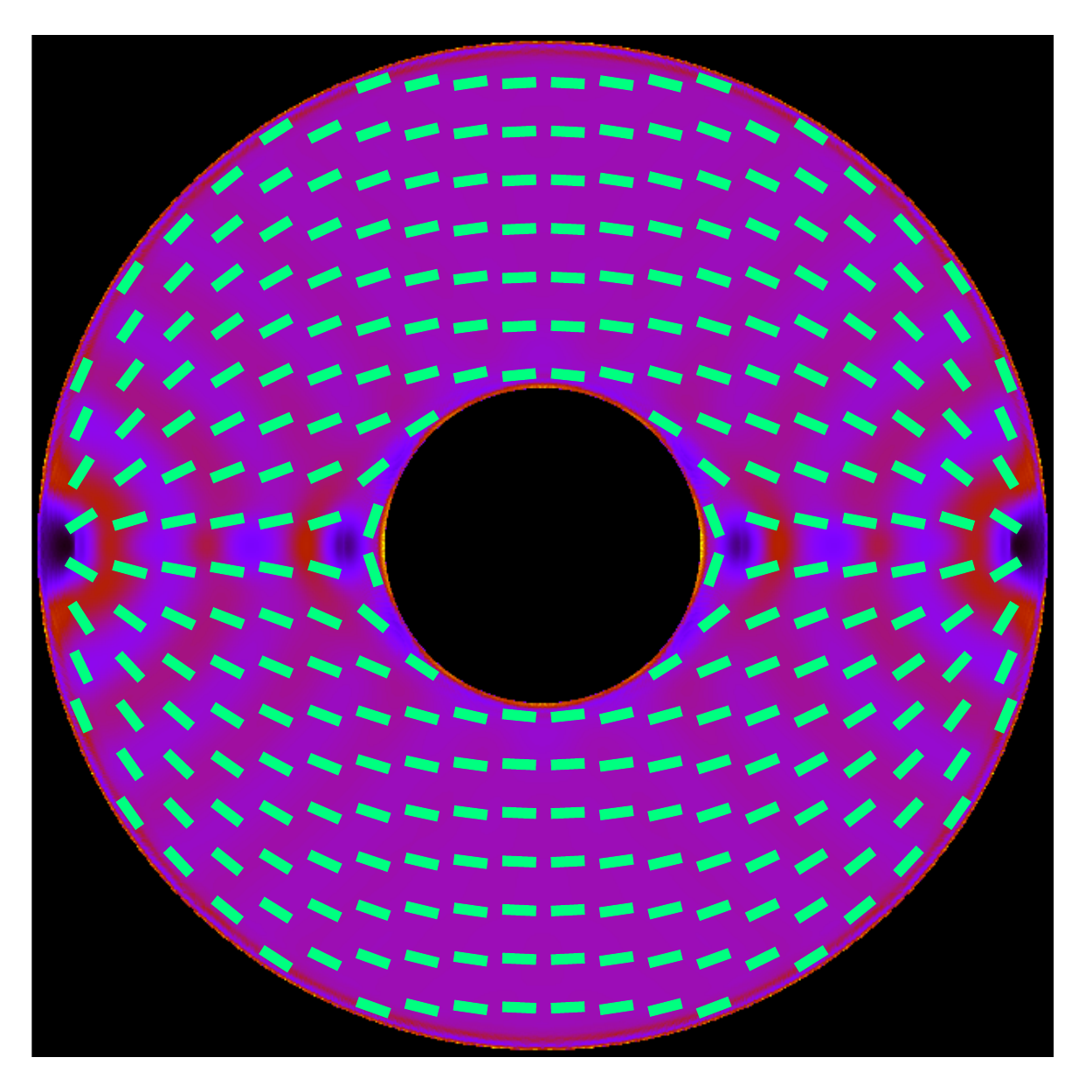}\hfill\parbox{0.015\linewidth}{\centering\vspace*{-2.85cm}$\rightarrow$}\hfill
\includegraphics[width=0.17\linewidth]{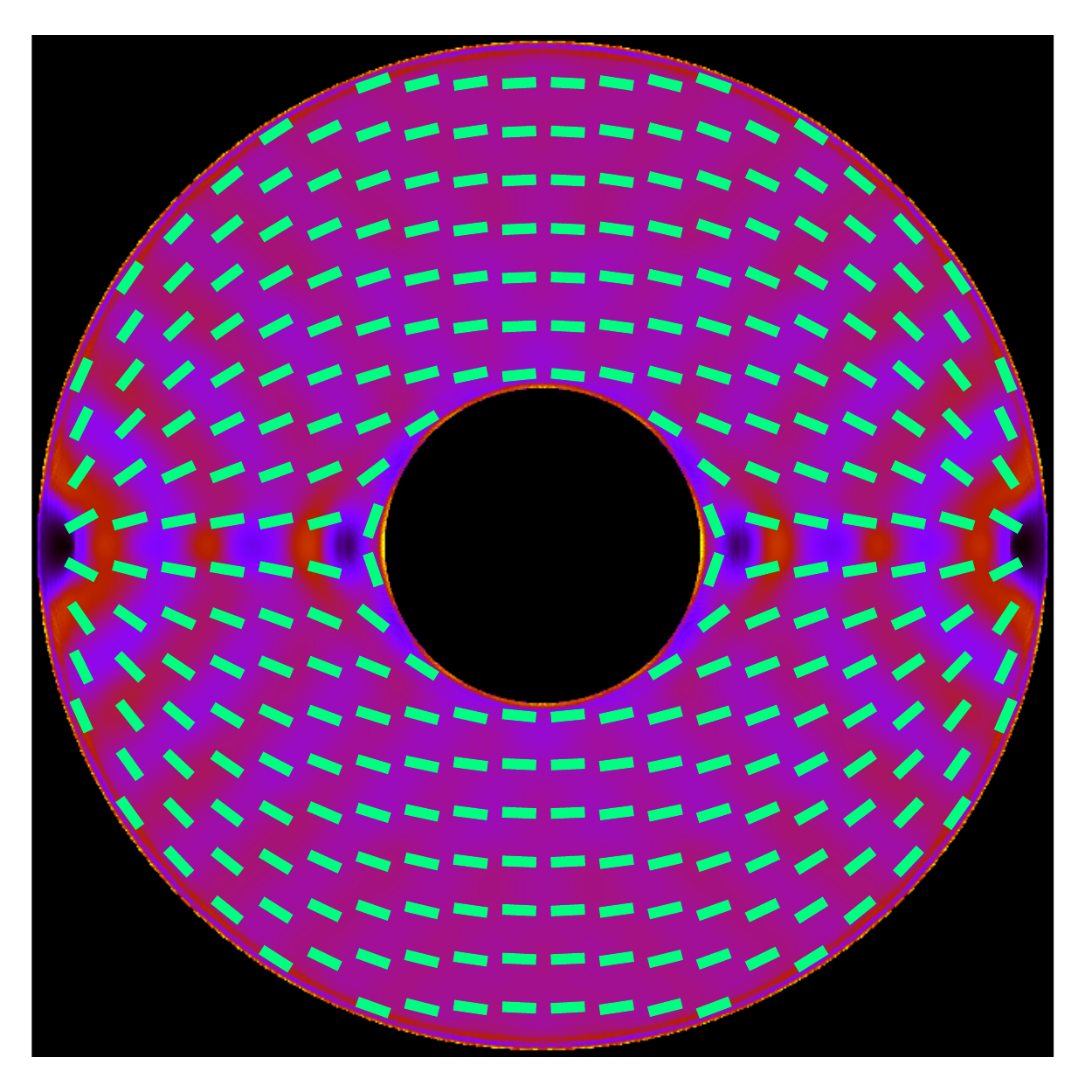}\hfill\parbox{0.015\linewidth}{\centering\vspace*{-2.85cm}$\rightarrow$}\hfill
\includegraphics[width=0.17\linewidth]{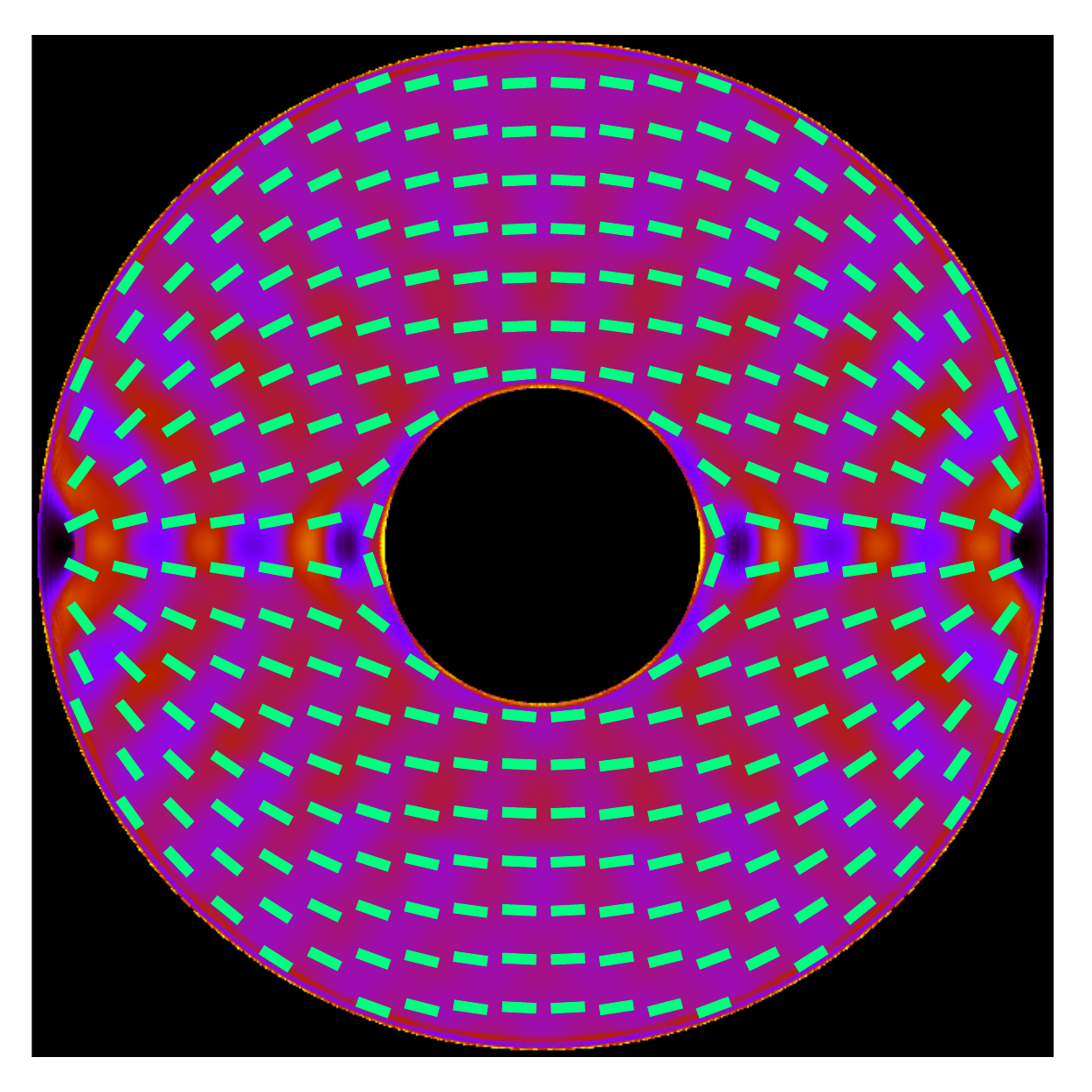}\hfill\parbox{0.015\linewidth}{\centering\vspace*{-2.85cm}$\rightarrow$}\hfill
\includegraphics[width=0.17\linewidth]{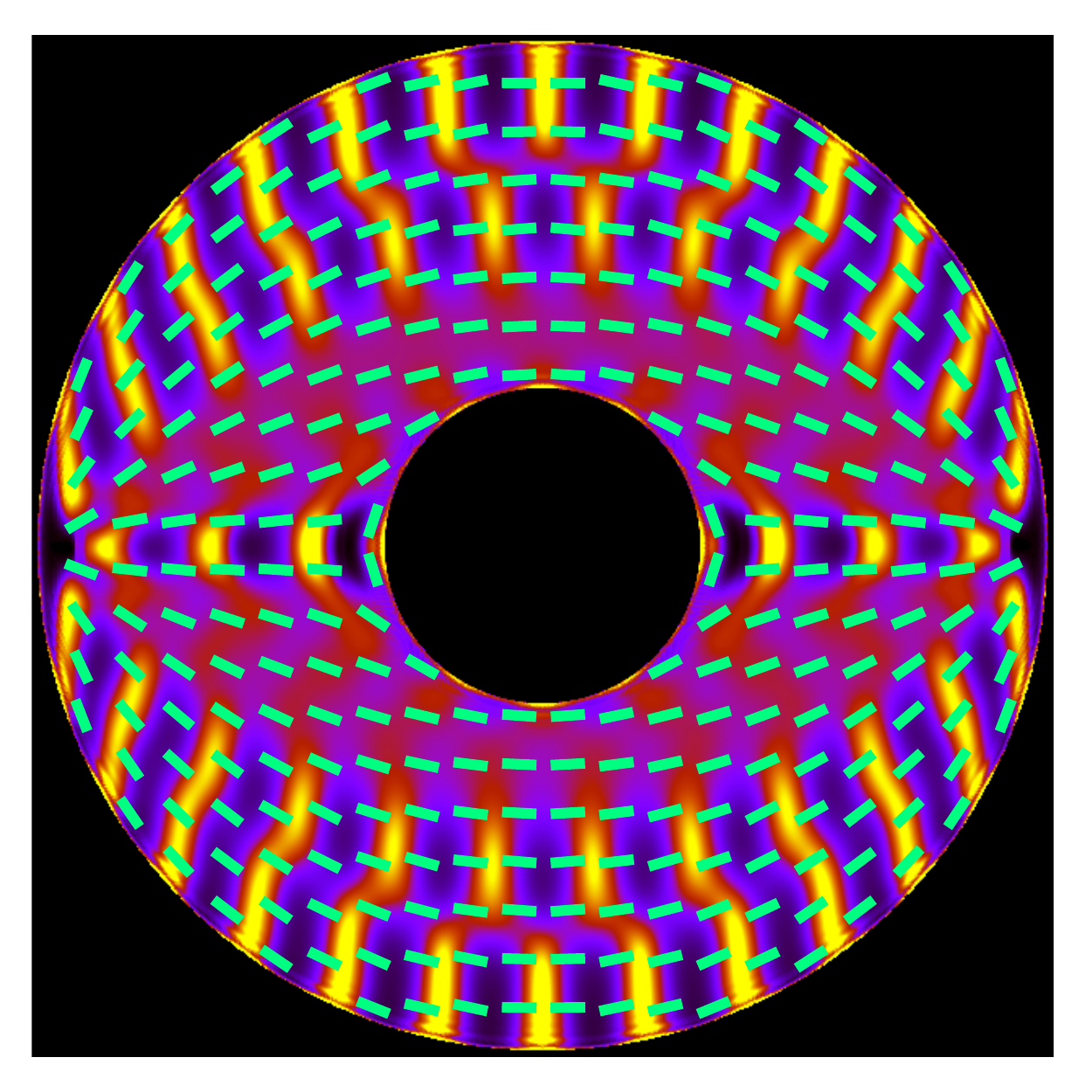}\hfill
\includegraphics[height=0.17\linewidth]{legend015new.pdf}\\\rotatebox{90}{\large$\ \ \ \ \ \ \ \ \ \: D_{3}$}\hfill
\includegraphics[width=0.17\linewidth]{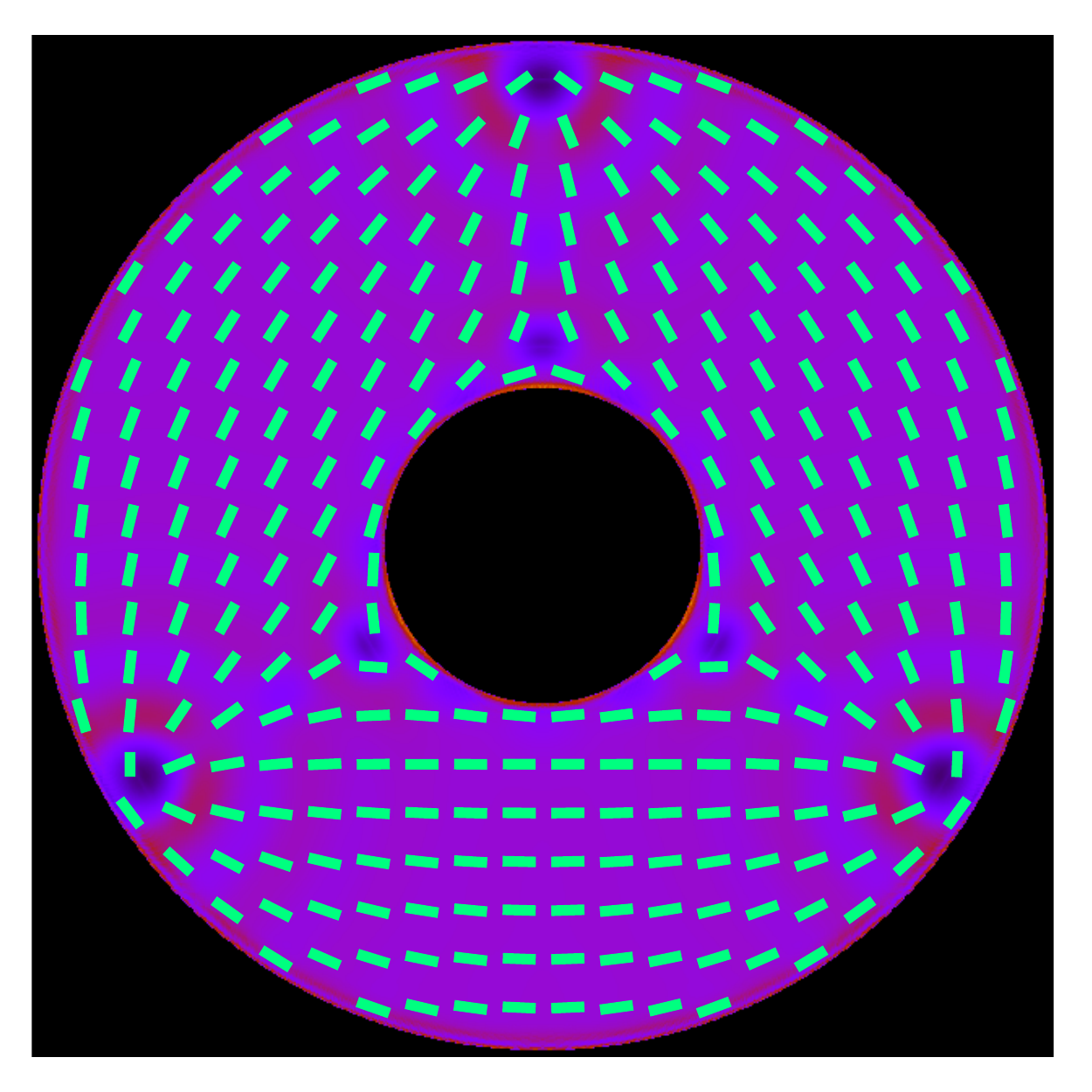}\hfill\parbox{0.015\linewidth}{\centering\vspace*{-2.85cm}$\rightarrow$}\hfill
\includegraphics[width=0.17\linewidth]{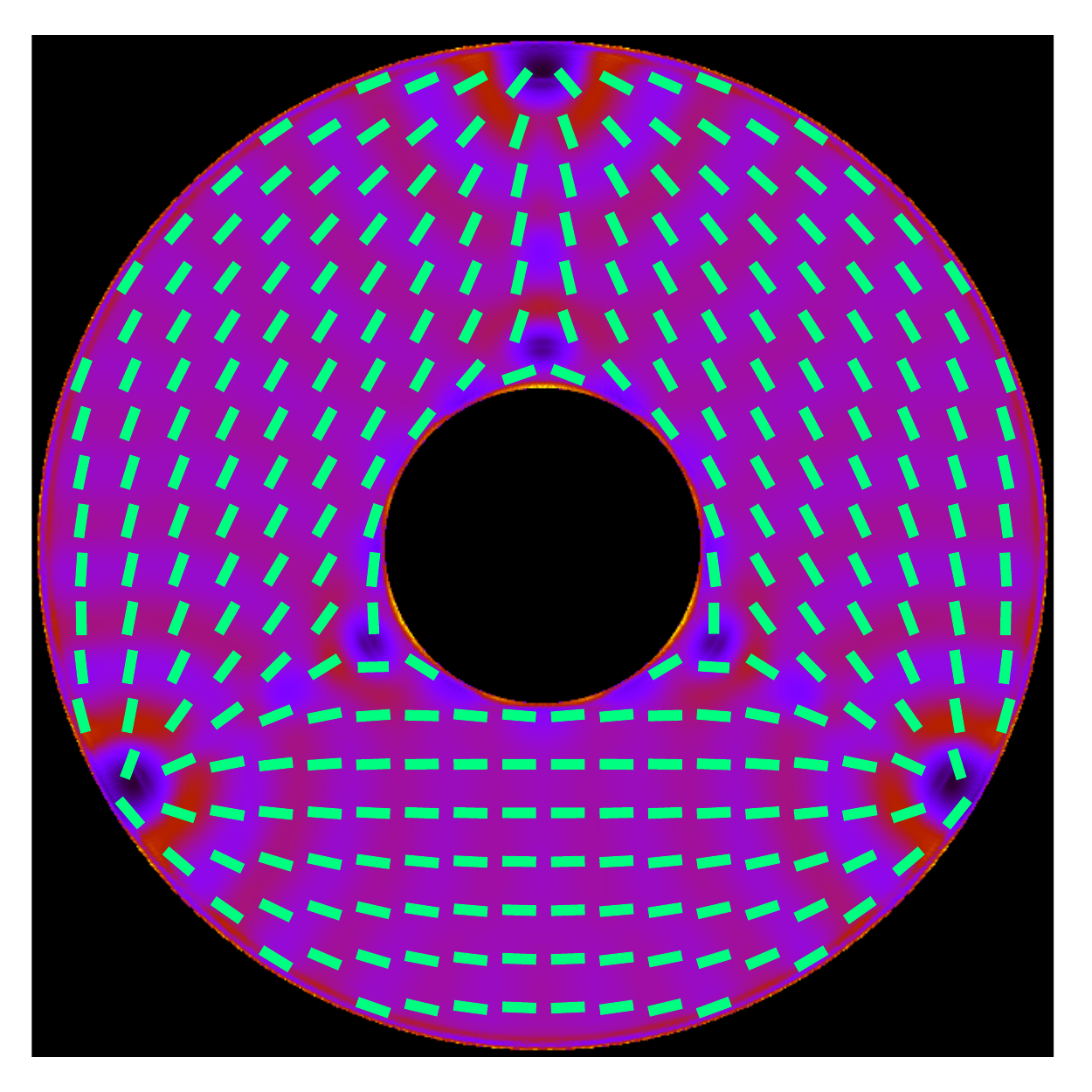}\hfill\parbox{0.015\linewidth}{\centering\vspace*{-2.85cm}$\rightarrow$}\hfill
\includegraphics[width=0.17\linewidth]{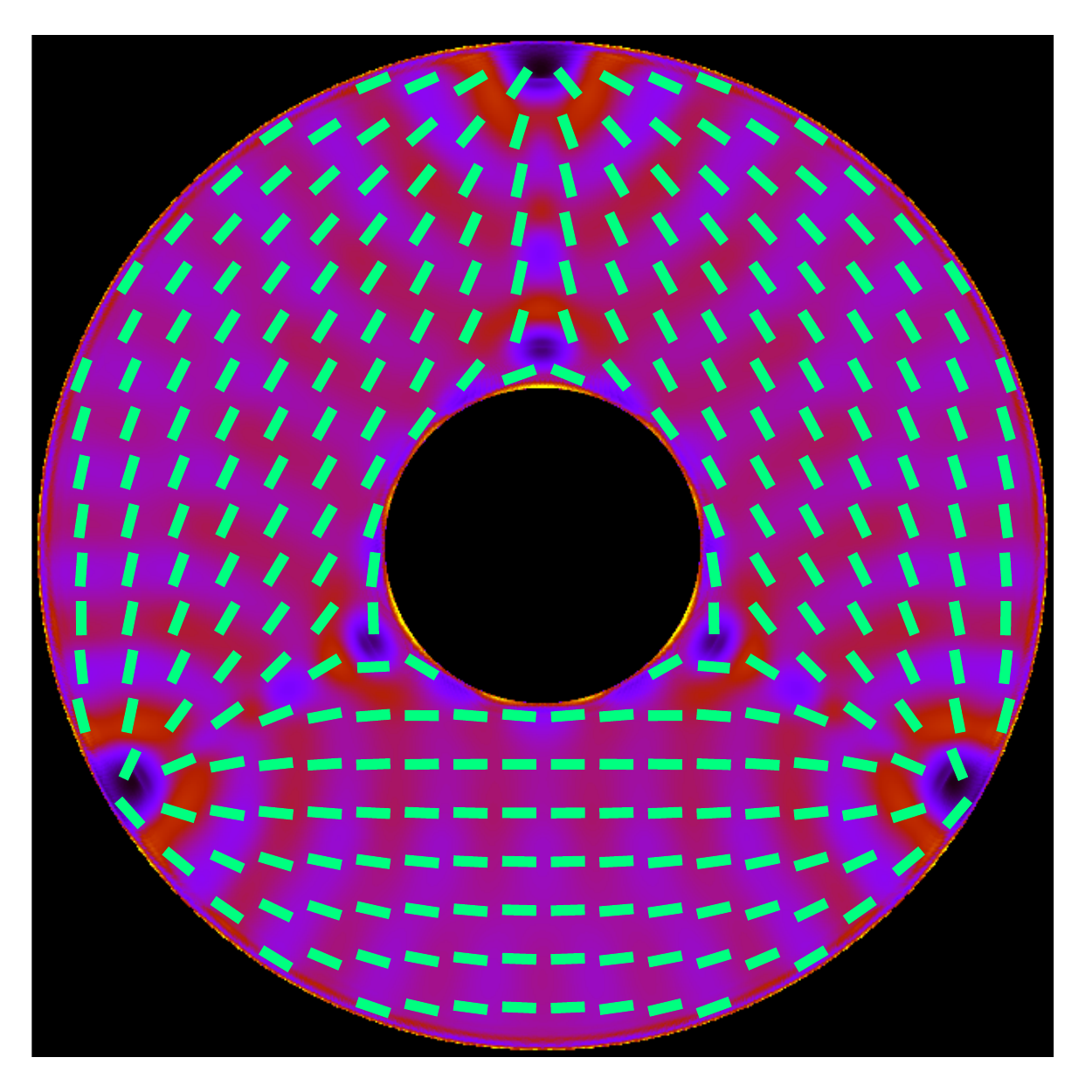}\hfill\parbox{0.015\linewidth}{\centering\vspace*{-2.85cm}$\rightarrow$}\hfill
\includegraphics[width=0.17\linewidth]{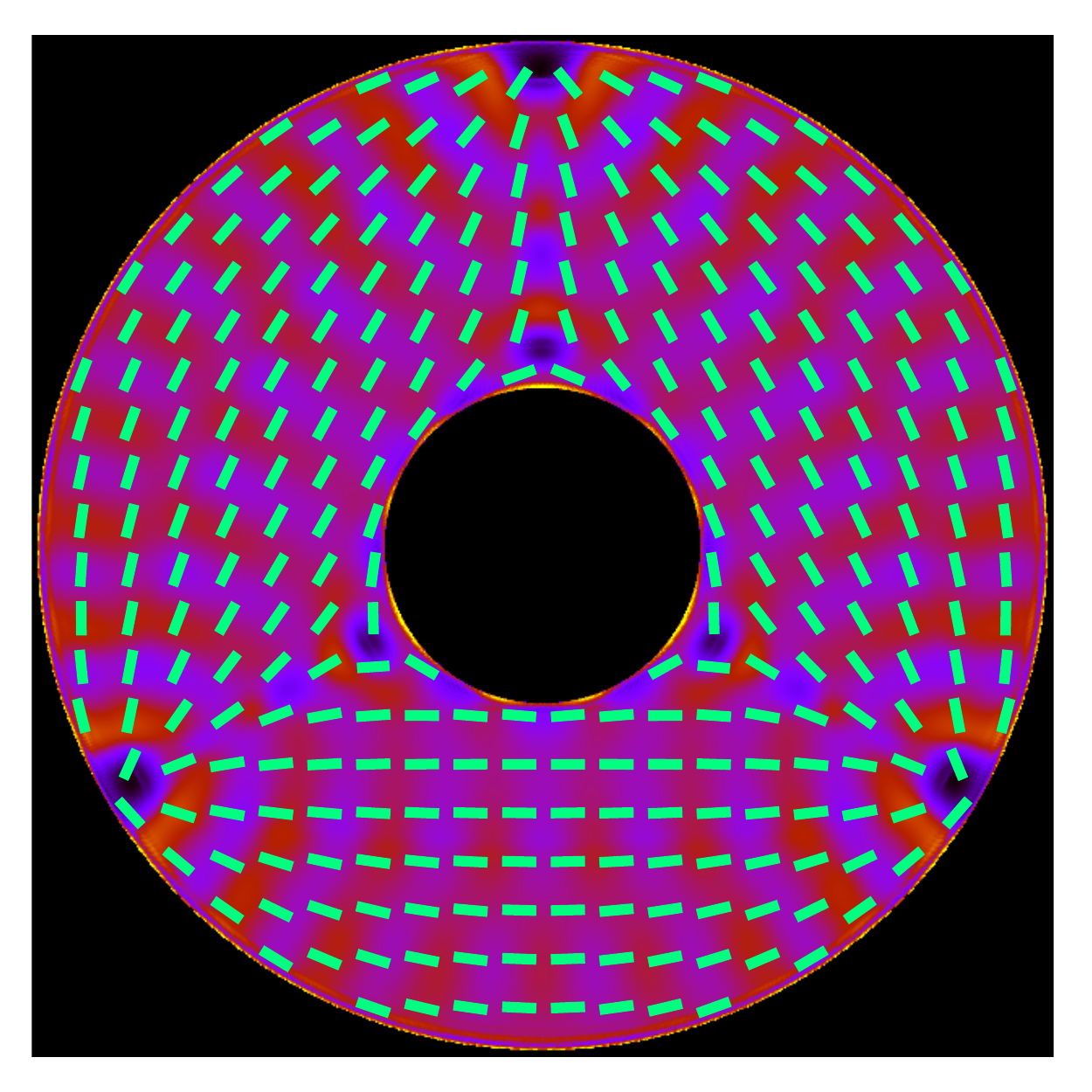}\hfill\parbox{0.015\linewidth}{\centering\vspace*{-2.85cm}$\rightarrow$}\hfill
\includegraphics[width=0.17\linewidth]{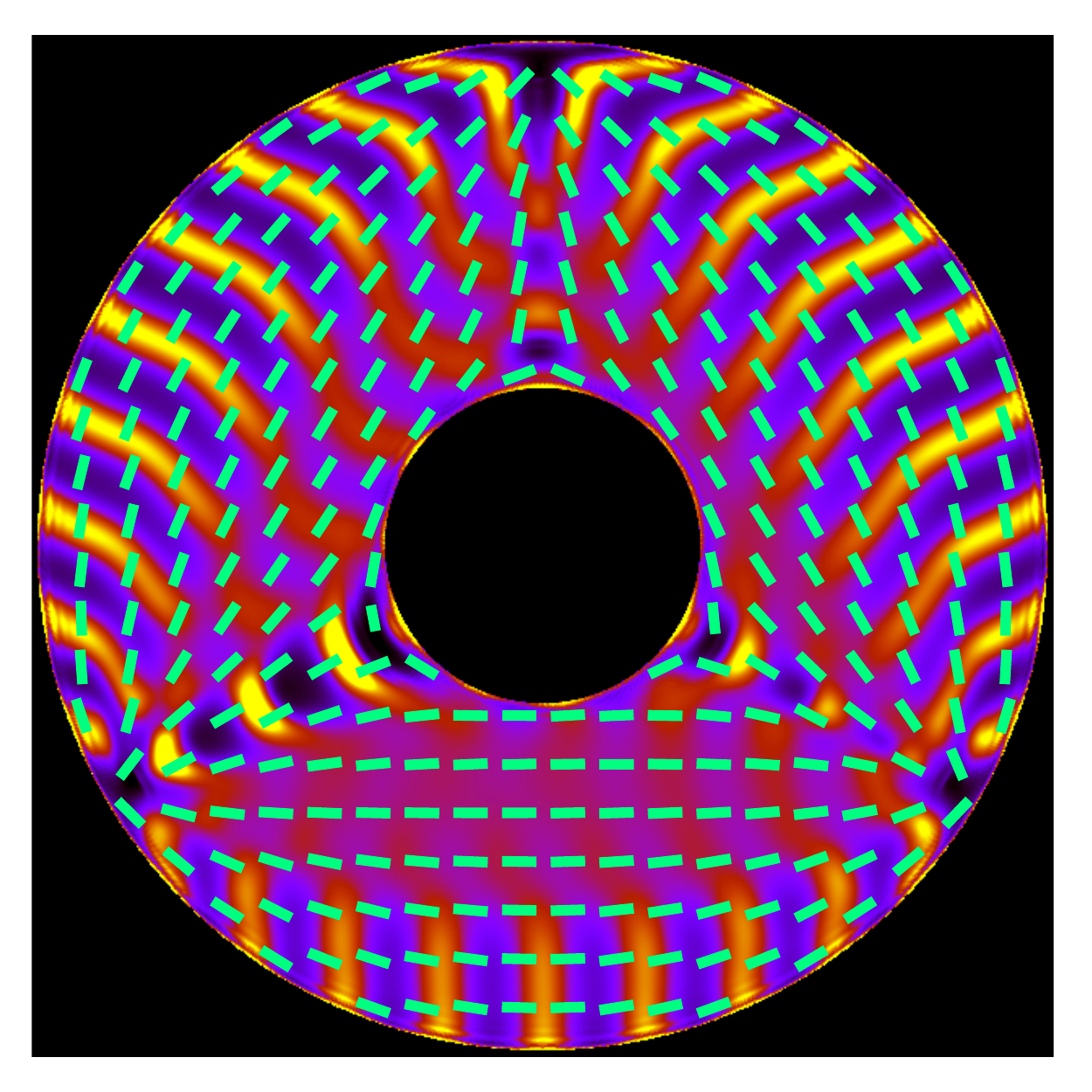}\hfill
\includegraphics[height=0.17\linewidth]{legend015new.pdf}\\\rotatebox{90}{\large$\ \ \ \ \ \ \ \ \ \: D_{4}$}\hfill
\includegraphics[width=0.17\linewidth]{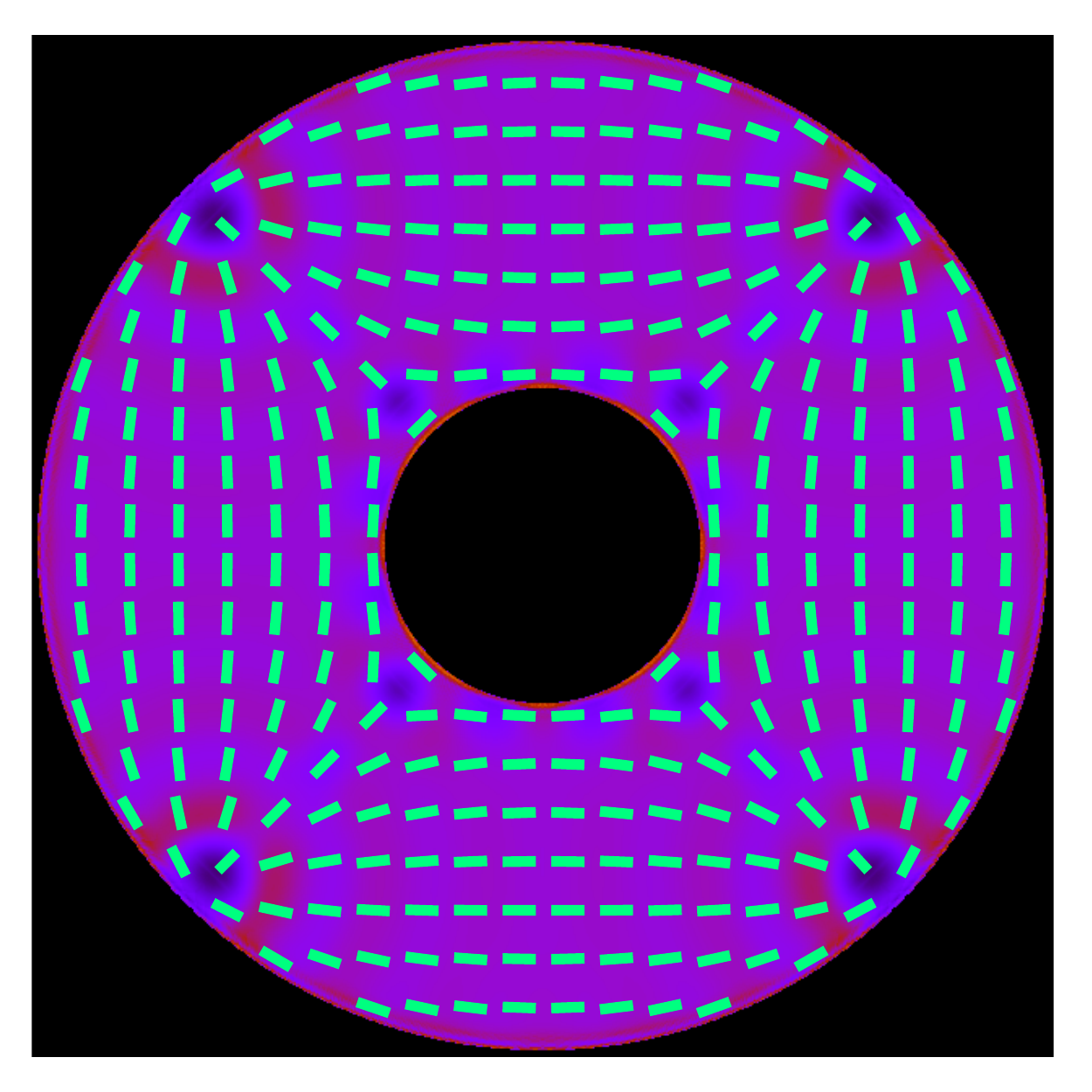}\hfill\parbox{0.015\linewidth}{\centering\vspace*{-2.85cm}$\rightarrow$}\hfill
\includegraphics[width=0.17\linewidth]{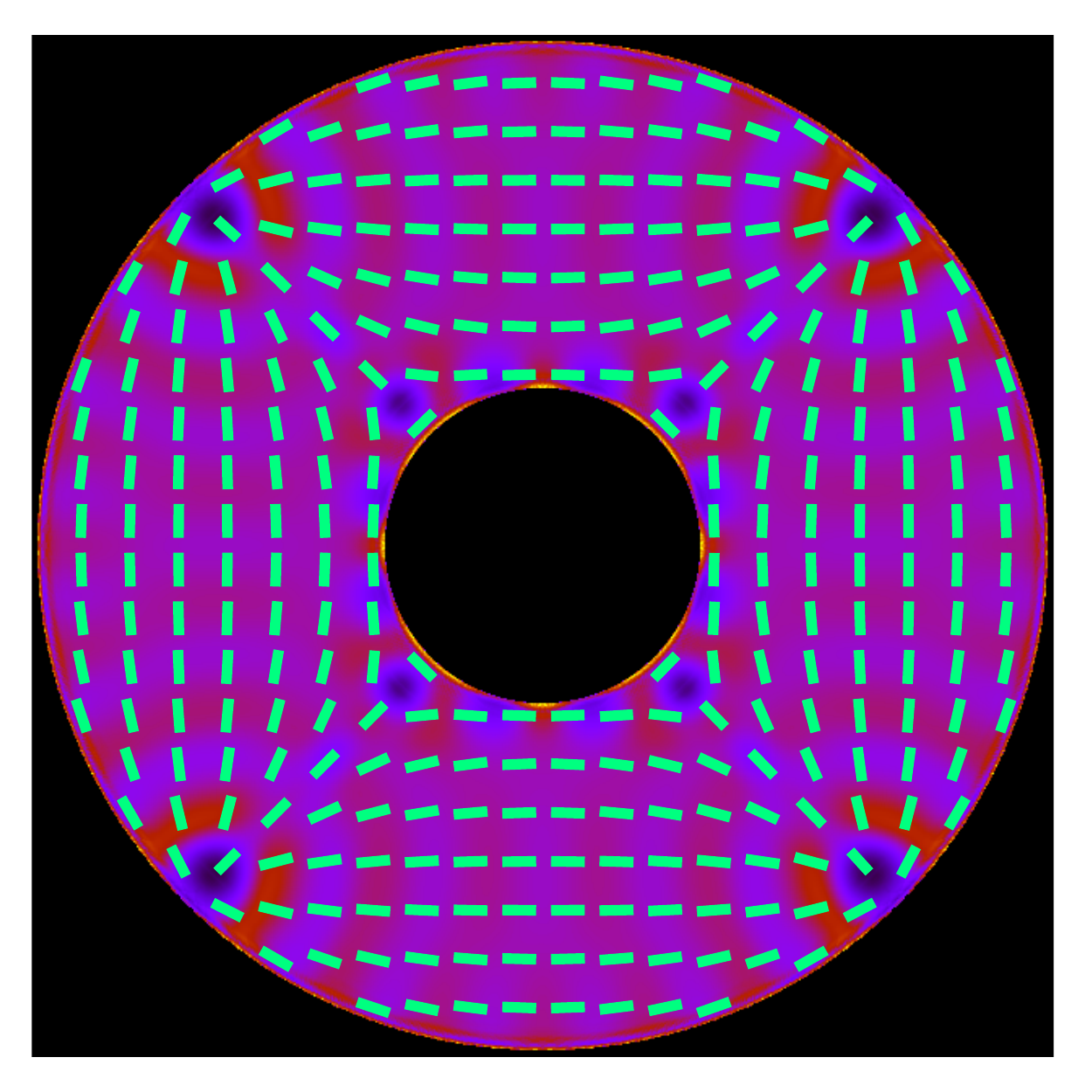}\hfill\parbox{0.015\linewidth}{\centering\vspace*{-2.85cm}$\rightarrow$}\hfill
\includegraphics[width=0.17\linewidth]{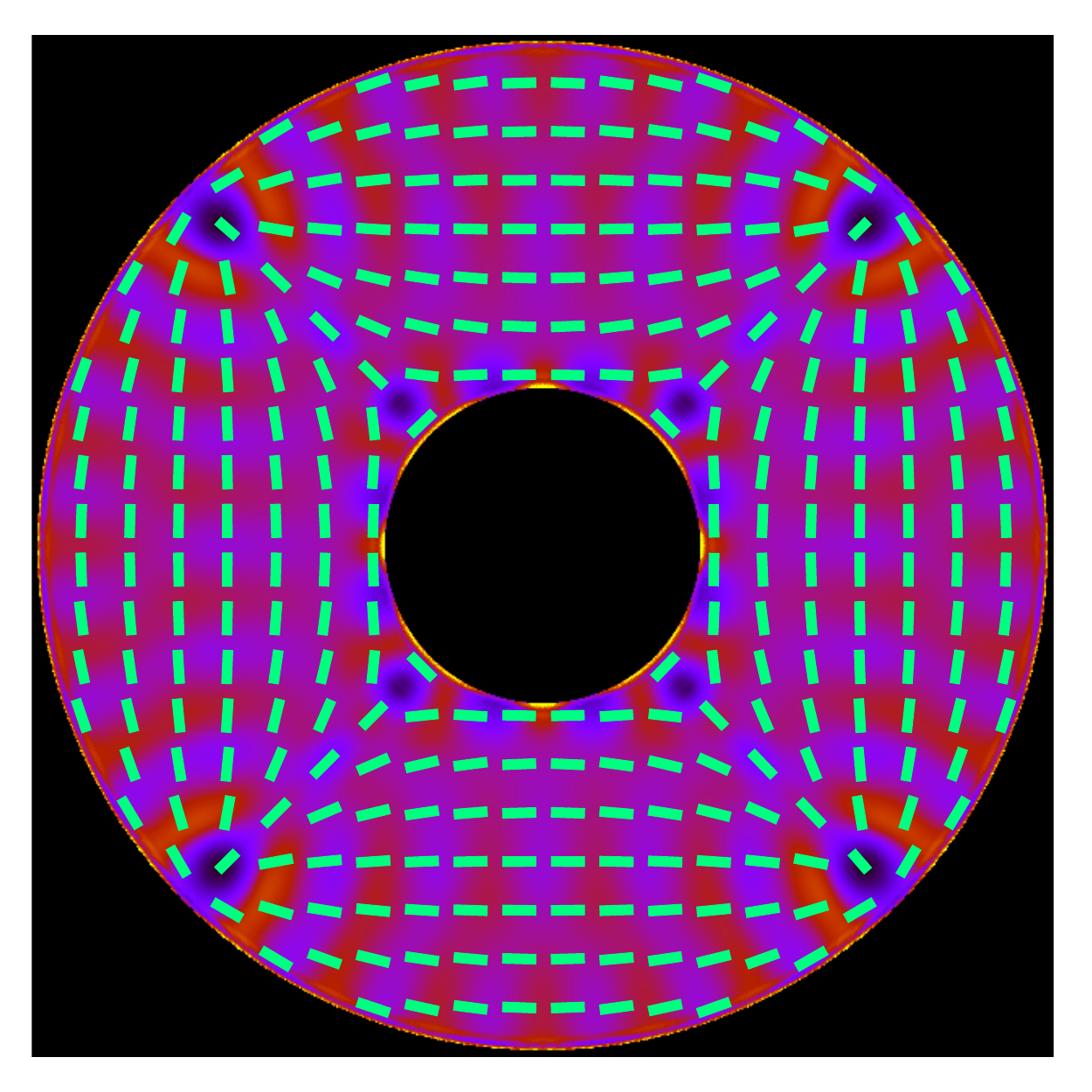}\hfill\parbox{0.015\linewidth}{\centering\vspace*{-2.85cm}$\rightarrow$}\hfill
\includegraphics[width=0.17\linewidth]{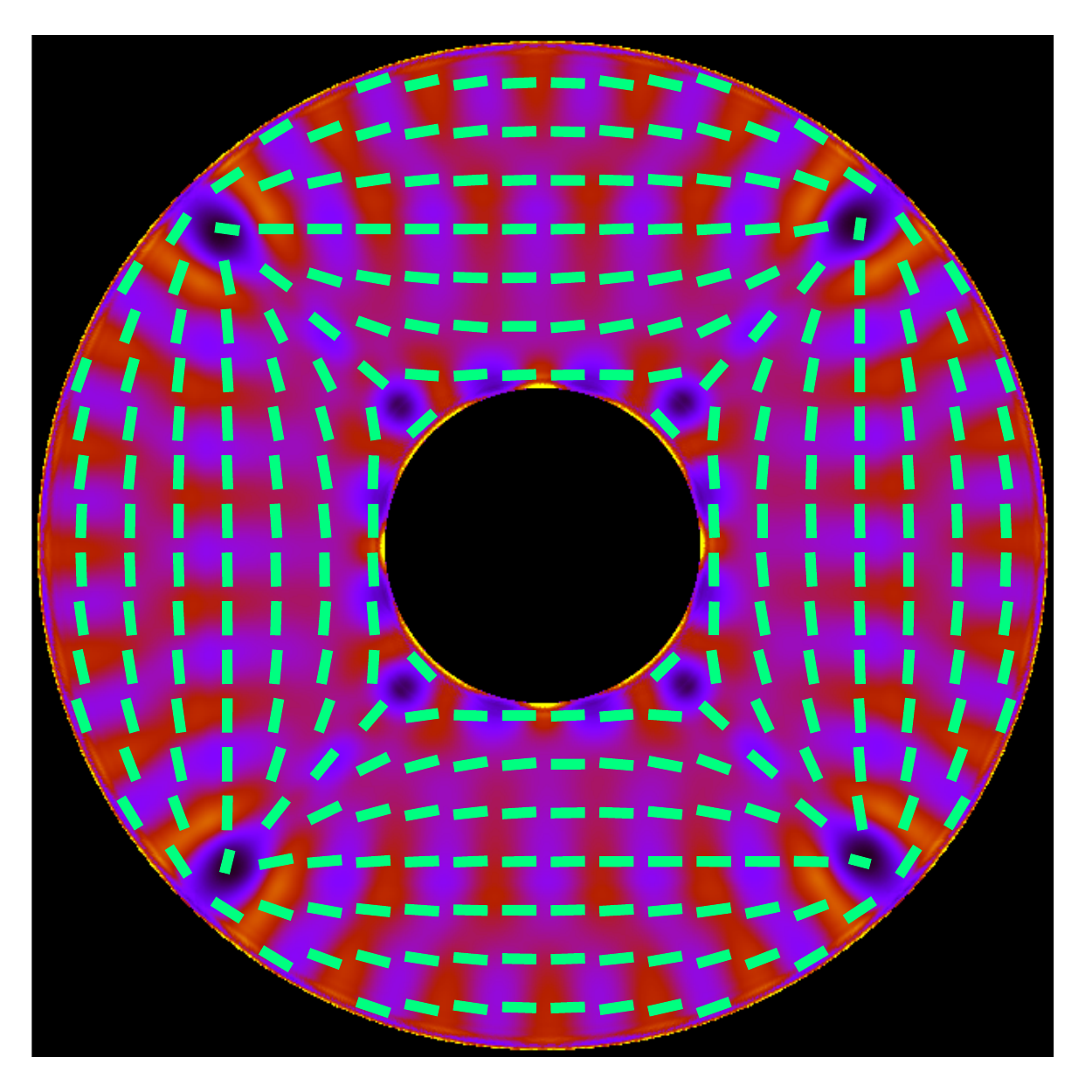}\hfill\parbox{0.015\linewidth}{\centering\vspace*{-2.85cm}$\rightarrow$}\hfill
\includegraphics[width=0.17\linewidth]{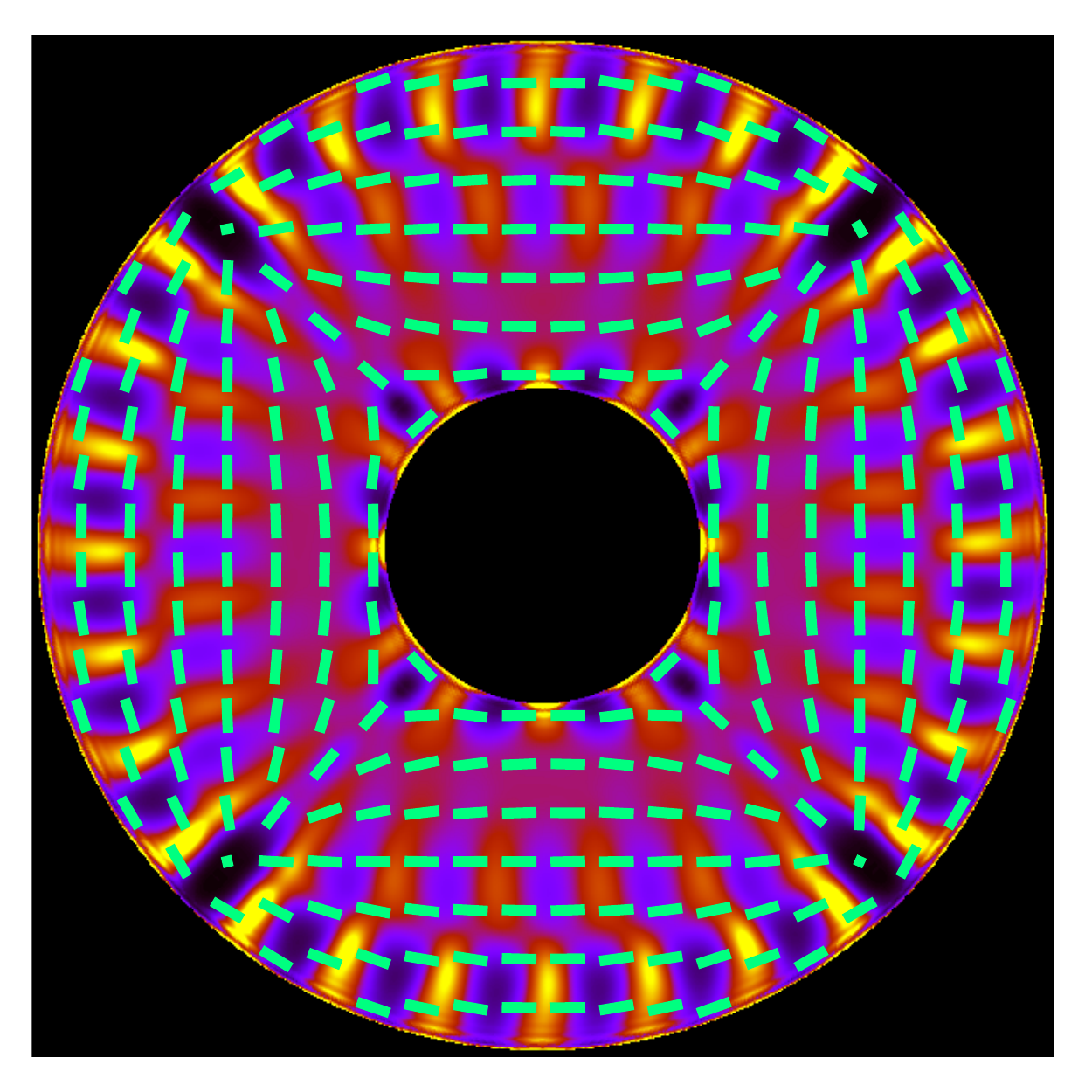}\hfill
\includegraphics[height=0.17\linewidth]{legend015new.pdf}\\\rotatebox{90}{\large$\ \ \ \ \ \ \ \ \ \: D_{5}$}\hfill
\includegraphics[width=0.17\linewidth]{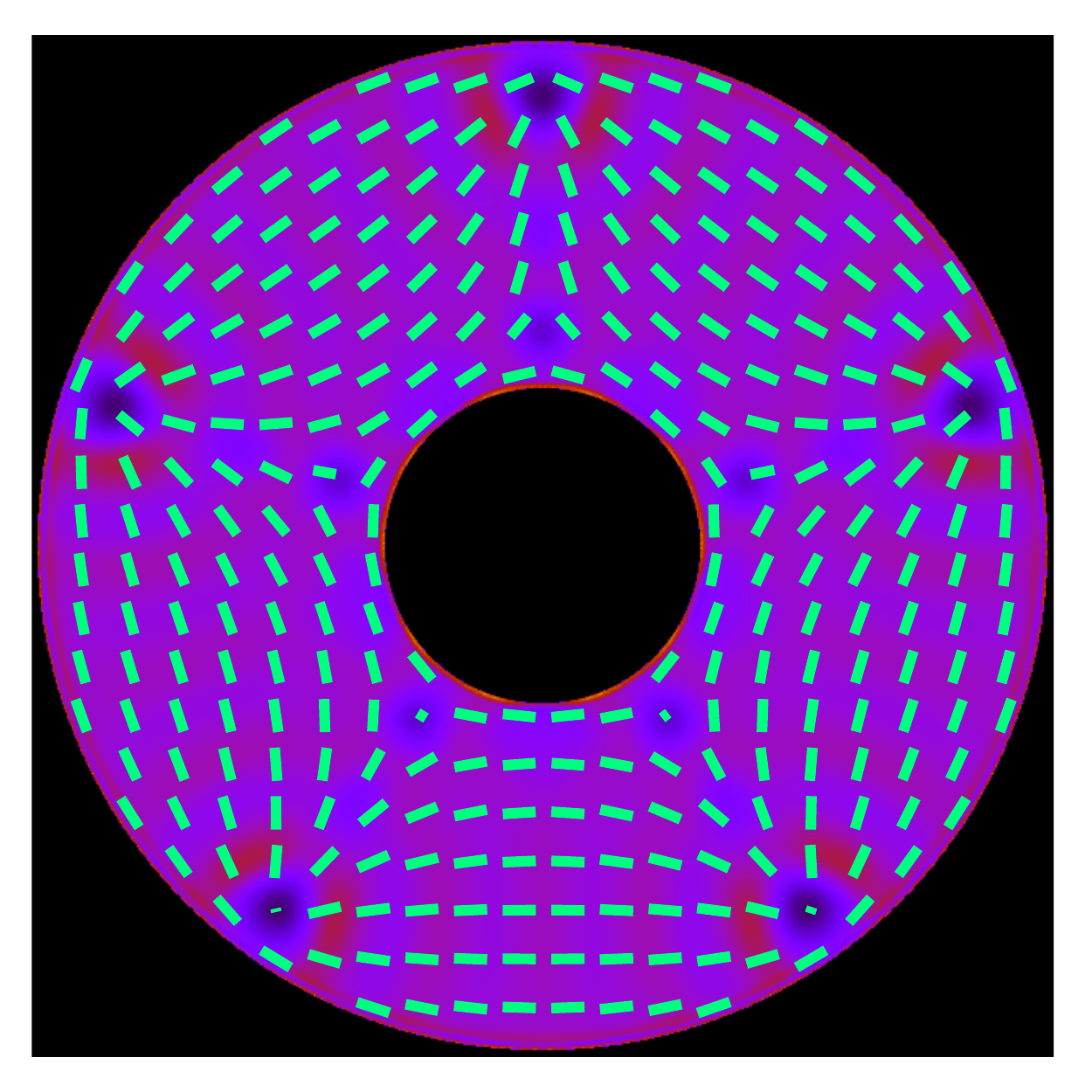}\hfill\parbox{0.015\linewidth}{\centering\vspace*{-2.85cm}$\rightarrow$}\hfill
\includegraphics[width=0.17\linewidth]{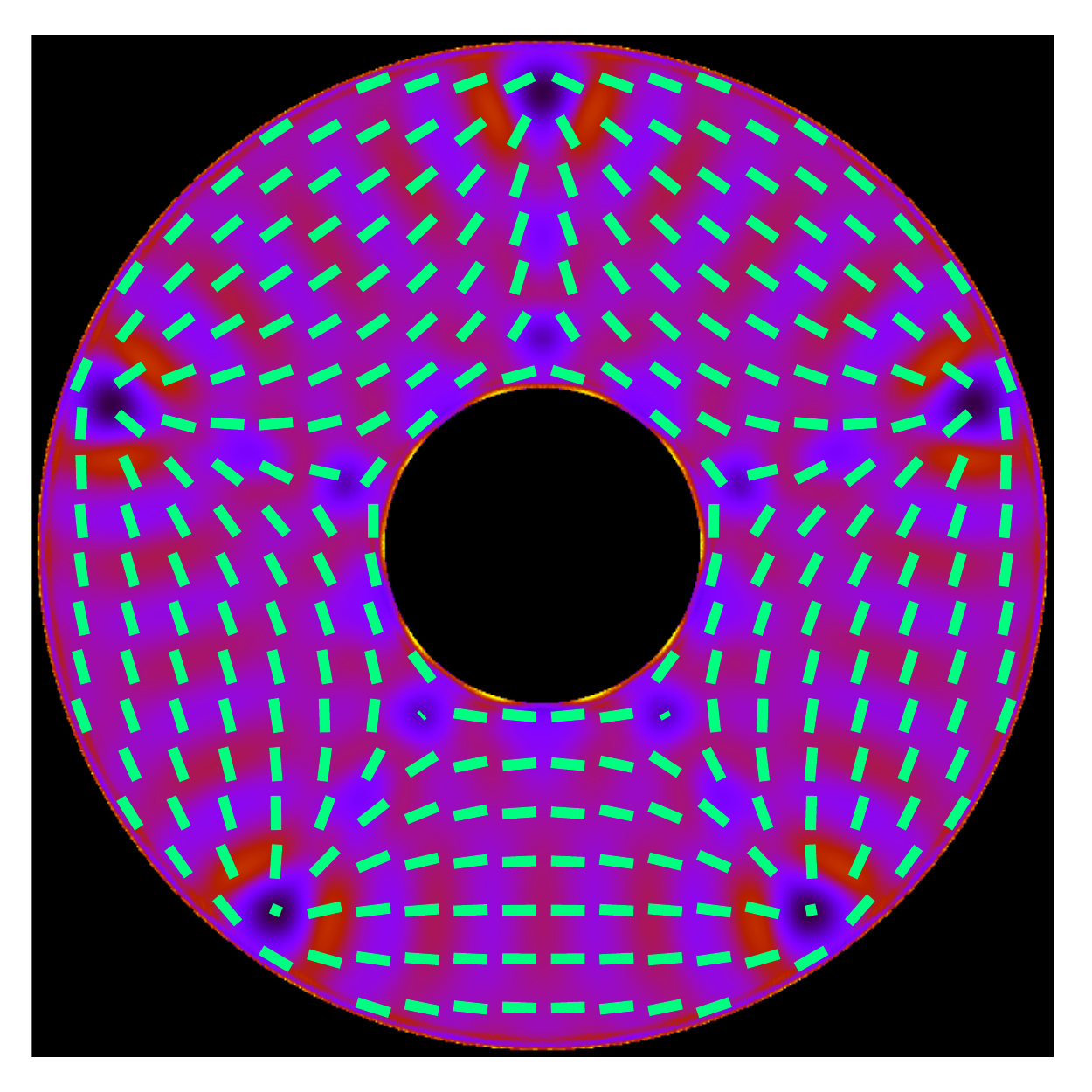}\hfill\parbox{0.015\linewidth}{\centering\vspace*{-2.85cm}$\rightarrow$}\hfill
\includegraphics[width=0.17\linewidth]{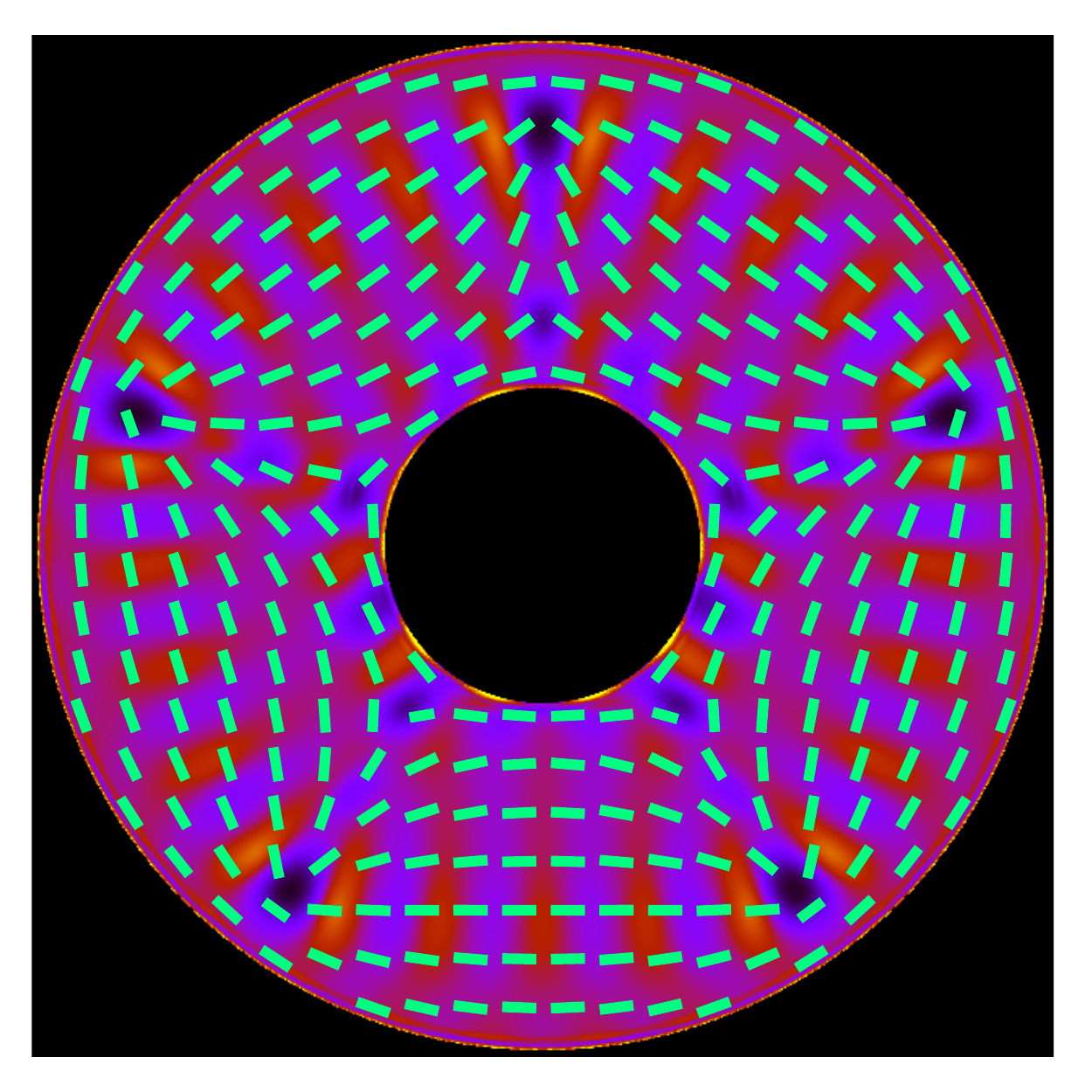}\hfill\parbox{0.015\linewidth}{\centering\vspace*{-2.85cm}$\rightarrow$}\hfill
\includegraphics[width=0.17\linewidth]{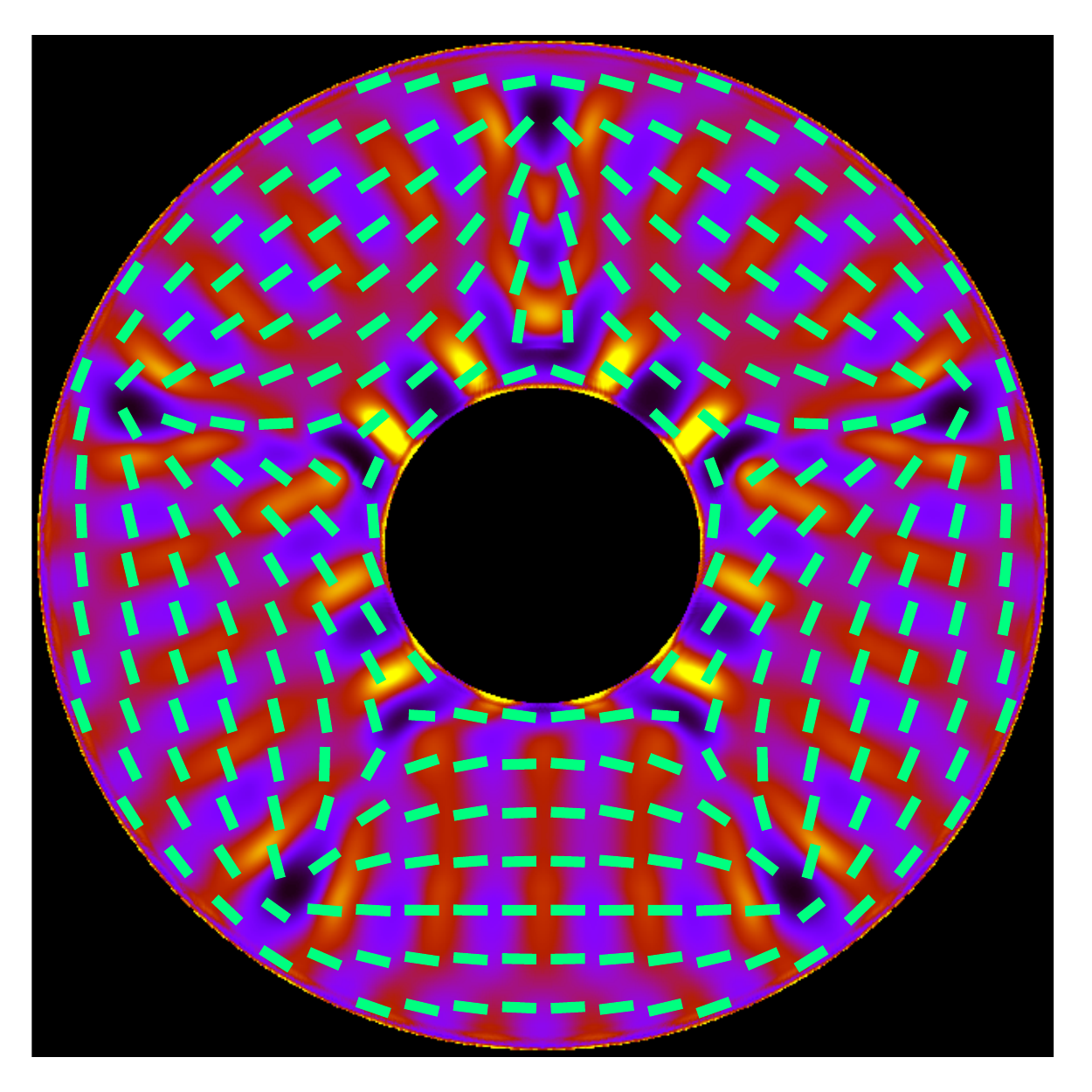}\hfill\parbox{0.015\linewidth}{\centering\vspace*{-2.85cm}$\color{white}\rightarrow$}\hfill
\hspace*{-0.3cm}\includegraphics[height=0.17\linewidth]{legend015new.pdf}\hspace*{0.3cm}\hfill
\includegraphics[width=0.17\linewidth]{DUMMY.pdf}\\
\hspace*{0.02875\linewidth} $\eta=0.5$ \hspace*{0.1125\linewidth} $\eta=0.55$ \hspace*{0.1125\linewidth} $\eta=0.58$ \hspace*{0.1125\linewidth} $\eta=0.6$ \hspace*{0.1125\linewidth} $\eta=0.62$ \hspace*{1.15cm}
\vspace*{0.025cm}\caption{ \textbf{Onset of smectic order emerging from nematic states.} Shown are the density profiles initialized for $R_\text{out}=6.3L$ and $b=0.3$ in different nematic states $D_n$ and those emerging upon increasing the density in discrete steps until the bulk transition density $\eta=0.62$ is reached (see labels). Not all structures are fully equilibrated, such that the otherwise unstable structures with $n=4$ and $n=5$ defect pairs could also be included. Color bar and arrows denote the orientationally averaged density and the director field, compare Eqs.~\eqref{eq_barrho} and~\eqref{eq_QQ}, respectively. Note that the color bar only has half the range as in the all other density plots throughout the manuscript. 
\label{fig_nem}}
\end{figure*}

{

\subparagraph*{Relation to nematic states in an annulus. }

As discussed in the main text, some of the observed smectic states possess topologically equivalent nematic states at lower density.
Here, we elaborate on two further questions.
 Can we infer the existence of additional smectic states from nematic ones with the same symmetry?
How does the transition from nematic to smectic states occur upon increasing the density?

To answer the first question, we undertook some efforts to stabilize a three-fold symmetric smectic structure
that behaves to the nematic $D_3$ state \cite{garlea2016finite} like the laminar state to the nematic $N_2$.
This means that there are three anti-radial disclination lines of charge $q=1/2$
and three regions of $q=-1/2$ charged misalignment at the inclusion,
whose winding number would be $k=-1/2$, compare appendix~\ref{SN5}.
We show in Fig.~\ref{fig_S3} that such a state indeed exists.
However, it is highly metastable as the layers are subject to a strong bend.
Even with this expensive deformation, there is not enough space for the layers hosting the rods oriented tangentially to the inclusion, 
to extend towards the outer wall without forming additional defects.
We can only speculate that such a structure can be stabilized in different geometries possessing a
threefold symmetry, as, for example, one with hexagonal walls.

The second question is addressed in Fig.~\ref{fig_nem}, where we created five distinct nematic states at an area fraction $\eta=0.5$ 
and subsequently increased the density toward the bulk nematic--smectic transition threshold $\eta=0.62$ in our theory.
Note that most of the shown structures are not fully equilibrated, as for example, the nematic states $D_4$ and $D_5$ would destabilize at the low nematic densities.
This procedure allows us to mimic, in a rough way, the nonequilibrium densification observed during the sedimentation process in the experiment.
We make two main observations.
First, the density at which we observe an onset of smectization depends on the nematic state.
More precisely, it appears that this density is lower if the deformations of the nematic director field are stronger, as in $D_0$ and $D_5$.
Second, the initially emerging smectic patterns still follow the orientational director field of the nematic states and thus become frustrated in their positional order.
This nicely underlines that the distinct director field of the equilibrated smectic structures is favorable
and underlines our suspicion that, in experimental sedimentation equilibrium, the structure of the nematic states is influenced by the smectic states below and not vice versa.

}

\section*{Acknowledgements}
We thank Christoph E.\ Sitta for implementing large parts of the DFT code, Paul A.\ Monderkamp for helpful discussions and providing a plotting tool, and Axel Voigt for helpful discussions. 
This work was supported by the German Research Foundation (DFG) within project LO 418/20-2.
 This project has received funding from the European Union’s Horizon 2020 research and innovation programme under the Marie Sk\l{}odowska-Curie Grant Agreement No 641839.

\bibliographystyle{naturemag}


\end{document}